\documentstyle[12pt]{report}

  \def\singlespacing{\parskip 5 pt plus 1 pt \baselineskip 13 pt
      \lineskip 7 pt \normallineskip 7 pt}
  \def\doublespacing{\parskip 5 pt plus 1 pt \baselineskip 25 pt
      \lineskip 13 pt \normallineskip 13 pt}

\begin{document}
\def\slash{\not{}{\mskip-3.mu}}
\def\ra{\rightarrow}
\def\lra{\leftrightarrow}
\def\bea{\begin{eqnarray}}
\def\eea{\end{eqnarray}}
\def\ena{\end{eqnarray}}
\def\beq{\begin{equation}}
\def\enq{\end{equation}}
\def\bec{\begin{center}}
\def\enc{\end{center}}
\def\cw{\cos \theta_W}
\def\sw{\sin \theta_W}
\def\tw{\tan \theta_W}
\def\eg{${\it e.g.}$}
\def\ie{${\it i.e.}$}
\def\etc{${\it etc}$}
\def\kln{\kappa_{L}^{NC}}
\def\krn{\kappa_{R}^{NC}}
\def\klc{\kappa_{L}^{CC}}
\def\krc{\kappa_{R}^{CC}}
\def\ttz{{\mbox {\,$t$-${t}$-$Z$}\,}}
\def\bbz{{\mbox {\,$b$-${b}$-$Z$}\,}}
\def\tta{{\mbox {\,$t$-${t}$-$A$}\,}}
\def\bba{{\mbox {\,$b$-${b}$-$A$}\,}}
\def\tbw{{\mbox {\,$t$-${b}$-$W$}\,}}
\def\tbW{{\mbox {\,$t$-${b}$-$W$}\,}}
\def\tltlz{{\mbox {\,$t_L$-$\overline{t_L}$-$Z$}\,}}
\def\blblz{{\mbox {\,$b_L$-$\overline{b_L}$-$Z$}\,}}
\def\brbrz{{\mbox {\,$b_R$-$\overline{b_R}$-$Z$}\,}}
\def\tlblw{{\mbox {\,$t_L$-$\overline{b_L}$-$W$}\,}}
\def\pbarp{ \bar{{\rm p}} {\rm p} }
\def\pp{ {\rm p} {\rm p} }
\def\ipb{ {\rm pb}^{-1} }
\def\ifb{ {\rm fb}^{-1} }
\def\stds{\strut\displaystyle}
\def\SST{\scriptscriptstyle}
\def\TT{\textstyle}
\def\ra{\rightarrow}
\def\cro{\cropen{12pt}}
\def\mf{m_f}
\def\mb{m_b}
\def\mt{m_t}
\def\MW2{M^2_W}
\def\MZ{M_Z}
\def\Cw{C_w}
\def\Sw{S_w}
\def\mHn{m_{\SST H^{\SST 0}} }
\def\mh0{m_{\SST h^{\SST 0}} }
\def\mHp{m_{\SST H^{\pm}} }
\def\mA0{m_{\SST A^{\SST 0}} }
\def\mH{m_{\SST H} }
\def\mHs{m^2_{\SST H} }
\def\qq{q_1 \bar{q}_2}
\def\jj{j_1 j_2}
\def\ee{e^+ e^-}
\def\ptW{P_{\SST T}^{\SST W} }
\def\DR{\Delta R}
\def\fL{f_{\SST L}}
\def\yW{y_{\SST W}}
\def\qs{\theta^\ast}
\def\MWW{M_{\SST WW}}
\def\EWQCD{(1.1)~}
\def\eg{${\it e.g.}$}
\def\ie{${\it i.e.}$}
\def\etc{${\it etc}$}
\def\etal{${\it et al.}$}
\def\hatt{ \hat {\rm T} }
\def\ETslash{\not{\hbox{\kern-4pt $E_T$}}}
\def\mynot#1{\not{}{\mskip-3.5mu}#1}
\def\sss{\scriptscriptstyle}
\def\rtS{\sqrt{S}}
\def\ra{\rightarrow}
\def\d{{\rm d}}
\def\M {{\cal M}}   
\def\qgtb{q' g \ra q t \bar b}
\def\Wgtb{q' g (W^+ g) \ra q t \bar b}
\def\ubdt{q' b \ra q t}
\def\udbt{q' \bar q \ra W^* \ra t \bar b}
\def\Wbt{W^+ b \ra t}
\def\ggtt{q \bar q, \, g g \ra t \bar t}
\def\ttb{t \bar t}
\def\Wt{W t}
\def\gbtW{g b \ra W^- t}
\def\width{\Gamma(t \ra b W^+)}
\def\tevs{Di-TeV}
\def\flong{f_{\rm Long}}
\def\ltap{\;\centeron{\raise.35ex\hbox{$<$}}{\lower.65ex\hbox{$\sim$}}\;}
\def\gtap{\;\centeron{\raise.35ex\hbox{$>$}}{\lower.65ex\hbox{$\sim$}}\;}
\def\gsim{\mathrel{\gtap}}
\def\lsim{\mathrel{\ltap}}
\def\del{\partial }
\def\D0{D\O~}

\def\bec{\begin{center}}
\def\enc{\end{center}}
\def\tbw{{\mbox {\,$t$-${b}$-$W$}\,}}
\def\tbW{{\mbox {\,$t$-${b}$-$W$}\,}}
\def\klc{{\kappa_{L}}^{CC}}
\def\krc{{\kappa_{R}}^{CC}}
\def\ra{\rightarrow}
\def\st{{\sin\theta}}
\def\ct{{\cos\theta}}
\def\sp{{\sin\phi}}
\def\cp{{\cos\phi}}
\def\M {{\cal M}}
\def\veps{{\varepsilon}}
\def\slash{\not{}{\mskip-3.mu}}
\newcommand{\myabs}[1]{{| {\vec{#1}} \, |}}
\newcommand{\br}[1]{{\langle {#1} |}}
\newcommand{\kt}[1]{{| {#1} \rangle}}
\newcommand{\bra}[2]{{\langle \, {\hat{#1}} \, {#2} |}}
\newcommand{\ket}[2]{{| \, {\hat{#1}} \, {#2} \rangle}}
\newcommand{\bk}[2]{{\langle \, {#1} | \, {#2} \rangle}}
\newcommand{\braket}[4]{{\langle \, {\hat{#1}} \, {#2} | \, {\hat{#3}} \, {#4} \rangle}}
\newcommand{\bsk}[5]{{\langle \, {\hat{#1}} \, {#2} | \slash {#3} \,
| \, {\hat{#4}} \, {#5} \rangle}}
\newcommand{\bssk}[6]{{\langle \, {\hat{#1}} \, {#2} | \slash {#3} \slash {#4} \, 
| \, {\hat{#5}} \, {#6} \rangle}}
\newcommand{\bsssk}[7]{{\langle \, {\hat{#1}} \, {#2} | \slash {#3} \slash {#4}\slash {#5} \, | \, {\hat{#6}} \, {#7} \rangle}}
\newcommand{\vect}[2]{\pmatrix{{#1}\cr{#2}\cr}}
\newcommand{\mat}[4]{\pmatrix{{#1}&{#2}\cr{#3}&{#4}\cr}}
\def\sone{\mat{0}{1}{1}{0}}
\def\stwo{\mat{0}{-i}{i}{0}}
\def\sthr{\mat{1}{0}{0}{-1}}
\def\gamz{\mat{0}{1}{1}{0}}
\def\gamj{\mat{0}{-\sigma_j}{\sigma_j}{0}}
\def\gam5{\mat{1}{0}{0}{-1}}
\def\Ppm{{1 \over 2} (1 {\pm} \gamma^5)}
\def\Pp{\mat{1}{0}{0}{0}}
\def\Pm{\mat{0}{0}{0}{1}}

\def\klc{{\kappa_{L}}^{CC}}
\def\krc{{\kappa_{R}}^{CC}}
\def\ra{\rightarrow}
\def\st{{\sin\theta}}
\def\ct{{\cos\theta}}
\def\sp{{\sin\phi}}
\def\cp{{\cos\phi}}
\def\M {{\cal M}}
\def\slash{\not{}{\mskip-3.mu}}
\def\gamz{\mat{0}{1}{1}{0}}
\def\gamj{\mat{0}{-\sigma_j}{\sigma_j}{0}}
\def\gam5{\mat{1}{0}{0}{-1}}
\def\Ppm{{1 \over 2} (1 {\pm} \gamma^5)}
\def\Pp{\mat{1}{0}{0}{0}}
\def\Pm{\mat{0}{0}{0}{1}}
\def\rts{\sqrt{s}}
\def\sh{\hat{s}}
\def\rsh{\sqrt{\hat{s}}}

\def\phip{e^{i\phi}}
\def\phipsq{e^{2i\phi}}
\def\phim{e^{-i\phi}}
\def\phimsq{e^{-2i\phi}}
\def\vtwo{\sqrt{2}\, }
\def\vmt{\sqrt{m_t}\, }
\def\vmttwo{\sqrt{2m_t}\, }
\def\vbp{\sqrt{E_b+p}\, }
\def\vbm{\sqrt{E_b-p}\, }
\def\epplus{(E_b+p)}
\def\epminus{(E_b-p)}
\def\ra{\rightarrow}
\def\Wtb{W-t-b}
\def\Lag{L}
\def\MWW{M_{WW}}
\def\tow{{m_t \over M_W}}
\def\fol{f_1^L}
\def\for{f_1^R}
\def\cpyftl{f_2^L}
\def\cpyftr{f_2^R}
\def\ctt{\cos{\theta \over 2}}
\def\stt{\sin{\theta \over 2}}
\def\rt{\sqrt{2}}
\def\klc{{\kappa_{L}}^{CC}}
\def\krc{{\kappa_{R}}^{CC}}
\def\ra{\rightarrow}
\def\st{{\sin\theta}}
\def\ct{{\cos\theta}}
\def\sp{{\sin\phi}}
\def\cp{{\cos\phi}}
\def\M {{\cal M}}
\def\flong{f_{\rm Long}}
\def\veps{{\varepsilon}}
\def\slash{\not{}{\mskip-3.mu}}
\def\gamz{\mat{0}{1}{1}{0}}
\def\gamj{\mat{0}{-\sigma_j}{\sigma_j}{0}}
\def\gam5{\mat{1}{0}{0}{-1}}
\def\Ppm{{1 \over 2} (1 {\pm} \gamma^5)}
\def\Pp{\mat{1}{0}{0}{0}}
\def\Pm{\mat{0}{0}{0}{1}}

\def\klc{{\kappa_{L}}^{CC}}
\def\krc{{\kappa_{R}}^{CC}}
\def\ra{\rightarrow}
\def\st{{\sin\theta}}
\def\ct{{\cos\theta}}
\def\sp{{\sin\phi}}
\def\cp{{\cos\phi}}
\def\M {{\cal M}}
\def\ugtb{q' g \ra q t \bar b}
\def\gutb{g q' \ra q t \bar b}
\def\ubdt{q' b \ra q t}
\def\budt{b q' \ra q t}
\def\veps{{\varepsilon}}
\def\slash{\not{}{\mskip-3.mu}}

\def\ra{\rightarrow}
\def\st{{\sin\theta}}
\def\ct{{\cos\theta}}
\def\sp{{\sin\phi}}
\def\cp{{\cos\phi}}
\def\M {{\cal M}}
\def\shat{\hat{s}}
\def\that{\hat{t}}
\def\uhat{\hat{u}}
\def\shat2{\hat{s}^2}
\def\that2{\hat{t}^2}
\def\uhat2{\hat{u}^2}

\newcommand{\xt}{{x_t=2p_t/\sqrt{s}}}

\def\be{\begin{equation}}
\def\ee{\end{equation}}
\def\bea{\begin{eqnarray}}
\def\eea{\end{eqnarray}}
\def\pbarp{ \bar{{\rm p}} {\rm p} }
\def\pp{ {\rm p} {\rm p} }
\def\ifb{ {\rm fb}^{-1} }
\def\del{\partial }
\def\to{\rightarrow}
\def\To{\Rightarrow}
\def\dis{\displaystyle}
\def\f{\frac}
\def\Psibar{\bar{\Psi}}
\def\ppbar{p\bar{p}}
\def\bbar{\bar{b}}
\def\tbar{\bar{t}}
\def\phibb{\phi b \bar{b}}
\def\hbb{\phi b \bar{b}}
\def\bbbb{b \bar{b} b \bar{b}}
\def\bbjj{b \bar{b} j j}
\def\zbb{Z b \bar{b}}
\def\ifb{${\rm fb}^{-1}$}
\newcommand{\tanb}{\tan\hspace*{-0.9mm}\beta}
\newcommand{\cotb}{\cot\beta}
\newcommand{\sinb}{\sin\beta}
\newcommand{\cosb}{\cos\beta}
\newcommand{\cosa}{\cos\alpha}
\newcommand{\sina}{\sin\alpha}
\newcommand{\cosba}{\cos (\beta -\alpha)}
\newcommand{\sinba}{\sin (\beta -\alpha)}

\newcommand{\lae}{\stackrel{<}{\sim}}
\newcommand{\gae}{\stackrel{>}{\sim}}

\input epsf
\pagestyle{plain}
\pagenumbering{roman}

\pagestyle{empty}

\begin{center}
\singlespacing

\vspace*{1.0in}

SIGNALS FOR THE ELECTROWEAK SYMMETRY BREAKING ASSOCIATED WITH THE 
TOP QUARK

\doublespacing

By

Timothy Maurice Paul Tait

\vspace{2.5in}
A DISSERTATION \\
\vspace{0.3in}

Submitted to
\singlespacing

Michigan State University

in partial fulfillment of the requirements

for the degree of

\doublespacing
DOCTOR OF PHILOSOPHY

\vspace{.2in}
Department of Physics and Astronomy
\doublespacing

1999
\end{center}



\newpage
\doublespacing

\pagestyle{empty}
\begin{center}
ABSTRACT \\ [15pt]

{
SIGNALS FOR THE ELECTROWEAK SYMMETRY BREAKING ASSOCIATED WITH THE
TOP QUARK } \\[30pt]

By \\ [12pt]

Timothy Maurice Paul Tait \\ [30pt]

The mechanism of the electroweak symmetry-breaking (EWSB) is studied in the
context of the heavy top quark, whose large mass may provide a clue
as to the mechanism which generates the mass of the $W^\pm$ and $Z$
bosons.  As a result, it seems quite likely that the top quark may be special
in the sense that it is involved in dynamics not experienced by the light
fermions.  Examples of this include models such as the top-condensate model
in which a bound state of top quarks condenses, generating both the top mass
and the gauge boson masses, and supersymmetric models in which the large
top Yukawa coupling naturally explains the EWSB by radiatively driving the
squared mass of a scalar particle (which is positive at a large energy
scale) negative at low energies.  Specific collider signatures of the
third family result from such scenarios, and can be used to test the
hypothesis that the top plays a role in the EWSB.  In particular, single top
production, as a measure of the top's weak interactions, provides an
excellent probe of nonstandard top quark properties.  The physics of single
top production at hadron colliders is carefully studied, with a particular
eye towards what can be learned from single top, including
the signs of new physics that may show up in the single top rate.
\end{center}

\newpage


\pagestyle{plain}
\pagenumbering{roman}
\vspace*{2.8in}
\begin{center}
This dissertation is dedicated to those who have contributed 
indirectly to its existence by
providing much-needed support and encouragement.  Among many others,
this includes 
my grand-parents, Maurice and Jean Jones, and Mimi,
who taught me to see the beauty inherent in the world;
and
my Parents, Susan and Peter Tait, and Rex and Maureen Daysh,
who taught me how to live in it.
Finally, this work
is for Simona; you are the light of my life.  Each day you are in my life
makes it brighter, and in the end it is this that makes it worth living.
\end{center}
\setcounter{page}{3}
\newpage

\begin{center} {\large ACKNOWLEDGEMENTS}  \end{center}

\bigskip

It is a pleasure to acknowledge those people who have helped me grow into
the physicist I am today.  Foremost is my thesis
advisor, C.--P. Yuan, who has shared with me many of his ideas
on which to pursue research, and encouraged me to learn interesting
and invaluable ideas, techniques, and skills.  
Ed Berger at Argonne National Lab has also provided invaluable guidance in 
research, encouragement to pursue technically challenging calculations, 
and has supported me financially when the means to do so did not exist
at Michigan State University.   I have further benefitted from the
exposure to high energy physicists which have given me very important
information, advice, and a context in which to fit my own work.
At Michigan State University
this includes Carl Schmidt, Wu-Ki Tung, Wayne Repko, Dan Stump, 
Jon Pumplin, Maris Abolins, Raymond Brock, and Joey Huston.
At Argonne, it includes Cosmas Zachos, Alan White,
Geoff Bodwin, Don Sinclair,
and Eve Kovacs.  All have had a large impact on my research.
I am also grateful to my committee members Vladimir Zelevinsky and 
William Hartmann for useful guidance.

I am further grateful
to Michael Klasen, Stephen Mrenna, Francisco Larios, Lorenzo Diaz-Cruz, 
Ehab Malkawi, Hong-Jian He, and Csaba Balazs, my collaborators,
for fun and interesting work, and to
my colleagues in physics, Simona Murgia, Phil Tsai, Andy Lloyd,
Hal Widlansky,
Doug Carlson, Mike Wiest, Xiaoning Wang, Hung-Liang Lai, David Bowser-Chao,
Kate Frame, Jim Amundson, Glenn Ladinsky, Chris Glosser, Pavel Nadolsky,
Rocio Vilar, Gervasio Gomez, Miguel Mostafa, Juan Valls, Andrea Petrelli,
Lionel Gordon, Jean-Fran\c{c}ios Lag\"ae, Carmine Pagliarone,
Zack Sullivan, Brian Harris, and Gordon Chalmers
for many hours of interesting discussions.

\newpage


\singlespacing
\tableofcontents
\clearpage

\addcontentsline{toc}{chapter}{LIST OF TABLES}
\listoftables
\clearpage

\addcontentsline{toc}{chapter}{LIST OF FIGURES}
\listoffigures
\clearpage


\doublespacing
   \pagestyle{plain}
   \pagenumbering{arabic}
   \setcounter{page}{1}
   \makeatletter
   \def\@evenfoot{}
   \def\@evenhead{\hfil\thepage\hfil}
   \def\@oddhead{\@evenhead}
   \def\@oddfoot{\@evenfoot}
   \makeatother

\chapter{Introduction : The Standard Model}
\label{intro}

The Standard Model of Particle Physics \cite{smref, smtasi}
(SM) contains, in principle,
a complete description of all of the phenomena currently observed in
high energy physics experiments, including the strong and electroweak
interactions.  However, as will be explained below, the model contains
a number of theoretical puzzles which indicate that it is not the
``ultimate" theory, but 
instead should be replaced by some more fundamental
theory at higher energy scales.  In fact, the striking success of the
SM at explaining the currently 
available data places strong constraints on
the nature of any theory that hopes to extend or 
supplant it at higher energies.  As this work is an examination of
several such models, we will begin with a presentation of the SM,
examining its strengths and short-comings, in order to better understand
these theories which hope to replace it.  We shall see that the one of the
great mysteries of the SM is the mechanism for the electroweak symmetry
breaking (EWSB) that provides masses for both the weak bosons and the
fermions.  Because of their large masses, the fermions of the third
family, and the top quark in particular, provide a natural place to
explore hypotheses concerning the EWSB.  In the succeeding chapters
we examine specific ways in which experiments at supercolliders can study
the possibility of a connection between the EWSB and heavy top through
single top production,
production of scalars in association with bottom quarks,
and supersymmetric particle production.

The SM is a quantum field theory of fermionic matter particles interacting 
with bosonic vector particles.  The interactions are fixed by requiring
the theory to be locally gauge invariant under transformations in 
the group
${\rm SU(3)}_{C} \times {\rm SU(2)}_{L} \times {\rm U(1)}_{Y}$.
It is quite remarkable that this condition is enough to uniquely fix 
the structure
of the renormalizable interactions between fermions and vector
particles.

\section{Yang-Mills Gauge Theory}
\label{yangmills}

We begin with a brief presentation of the construction of a Lagrangian
invariant under local gauge transformations, as these ideas form
the basic building blocks of the SM.
In the Yang-Mills gauge theory \cite{yangmills} invariant under
Lie group ${\cal G}$ with $N$ group generators $T^a$ ($a=1..N$),
we can express the generators as Hermitian matrices with commutators,
\bea
   [ T^a , T^b ] = i \; f^{a b c} \; T^c ,
\eea
where the $f^{a b c}$ are the structure constants of ${\cal G}$.  An element
of the local gauge transformation acting on a set of Dirac fermions
may be expressed as
\bea
  \Psi (x) \ra e^{i \alpha^a(x) T^a} \; \Psi (x),
\eea
where the real function $\alpha^a(x)$ is the
local transformation parameter.  Clearly the
usual free field Lagrangian density for a set of Dirac spinors 
is not invariant under this transformation, because the transformation
of $\partial_\mu \Psi(x)$ will generate a term in which the derivative
acts on $\alpha^a(x)$.  This is remedied by introducing a covariant derivative,
\bea
  D_\mu = \partial_\mu + i \; g \; T^a \, A^a_\mu (x) ,
\eea
which insures that $D_\mu \Psi$ transforms like $\Psi$ under the gauge
group provided that the real vector field $A^a_\mu (x)$ 
transforms according to,
\bea
  T^a \, A^a_\mu (x) \ra e^{i \alpha^a(x) T^a} 
  \left\{ T^b \, A^b_\mu (x) - \frac{i}{g} \partial_\mu \right\}
  \left( e^{i \alpha^c(x) T^c} \right)^\dagger .
\eea
This allows us to write down the gauge-invariant kinetic terms for the Dirac
fermion, 
\bea
   \label{FK}
   {\cal L}_{\rm FK} = i \, \overline{\Psi} \, \gamma^\mu \, D_\mu \, \Psi 
   - m \, \overline{\Psi} \, \Psi ,
\eea
where for brevity we no longer explicitly write the fields
as functions of space-time.
The presence of the covariant derivative, dictated by the gauge invariance,
has thus forced us to include an interaction between the fermion fields
$\Psi$ and the vector fields $A^a$.
It should be noted that
a four-component Dirac spinor may be written in terms of a
left-handed and a right-handed two-component
Weyl spinor \cite{peskin}.  
One can formulate a theory of massless fermions in
two-component form, which proceeds much as it is described above, but with
no mass term in ${\cal L}_{\rm FK}$.

In order for the vector field to be dynamical it must also have kinetic
terms in the Lagrangian.  It is easy to verify that,
\bea
 {\cal L}_{\rm GK} = -\frac{1}{4} {F^a}^{\mu \nu} \, {F^a}_{\mu \nu}
\eea
with,
\bea
 {F^a}_{\mu \nu} = \partial_\mu A^a_\nu - \partial_\nu A^a_\mu
                   - g \, f^{a b c} \, A^b_\mu \, A^c_\nu
\eea
will serve\footnote{A term of the form 
${\theta \over 2} {{F}^a}^{\mu \nu} {\tilde{F}}^a_{\mu \nu}$ with 
${\tilde{F}^a}_{\mu \nu} = 1 / 2 \epsilon_{\mu \nu \alpha \beta}{F^a}^{\alpha \beta}$ 
is also gauge 
invariant, and could be included in ${\cal L}_{\rm GK}$.  For an Abelian
group ${\cal G}$ this term corresponds to a total derivative, and thus
does not contribute to the dynamics.  In the case
of a non-Abelian group this term is
related to the $CP$ properties of the theory.
For simplicity we will not consider
such a term here.}, 
and itself respects gauge invariance, with $g$ the same coupling
that appears in the covariant derivative for the fermion.
In the case in which ${\cal G}$ is a non-Abelian group (and thus the
structure constants do not vanish), this term will contain cubic and 
quadratic interactions of the gauge field, in addition to the kinetic
terms.  It is important to note that ${\cal L}_{\rm GK}$ does not contain
a mass term for the vector field.  In fact, such a term is forbidden by
gauge invariance, and thus the vector fields are necessarily massless as a
result of the gauge symmetry.  This theory may now be quantized by
employing, i.e., the Faddeev-Popov formalism \cite{faddeevpopov}
to quantize only the physical degrees of freedom.

In addition to fixing the form of the renormalizable interactions,
the gauge symmetry plays a further role in the construction of a model in that
it corresponds to conserved currents as predicted by Noether's 
theorem\footnote{Noether's theorem guarantees that a continuous
symmetry corresponds to a conserved current at the classical level.
Some symmetries, known as anomalous symmetries, are broken by quantum
effects.  The requirement that a desired classical symmetry survives
quantization can provide non-trivial
constraints on the theory.} \cite{noether},
and implies relations among the Green's functions known as Ward-Takahashi
identities \cite{wardid}.
The Ward identities, and thus the gauge symmetry itself,
play an important role in the proof of the 
decoupling of the ghost states from physical amplitudes \cite{wdecoup},
and in proving the unitarity and
renormalizability \cite{wardid, wdecoup, wrenorm} of the Yang-Mills theory.
The gauge symmetry is therefor
seen as an essential ingredient for a theory of
vector particles.

\section{The Standard Model}

\subsection{Spin $1$ Gauge Boson Fields}
\label{vbosons}

Having briefly reviewed the general Yang-Mills theory of a 
set of fermions interacting with gauge bosons as is dictated by local
invariance under a symmetry group ${\cal G}$, we now specify to the
SM, with symmetry group
${\cal G} = {\rm SU(3)}_C \times {\rm SU(2)}_L \times {\rm U(1)}_Y$.
The ${\rm SU(3)}_C$ gauge symmetry corresponds to the strong interaction
of quantum chromodynamics (QCD).  Its eight
gauge bosons $G^a_{\mu}$ are known as
gluons.  The ${\rm SU(2)}_L \times {\rm U(1)}_Y$ sector contains the
combined
electromagnetic and weak interactions, generally referred to as the
electroweak symmetry.  The three ${\rm SU(2)}_L$ bosons are
denoted $W^i_\mu$ and couple to the weak iso-spin, whereas
the ${\rm U(1)}_Y$ gauge boson, $B_\mu$ couples to hypercharge.
Since the gauge bosons of one of the symmetry subgroups 
in ${\cal G}$
do not transform under the other gauge symmetries
in the product of groups, the gauge kinetic term
may be simply written as a sum of the individual gauge kinetic terms,
\bea
{\cal{L}}_{GK} = -{1\over 4} {B}_{\mu \nu} {B}^{\mu \nu}
       -{1\over 4} {W}^i_{\mu \nu} {W}^{i \mu \nu}
       -{1\over 4} {G}^a_{\mu \nu} {G}^{a \mu \nu},
\eea
where,
\bea
 {B}_{\mu \nu} &=& \partial_\mu B_\nu - \partial_\nu B_\mu ,
 \\
 {W^i}_{\mu \nu} &=& \partial_\mu W^i_\nu - \partial_\nu W^i_\mu
                   - g_2 \, \epsilon^{i j k} \, W^j_\mu \, W^k_\nu ,
 \nonumber \\
 {G^a}_{\mu \nu} &=& \partial_\mu G^a_\nu - \partial_\nu G^a_\mu
                   - g_3 \, f^{a b c} \, G^b_\mu \, G^c_\nu ,
  \nonumber
\eea
with $g_i$ the gauge couplings, and $\epsilon^{i j k}$ and
$f^{a b c}$ the structure constants for ${\rm SU(2)}$
and ${\rm SU(3)}$, respectively.

As will be explained in detail below, the electroweak
symmetry is spontaneously broken, resulting in mixing between the
$B_\mu$ and $W^3_\mu$ fields, and
non-zero masses for three of the
gauge bosons ($W^\pm$ and $Z^0$). The photon ($A$) remains
massless, due to a residual ${\rm U(1)}_{\rm EM}$ gauge
symmetry that remains unbroken.
The physical (mass-eigenstate) gauge bosons 
and their masses are shown in Table~\ref{bm}.

\begin{table}
\caption{Vector Boson Masses}
\label{bm}
\begin{center}
\vspace{.5cm}
\begin{tabular}{lccl}
  Particle &  Symbol   & Mass (GeV)       & \\ \hline \hline \\
  Photon   & $A$       & 0                & Electromagnetic Force\\
  W Boson  & $W^{\pm}$ & 80.33            & Charged Weak Force\\
  Z Boson  & $Z^0$     & 91.187           & Neutral Weak Force\\
  Gluon    & $G^a$     & 0                & Strong Force\\ \\
\hline\hline
\end{tabular}
\end{center}
\end{table}

\subsection{Spin $1 \over 2$ Matter Fields}
\label{matter}

The SM contains three families (also called generations)
of spin $1 \over 2$ matter fields,
in the fundamental representation of the gauge groups.
Each family is a ``copy" of the other families with respect to gauge
quantum numbers, but have diverse masses.  Each generation contains
a charged and a neutral lepton, which interact electroweakly,
and an up-type and a down-type quark, which interact both electroweakly and
with the gluons.  A list\footnote{It is worth mentioning
that the association of a particular doublet of leptons with
a particular doublet of quarks in order to form a generation is arbitrary
in the SM, because there are no local interactions between quarks and
leptons.  The SM makes this identification in a natural way by identifying
the doublet containing the heaviest charged lepton with the doublet
containing the heaviest quarks (and so on), 
but one could in principle associate
any quark doublet with any lepton doublet and call that a family.}
of the fermions, including their
masses, is presented in Table~\ref{fm}.

\begin{table}
\caption{Lepton and Quark Masses}
\label{fm}
\begin{center}
\vspace{.5cm}
\begin{tabular}{lccc}
  Particle          &  Symbol    & Mass (GeV) &\\ \hline \hline \\
  Electron neutrino & $\nu_e$    & 0          &\\
  Electron 	    & $e$        & 0.00051    & First\\
  Up quark 	    & $u$	 & 0.002 to 0.008 & Generation\\
  Down quark        & $d$        & 0.005 to 0.015 &\\ \\
  Muon neutrino     & $\nu_\mu$  & 0          &\\
  Muon 	            & $\mu$      & 0.106      & Second\\
  Charm quark 	    & $c$	 & 1.0 to 1.6 & Generation\\
  Strange quark     & $s$        & 0.1 to 0.3 &\\ \\
  Tau neutrino      & $\nu_\tau$ & 0          &\\
  Tau 	            & $\tau$     & 1.78       & Third\\
  Top quark 	    & $t$	 & 175        & Generation\\
  Bottom quark      & $b$        & 4.1 to 4.5 & \\ \\ \hline\hline
\end{tabular}
\end{center}
\end{table}

In Table~\ref{qn} can be found the
transformation properties of the fermions of the first family
under the gauge groups.  Since the second and third families are copies of
the first family as far as the quantum number assignment is concerned,
only the first family is presented.
The left-
and right-chiral fermions have different quantum numbers, with the left-chiral
fields arranged in doublets,
\bea
  L_L = {\pmatrix{{\nu_e}\cr{e}\cr}}_L, \quad
  Q_L = {\pmatrix{{u}\cr{d}\cr}}_L, 
\label{doublet}
\eea
and the right-chiral fermions are in singlets,
\bea
   e_R,\quad u_R,\quad d_R.
\eea
The gauge invariance under ${\rm SU(2)}_L$
thus forbids the presence of a mass term for the 
fermions.  As we will see below, 
in the SM, fermion masses are generated by the same spontaneous
symmetry-breaking Higgs mechanism that provides mass for the weak bosons.
There is no right-handed neutrino in the SM, and thus the neutrino is
a massless Dirac field.
With respect to the color gauge group of quantum chromodynamics the
quark fields are arranged in triplets,
\bea
   q = {\pmatrix{{q_r}\cr{q_g}\cr{q_b}\cr}} ,
\eea
where we have used the common convention of referring to the ${\rm SU(3)}_C$
indices as red ($r$), green ($g$), and blue ($b$).  

The gauge invariant kinetic Lagrangian for a particular fermion,
$\Psi$, is given in Equation~\ref{FK} with
the covariant derivative given by,
\bea
   D_{\mu} = \partial_{\mu} + i \, g_1 \, {Y\over 2} \, B_{\mu} +
   i \, g_2 \, {\tau^j} \, W^j_{\mu} 
   + i \, g_3 \, {\lambda^a} \, G^a_{\mu},
\label{cov}
\eea
with $Y$ the hypercharge of the fermion, and ${\tau^j}$ and
${\lambda^a}$ the generators of ${\rm SU(2)}$ and ${\rm SU(3)}$
in the representation appropriate for $\Psi$.  The hypercharge has been
normalized such that the electric charge of the fermion is
given by $Q = T^3_L + {Y \over 2}$.

\begin{table}
\caption{Quantum Numbers of the Fermions}
\label{qn}
\begin{center}
\vspace{.5cm}
\begin{tabular}{lcccc}
Chirality   & $Q$ &$T^3_{L}$& $Y$ & $C$\\ \hline \hline \\
${\nu_e}_L$ &  0  &  1/2 & -1   & 0        \\
$e_L$       & -1  & -1/2 & -1   & 0        \\ \\
${u}_L$     & 2/3 &  1/2 & 1/3  & $r,g,b$  \\
$d_L$       &-1/3 & -1/2 & 1/3  & $r,g,b$  \\ \\
$e_R$       & -1  &   0  & -2   & 0        \\ \\
$u_R$       & 2/3 &   0  & 4/3  & $r,g,b$  \\ \\
$d_R$       &-1/3 &   0  &-2/3  & $r,g,b$  \\ \\ \hline\hline
\end{tabular}
\end{center}
\end{table}

\subsection{Masses and the Higgs Mechanism}
\label{higgs}

As we have seen, the gauge symmetries of the SM forbid explicit masses for 
vector bosons (as is true for any gauge theory) and fermions (as is true
in the case of the SM, in which left- and right-chiral fermions transform
differently).  In order to describe the world seen in particle physics
experiments, these objects must acquire masses.  This may be resolved by
introducing a spontaneous breaking of the electroweak symmetry,
through the Higgs Mechanism \cite{higgsmech}.
The spontaneous symmetry-breaking (SSB) occurs when the Lagrangian is 
invariant under the gauge transformations, but the vacuum state does
not respect the symmetry.  In the SM this is accomplished by introducing
a weak iso-spin doublet of complex scalar fields, the Higgs doublet.
This doublet carries hypercharge $+1$, and thus can be expressed as,
\bea
  \label{higgsdef}
  \Phi = \frac{1}{\sqrt{2}} {\pmatrix{{\phi_1 + i \, \phi_2}\cr
				  {\eta + i \, \phi_3}\cr}}
     = \left( e^{i \, \theta^i(x) \tau^i} \right)^\dagger
         {\pmatrix{{0}\cr
         {\sigma(x)}\cr}} .
\eea
where the second form (which displays the space-time dependence
of the four fields $\theta^i$ and $\sigma$ explicitly for clarity)
illustrates an interesting property of the Higgs doublet,
which can be seen by noting that under a
${\rm SU(2)}_L \times {\rm U(1)}_Y$ gauge transformation, the doublet
transforms as,
\bea
 \label{higgstransform}
 \Phi \ra e^{i \, {\beta(x) \over 2} } \, 
 e^{i \, \alpha^i(x) \tau^i} \; \Phi ,
\eea
and thus comparing Equations~\ref{higgsdef} and \ref{higgstransform}
indicates that provided the expectation value of $\sigma$ is 
non-zero (which indicates that SSB has occurred),
one may choose a particular gauge in which three of the four
real degrees of freedom of the Higgs doublet 
vanish.
Since on the one hand
it is possible to ``gauge away'' these fields, while on the other
physical quantities are independent of the choice of gauge,
this indicates that these degrees of freedom are unphysical.
As we shall see below, under SSB these unphysical scalars reappear
as the longitudinal polarizations of the weak bosons.

The scalar field can be given gauge invariant terms in the Lagrangian,
\bea
  \label{lphi}
  {\cal L}_{\Phi}={({D_\mu} \Phi )^{\dagger}}\, ({D^\mu} \Phi )
  - \mu^2 \Phi^\dagger \, \Phi 
  - \lambda \left( \Phi^\dagger \, \Phi \right)^2,
\eea
with,
\bea
  D_\mu = \partial_{\mu} + i \, g_1 \, {1 \over 2} B_{\mu} +
               i \, g_2 \, {\tau^j} W^j_{\mu} ,
\eea
where the first (kinetic) term in Equation~\ref{lphi}
is required by gauge invariance, and the remaining terms
correspond to a mass-like term and a self-interaction of the $\Phi$ field.
These latter two terms together are
generally referred to as the Higgs potential.
One can also construct gauge invariant Yukawa couplings between the doublet
and the fermions,
\bea
  \label{lyukawa}
  {\cal L}_{Yukawa} &=& \sum_{n = 1}^3 \, \sum_{m = 1}^3 \:
    \left( \: y^{n m}_e \; \overline{e}^n_R \; \Phi^\dagger \; L^m_L \;
            + \; {y^{n m}_e}^* \; \overline{L}^m_L \; \Phi \; e^n_R \: \right)
  \\[0.2cm]
  & & + \left( \: y^{n m}_d \; \overline{d}^n_R \; \Phi^\dagger \; Q^m_L \;
            + \; {y^{n m}_d}^* \; \overline{Q}^m_L \; \Phi \; d^n_R \: \right)
  \nonumber \\[0.2cm]
  & & + \left( \: y^{n m}_u \; \overline{u}^n_R \;
              \tilde{\Phi}^\dagger \; Q^m_L \;
            + \; {y^{n m}_u}^* \; \overline{Q}^m_L \; 
              \tilde{\Phi} \; u^n_R \: \right) , \nonumber
\eea
with,
\bea
  \tilde{\Phi} = i \, \tau^2 \, \Phi^* ,
\eea
and the sum over $n$ and $m$ is over the three families of fermions.

Spontaneous symmetry-breaking is exhibited 
by assuming\footnote{It is important to note that $\lambda$ must be positive
in Equation~\ref{lphi} in order for the theory to
possess a stable vacuum.}
$\mu^2 < 0$.
Under these conditions the minimum of the Higgs potential shifts 
(in field space) from $\Phi = 0$ to,
\bea
   \Phi^\dagger \Phi &=& \phi_1^2 + \phi_2^2 + \phi_3^2 + \eta^2
   = \frac{-\mu^2}{\lambda} = v^2 .
\eea
The field thus acquires a non-zero vacuum expectation value (VEV).
Expressing this
condition in terms of the real scalars $\phi^{1..3}$
and $\eta$, we see that
the minimization condition allows for any of these (or a  combination
of them) to carry the VEV.  Choosing $< \eta > = v$,
we expand about $v$,
\bea
\Phi = \pmatrix{{\phi^+}\cr
               {\frac{v + h + i \, \phi_3}{\sqrt{2}} }\cr},
\eea
with $\phi^+ = (\phi^1 + i \, \phi^2) / \sqrt{2}$ and $\eta = v + h$.
Inserting this into Equation~\ref{lphi}, we see after some algebra 
(which can be simplified by working in the unitary gauge, 
$\phi^{1..3} = 0$)
that tree-level
mass terms for the $h$ field and gauge bosons are present,
\vspace{1cm}
\bea
   \label{bosonmass}
   {\cal L}_{\rm MB} &=& - \left(
    \frac{2 \, \lambda \, v^2}{2} \right) \; h^2 \:
     + \: \left( \frac{g_2 \, v}{2} \right)^2 {W^+}^{\mu} \; W^-_{\mu}
   \\
   & & + \left( \frac{v}{2} \right)^2 \left( g_2 W^3_\mu - g_1 B_\mu \right)
     \left( g_2 {W^3}^\mu - g_1 B^\mu \right) , \nonumber
\eea
along with many other interaction terms.  The fields $W^\pm_\mu$ are defined
as $W^\pm_\mu = (W^1_\mu \mp i \, W^2_\mu) / \sqrt{2}$ to 
be electric charge eigenstates.
Physically, the appearance of mass terms for the gauge bosons after SSB can
be explained by the gauge bosons absorbing (``eating'') the
unphysical would-be Goldstone bosons,
$\phi^{1..3}$, which serve as the longitudinal degrees of freedom
that distinguish massive from massless vector fields.
This is referred to as the electroweak symmetry-breaking
and can be denoted ${\rm SU(2)}_L \times {\rm U(1)}_Y \ra 
{\rm U(1)}_{\rm EM}$, because the ${\rm U(1)}$ to which the photon 
corresponds remains unbroken.
An interesting heuristic picture for the
Higgs mechanism is that the Higgs potential generates dynamics which
``fills'' the vacuum with Higgs field.  The resulting masses for the
vector bosons and fermions are then seen as a result of these particles
interacting with this ``medium'' as they move through the vacuum.

As was mentioned previously, the SSB has mixed the $B_\mu$ and $W^3_\mu$
gauge bosons.  The mass eigenstates thus consist of the 
massive $Z$ boson and massless photon,
\bea
 \ \pmatrix{{Z_\mu}\cr {A_\mu}\cr} = \ \pmatrix {{\cw} & {-\sw}\cr
{\sw} & {\cw}\cr}\ \pmatrix {{W_\mu^3}\cr {B_\mu}\cr},
\label{gaugemix}
\eea
with the weak mixing angle $\theta_W$ given by,
\bea
 \tan \theta_W = \frac{ g_1 }{ g_2 }.
\eea
It is conventional to discuss the SM couplings in terms of the
coupling of the photon to the electron, $e$, the weak mixing
parameter, $\sin^2 \theta_W$, and the masses of the $Z$ and Higgs bosons,
$M_Z$ and $M_h$, as opposed to the original couplings, $g_1$, $g_2$,
$\mu^2$, and $\lambda$ in which the theory was formulated.  From the
presentation above, it should be clear how to relate these two sets of
parameters at tree-level.  At higher orders in perturbation theory,
the relations depend on the renormalization scheme.

It is worth noting that the $W^\pm$ and $Z$ mass 
terms in Equation~\ref{bosonmass}
arose from the covariant derivative part of Equation~\ref{lphi} (the term that
was fixed by gauge invariance).  This has two interesting consequences
for the masses generated.  The first is that once the gauge couplings $g_1$
and $g_2$ are specified, the $W^\pm$ and $Z$ masses are determined
by $v$ and the representation of $\Phi$
(this property is general for the Higgs mechanism).  The second
property is specific to the particular quantum numbers assigned to the SM Higgs
doublet; the quantity,
\bea
   \rho = \frac{M_W}{M_Z \cos \theta_W},
\eea
is equal to one at tree level in the SM, and thus provides a test of the SM
realization of SSB compared to other models.

\subsection{ Fermion Masses and the CKM Matrix}
\label{CKM}

We saw in the previous section how the SSB provides masses for the gauge
bosons.  In the SM, the same mechanism provides masses for the fermions
through the Yukawa interactions in Equation~\ref{lyukawa}.  As the
values of these couplings are not fixed by the gauge symmetry, they can
be tuned to correspond to the particular fermion masses observed in nature.
This is complicated by the fact that in general
the interaction eigenstates need not be the same as the
mass eigenstates because of the off-diagonal (in family-space) interactions
between fermions and the Higgs doublet in Equation~\ref{lyukawa}.
In terms of the $3 \times 3$ interaction eigenstate mass matrices,
$M_u$, $M_d$, and $M_e$, the fermion mass terms can be expressed,
\bea
{\cal L}_{\rm MF} =  \bar{\bf u}_R \, M_u \, {\bf u}_L 
               + \bar{\bf d}_R \, M_d \, {\bf d}_L
               + \bar{\bf e}_R \, M_e \, {\bf e}_L + H.c.,
\eea
where $+ H.c.$ indicates the Hermitean conjugate of the preceeding terms.
The bold-faced fermion fields now indicate a vector containing the fields
of a given type for all three families,
\bea
   {\bf u}_L = \pmatrix{{u_L}\cr{c_L}\cr{t_L}\cr}, \qquad
   {\bf d}_L = \pmatrix{{d_L}\cr{s_L}\cr{b_L}\cr}, \qquad
   {\bf e}_L = \pmatrix{{e_L}\cr{\mu_L}\cr{\tau_L}\cr},
\eea
and similar notation for ${\bf u}_R$, ${\bf d}_R$, and ${\bf e}_R$.
To make the connection between ${\cal L}_{\rm MF}$ and 
Equation~\ref{lyukawa} explicit, we present as an example the
the mass matrix for up-type quarks after SSB,
\bea
  M_u = \frac{v}{\sqrt{2}}
  \pmatrix { {y_u^{11}} & {y_u^{12}} & {y_u^{13}} \cr 
             {y_u^{21}} & {y_u^{22}} & {y_u^{23}} \cr
             {y_u^{31}} & {y_u^{32}} & {y_u^{33}} \cr }.
\label{upmass}
\eea

To express the fermion masses in terms of mass eigenstates, one uses
the fact that it is possible to rotate the left- and 
right-chiral fields among the three generations.
A general unitary rotation of the fields may be denoted,
\bea
   {\bf u}_R \ra R_u \; {\bf u}_R, \qquad {\bf d}_R \ra R_d \; {\bf d}_R, 
   \qquad {\bf e}_R \ra R_e \; {\bf e}_R, \\
   {\bf u}_L \ra L_u \; {\bf u}_L, \qquad {\bf d}_L \ra L_d \; {\bf d}_L,
   \qquad {\bf e}_L \ra L_e \; {\bf e}_L , \nonumber
\eea
and the condition for mass eigenstates may be expressed by requiring that
these transformations diagonalize the interaction eigenstate mass matrices.
Employing $D_i$ (with $i = u, d, e$) to indicate
the diagonalized matrix, this may be written,
\bea
   D_u = R_u^\dagger \; M_u \; L_u, \qquad D_d = R_d^\dagger \; M_d \; L_d,
   \qquad D_e = R_e^\dagger \; M_e \; L_e.
\eea
The requirement that the nine free entries in $D_u$, $D_d$, and $D_e$
correspond to the fermion masses observed in nature provides some information
about the Yukawa interactions in Equation~\ref{lyukawa}.  The remaining
information must come from studying the effect of the quark mixing on the
quark interactions with other particles.

Having transformed to mass eigenstates, it is still necessary to examine the
effect of these rotations on the interactions of the fermions with the
vector and Higgs bosons.  From the fact that the mass terms came from the
fermion interactions with the Higgs doublet, it is clear that the interactions
of $h$ with the fermions are diagonal in the mass eigenbasis.
For the gauge bosons, it is simple to show that the left- and right-handed
pieces of the $A$, $G$,
and $Z$ coupling to fermion $f$ are proportional to 
$L_f^\dagger L_f = 1$ and $R_f^\dagger R_f = 1$, respectively.  Thus,
these interactions are the same in mass and interaction eigenbasis.
On the other hand, the $W^\pm$ interactions with the quarks pick up a factor
of $L_u^\dagger L_d$, which allows the $W^\pm$ interactions to couple up-
and down-type quarks of different generations.  The lepton sector has no
equivalent effect, because in the SM the massless neutrinos have no
mass diagonalization requirement, and thus
may always be
rotated such that the $W^\pm$ couplings are diagonal in the generations.

Thus, the only observable matrix related to the generational rotations is
$V = L_u^\dagger L_d$, the Cabibbo-Kobayashi-Maskawa (CKM) matrix
\cite{ckm}.  By convention, the matrix $L_u$ is set to 
the unit matrix, and in that case
$V = L_d$.  Since in the SM only the combination is of physical relevance,
this does not result in a loss of generality (though it should be kept in
mind that for a more general model it may be important to recall that
$V = L_u^\dagger L_d$ is the true relation).
Thus we write,
\bea
\pmatrix{{d}\cr{s}\cr{b}\cr}_{\rm Weak} =
\pmatrix{{V_{ud}}&{V_{us}}&{V_{ub}}\cr{V_{cd}}&{V_{cs}}&{V_{cb}}\cr
{V_{td}}&{V_{ts}}&{V_{tb}}\cr}\,
\pmatrix{{d}\cr{s}\cr{b}\cr}_{\rm Mass} ,
\eea
with, in the ``standard parameterization'' 
advocated by the Particle Data Group \cite{pdg},
\bea
V =
\pmatrix{{c_{12}c_{13}}&{s_{12}c_{13}}&{s_{13}e^{-i\delta_{13}}}\cr
        {-s_{12}c_{23}-c_{12}s_{23}s_{13}e^{-i
        \delta_{13}}}&{c_{12}c_{23}-s_{12}s_{23}s_{13}
        e^{-i\delta_{13}}}&{s_{23}c_{13}}\cr
        {s_{12}s_{23}-c_{12}c_{23}s_{13}
        e^{-i\delta_{13}}}&{-c_{12}s_{23}-s_{12}c_{23}s_{13}
        e^{-i\delta_{13}}}&{c_{23}c_{13}}\cr}.
\eea
In this equation, $c_{ij} = \cos \theta_{ij}$ and
$s_{ij} = \sin \theta_{ij}$, with $i$ and $j$ labeling the generations.
$\delta_{13}$ is a complex phase that can induce $CP$ violating effects.
While a general $3 \times 3$ unitary matrix has three independent phases,
this parameterization has used the fact that we may redefine the quark fields
to include a complex phase such that only $\delta_{13}$ remains.
In a model containing physics beyond the SM, extended fermion interactions
may allow for effects related to more of these mixing matrices than the
single combination that is the CKM matrix.  
For example, a model that includes masses
for the neutrinos, it may be appropriate to introduce a CKM-like matrix to
include a $W^\pm$ coupling to leptons of various mixed generations.

Experimental measurements of the hadrons containing various types of
quarks provide information about all of the CKM elements except $V_{tb}$.
As we will see in Chapter~\ref{chapSingletop}, $V_{tb}$ can be measured
by studying single top production.
In fact, in the SM there are only three generations of fermions and thus
the requirement that the CKM matrix be unitary already provides strong limits
that $V_{tb}$ be close to one.  Nonetheless, it is important to directly
measure $V_{tb}$, since a deviation from the SM limit on $V_{tb}$ would
be a signal of physics beyond the Standard Model.  In the very least, one
could find an indication of a fourth generation of quarks that is strongly
mixed with the third family, but almost unmixed with the first two families
by measuring $V_{tb}$ to be considerably smaller than unity.

\section{ Theoretical Puzzles of the Standard Model }
\label{theorypuzz}

\subsection{General Considerations}

In spite of its enormous success in explaining high energy physics
phenomena, the SM still contains a number of theoretical flaws
and puzzles that lead us to believe it should be replaced by a more
fundamental theory at higher energy scales.  As there are a large
number of opinions and approaches to this question,
the discussion below will necessarily be somewhat personalized
and incomplete.  In this section we will discuss some general puzzles
of the SM, followed by more detailed discussion of
issues concerning the EWSB and the fermion mass hierarchy that will
be explored in the rest of this work.

In fact it is quite obvious that the SM is ``only an effective theory''
because it does not include gravity.  A
truly fundamental theory should explain all four of the forces observed
in nature.  The SM contains a description of the electromagnetic, weak,
and strong nuclear forces, but does not address how to include gravitational
interactions within its framework.  
In fact, because of the negative
mass dimension of the
gravitational coupling constant
($1 / M_{Planck}^2$),
simple power counting of loop diagrams
indicate that a field theory of gravity is not expected to be renormalizable.
Thus, there is no way to consistently include quantum corrections to 
gravitational phenomena within a field theory of point-like objects such
as the SM.  Further, the evolution of the structure of the universe
under gravitational interactions is sensitive to the cosmological constant,
which is observed to be very small (or zero).  Why this constant is so
small compared to typical particle physics energy scales remains a mystery.

Even if one were to focus on only a more modest goal of a theory
without gravity, the SM still contains many puzzling features.
For example, it includes 18 free parameters, including the three gauge
couplings $e$, $\sin^2 \theta_W$, and $g_3$;  the two Higgs potential
couplings that may be expressed as $M_Z$ and $M_h$; and 9 fermion masses
and 4 CKM mixing parameters that contain the physical information about
the Higgs Yukawa interactions with the fermions.  As the Higgs boson
has not been observed, its mass is the only undetermined quantity in the 
SM\footnote{Recall that in the SM $V_{tb}$ is fixed by the unitarity of
the CKM matrix.  It is worth noting that
precision measurements have become sensitive to
radiative corrections involving the Higgs.  Thus, indirect constraints
on $M_h$ already exist, as well as excluded regions
of $M_h$ that correspond to predictions for signals that have not been
observed at colliders \cite{lepdata, tevdata}.}.
This large
number of parameters can itself be seen as a drawback of the SM.  It would
seem more attractive if there were a deeper symmetry or structure that
could explain the origin of these seemingly arbitrary quantities
in terms of a smaller set of parameters.

A seemingly straight-forward means
to accomplish this would be to invoke a larger gauge symmetry,
with the quarks and leptons put together in its multiplets.
The SM contains a separate symmetry for the strong interactions, and a mixture
of two symmetries results in the weak and electromagnetic interactions.
Each symmetry has its own coupling constant, and thus the theory still
includes three separate forces.  Following a reductionist mentality,
it is an attractive idea that these three forces should be unified into
one single force.  This ``grand unified'' theory (GUT) could then be
spontaneously broken (by SSB similar to the EWSB, for example) at a high
energy scale
($M_{GUT}$), resulting in the three symmetry groups we see at low energies.

However, it is still not clear how this works (if it works at all).
A grand unified theory should have one coupling constant at $M_{GUT}$,
which indicates that the three couplings we see at low energies should
converge at some high energy scale.  By running the SM couplings, one finds
that they approximately unify at $\sim 10^{15}$ GeV, but do not quite
meet at a single energy scale.  Of course, in carrying out this computation
one must assume that there is no additional heavy matter between the weak
scale and the GUT scale, and so one must be careful in drawing conclusions.
Even if one were to resolve this issue, however, it would still raise
the question why the grand unified symmetry is apparently broken to
${\rm SU(3)}_C \times {\rm SU(2)}_L \times {\rm U(1)}_Y$ at 
$M_{GUT} \sim 10^{15}$ GeV,
whereas the electroweak symmetry is broken,
${\rm SU(2)}_L \times {\rm U(1)}_Y 
\ra {\rm U(1)}_{\rm EM}$ at the weak scale $v \sim 246$ GeV.  Such
a large hierarchy seems unnatural.

\subsection{The Electroweak Symmetry-Breaking}

As we saw above, the SM uses a Higgs doublet to generate the EWSB.
Thus, there is a physical Higgs boson $h$ remaining after generation
of the $W^\pm$ and $Z$ masses.  The Higgs boson has yet to be discovered
experimentally, and thus we still lack direct evidence
that the method of SSB employed by the SM is correct.  In fact, from the
discussion above it should be clear that the Higgs sector of the SM contains
most of the assumptions that went into the SM.  For example, the interactions
of gauge bosons with each other and the fermions is fixed in terms of
the three gauge couplings $g_1$, $g_2$, and $g_3$. On the other hand,
all fifteen of the remaining parameters in the SM are related to the
Higgs interactions with the fermions and with itself \cite{hhunter}.

\begin{figure}[t]
\epsfysize=0.8in
\centerline{\epsfbox{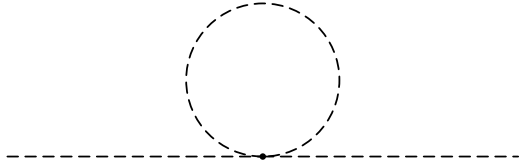}}
\caption{A Feynman diagram illustrating quantum corrections to the
scalar mass coming from a self-interaction.}
\label{hmasscorr}
\end{figure}

The Higgs boson is also a source of fine-tuning in the SM.  If one computes
the quantum corrections to its mass coming from the self-interaction in the
Higgs potential (from Feynman diagrams such as that shown in 
Figure \ref{hmasscorr}), one finds that the corrections have the form,
\bea
  \delta M_h^2 \sim \lambda \, \Lambda^2 ,
\eea
where $\Lambda$ is a cut-off introduced to represent the high energy
scale at which the SM ceases to be a good description of nature.
We have already seen that such a scale is expected to occur at
the Planck mass (though if there is new physics at lower energy scales
then it could also be lower).
This indicates that whatever the mass of the Higgs is at tree-level,
quantum corrections tend to push it to the scale $\Lambda$.  In order for
the Higgs to have mass around the weak scale, one must require an amazing
degree of cancellation between the bare Higgs mass and the quantum 
corrections to occur such that we subtract two quantities of order $\Lambda$
(possibly as high as $M_{Planck} \sim 10^{19}$ GeV) and arrive at
a difference on the order of the weak scale, $v$.
It is in this sense that the Higgs mass requires fine-tuning.  It seems
quite unnatural that such a delicate cancellation should occur between these
two a priori unrelated quantities.

It is not very satisfying to simply accept that the Higgs might be 
an extremely heavy
particle.  If that is the case, then one must ask the question
why the EWSB contains these two very different energy scales, $M_h$, and
$M_Z$.  This itself seems quite unnatural.
Further, if $M_h >$ about 1 TeV, the interaction between the 
longitudinal $W^\pm$ and $Z$ bosons becomes non-perturbative \cite{strongW},
and the perturbative expansion of the SM will no longer suffice to accurately
compute scattering amplitudes.  In fact, precision measurements at
LEP \cite{lepdata} and the Tevatron \cite{tevdata} indicate that
for consistency with the SM, the data prefers a Higgs boson
lighter than a few hundred GeV.

\begin{figure}[t]
\epsfysize=0.7in
\centerline{\epsfbox{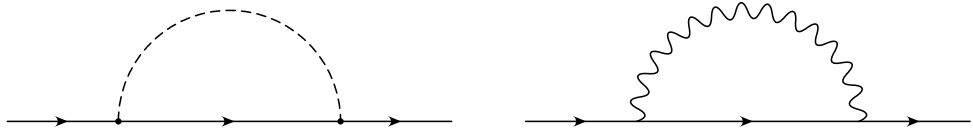}}
\caption{Feynman diagrams illustrating quantum corrections to the
fermion mass coming from interactions with a scalar or a vector boson.}
\label{fmasscorr}
\end{figure}

While on the subject of radiative corrections to particle masses, we
note that the situation is very different for
fermion masses, which have quantum corrections (from Feynman diagrams
such as the ones shown in Figure~\ref{fmasscorr}),
\bea
  \delta m_f \sim g^2 \, m_f \, \log \left( \frac{\Lambda}{m_f} \right),
\eea
where $g$ is the coupling of the fermion to the boson in
Figure~\ref{fmasscorr}.
In this equation, we see two features.  The first is that the correction
to $m_f$ is proportional to $m_f$ itself\footnote{This can be easily understood
from the fact that a theory of fermions with no masses contains a chiral
symmetry that protects the fermion mass from acquiring quantum corrections.
Introducing a fermion mass, $m_f$, breaks this symmetry, and since $m_f$
is now the order parameter that indicates that the symmetry is broken,
the quantum corrections are proportional to it.}.  The second occurs
because
the requirement that $\delta m_f$ be proportional to $m_f$ means any
function of lambda multiplying $m_f$ must be dimensionless (and because
of ultraviolet (UV) singularities in the loop integrals, divergent as
$\Lambda \ra \infty$).  Thus the correction depends only on $\log \Lambda$
and it is clear that the corrections to $m_f$ are naturally of the
same order as $m_f$ itself.

A further weakness of the SM coming from the Higgs sector is the problem of
``triviality''.
This problem arises in any theory of a scalar field
interacting with itself via a quartic interaction.  From Equation~\ref{lphi}
and the discussion following it, it is clear that such a term is vital
in the SM to induce SSB.  The problem can be studied by examining the
running coupling for the quartic scalar interaction.  From next-to-leading
order (NLO) in perturbation theory, the running coupling may be expressed,
\bea
   \lambda (\mu ) = \frac{ \lambda_0}
   { 1 - {3 \, \lambda_0 \over 4 \, 
    \pi^2} \log ( {\mu \over \mu_0} ) } ,
\eea
with $\lambda_0 = \lambda ( \mu_0 )$ the 
value of the coupling at some reference 
energy scale.  This expression shows that for a given $\lambda ( \mu_0 )$,
there is a large energy scale for which the denominator goes to zero,
and thus the coupling blows up.  If the SM is to remain perturbatively
valid all the way to the Planck scale, this limits the size of
$\lambda ( \mu_0 )$ at the weak scale.  As we saw above, $\lambda ( \mu_0 )$
is related to $M_h$, and so this statement can be reformulated as
saying that if the SM is to remain valid to the Planck scale, $M_h$ must
be smaller than about 1 TeV.  The precise $M_h$ for which the break-down
occurs also depends on contributions from the heavy top quark, and is
best studied non-perturbatively (i.e., on a lattice) because as the coupling
becomes large, results based on the perturbative expansion are not expected
to be very reliable.  This issue is generally referred to as ``triviality''
because the only way to guarantee that the SM is valid to an arbitrarily
high energy scale is to take $\lambda ( \mu_0 ) \ra 0$, which
results in a trivial, non-interacting scalar theory.

Because of these apparent problems (triviality and fine-tuning) associated
with scalar fields, there are a number of
proposed extensions of the SM that hope to address these weaknesses.  
There are two widely considered models that fall into this category.  The
first class of models, the supersymmetry (SUSY) models
\cite{susy}, invoke an additional
symmetry relating bosons and fermions to stabilize the Higgs sector of the SM.
Under SUSY, each field of the SM is given a partner with identical charges,
but spin differing by $1 \over 2$.  Loops of fermions appear in the
quantum corrections to the Higgs mass with a negative sign relative to the
scalar contributions, and cancel the quadratic divergences, thus removing
the fine-tuning problem.  
In the minimal supersymmetric standard model (MSSM) \cite{mssm},
the quartic Higgs interaction is related to the gauge interactions,
with a different behavior under the renormalization group than the
quartic interaction of the SM.  This takes care of the triviality problem.

The second class of model can be generically referred to as dynamical symmetry
breaking models.  There are many models of this kind, with the common feature
that the Higgs mechanism results not from a fundamental scalar field
acquiring a VEV, but from a composite scalar operator condensing.  This
composite operator may be built from heavy fermion fields whose masses,
as shown above, are not subject to large quantum corrections.  Thus, the
problems with a scalar field are side-stepped by requiring that the scalar is
not fundamental.  At high enough energies  the low energy effective theory
in terms of a scalar particle breaks down, and one is left with a theory
without scalar particles, whose high energy behavior is thus improved.
Examples of this type of theory include technicolor models \cite{technicolor},
top-condensate models \cite{topcondensate}, and top-color models 
\cite{topcolor}.

\subsection{The Fermion Mass Hierarchy}

Having discussed some of the issues involved in using the Higgs
mechanism to generate masses for the gauge bosons, we now
examine the fermions.  A deep puzzle of the SM is the question as
to why there exist three generations, interacting identically with
the gauge bosons, but very differently with the Higgs doublet,
as can be seen by the wide range of masses listed in Table~\ref{fm}.
Outstanding issues include the questions of why there are three
families (and not some other number), why neutrinos are massless
whereas the other fermions are massive, why the top quark
is so much heavier than the other fermions,
why the light fermions have masses so much smaller than $v$
(and masses that are so diverse from one another), and
why the CKM matrix
is almost diagonal and has such a small $CP$ violating phase.

A particular puzzle is the top quark.  The top is the only quark to
have a mass on the same order as $v$, and thus a Yukawa interaction close
to 1.  From that point of view, it seems that the top is the ``natural''
quark, while all of the other quarks are odd in that their Yukawa couplings
are very very small.  Another point of view is that the top quark is heavy 
because it is special in some way, perhaps having been given a special role
in the mechanism of EWSB
(as, for example, in the top-condensation models which provide the large
top mass and the EWSB through the same mechanism).  
Following this line of thought, it is very natural
to study the top quark very carefully.  If the top is special in some way,
then studies of top should reveal in what way the top is special, and
what that means for the EWSB. In the very least, careful study of the
top interaction with the Higgs would indicate whether or not the mechanism
that generates the top mass is identical to that which generates the boson
masses.

\subsection{The Electroweak Chiral Lagrangian}
\label{ewcl}

As we have seen, the Higgs sector of the SM represents the single largest
source of our ignorance concerning particle physics : the mechanism of the
EWSB.  Thus, it seems reasonable that one could expect new phenomena
to appear at energies not much greater than the weak scale, and thus
within the range of supercollider experiments currently underway
and planned for the future.
In order to search for signals of new physics effectively, there
are generally two sorts of deviations from the SM that one could consider
a sign of new physics.  The first is some sort of exotic particle beyond
those predicted by the SM.  The supersymmetric partners present in a
supersymmetrized SM are one example of this type of new phenomenon, and
composite scalar bound states of top quarks (or some other heavy fermion)
that often arise in dynamical EWSB models are another.  Searches for particles
of this type are necessarily model-dependent, because one must specify how
the ``new particle'' interacts with the known ones, thus determining how it
is produced, what (if anything) it decays into, and even how it interacts
with the material of a particle detector.  The second class of new phenomenon
involves modified properties of the known particles of the SM.  This
type of modification could be caused, for example,
by quantum effects from particles too heavy to be directly 
produced at colliders.  This kind of new phenomenon can be tested in a
model-independent way by carefully measuring various masses and interactions
of the known particles (and being careful to avoid ``assuming the SM'' in
interpreting the results).

A powerful tool with which one can examine new phenomena is the electroweak
chiral Lagrangian (EWCL) \cite{ewcl}.  The philosophy behind the EWCL
is that since we observe the masses of the $W^\pm$ and $Z$ bosons,
in some sense we have already seen the would-be Goldstone 
bosons \cite{kaplan}.  Using these ingredients, one can
construct the most general effective Lagrangian that realizes the
${\rm SU(2)}_L \times {\rm U(1)}_Y$ symmetry nonlinearly while preserving
${\rm SU(3)}_C \times {\rm U(1)}_{\rm EM}$.  
The result is an effective theory that is constructed
to encapsulate what is known about the presence of the gauge symmetry, while
allowing for more freedom in how the symmetry is spontaneously broken than the
particular realization of the Higgs mechanism employed in the SM.  Further,
such a construction allows one to search for new phenomena in a 
model-independent fashion.  The results may then be applied to learn something
about what sort of new physics is consistent with observed data, or to
confirm or rule out a given model of physics beyond the Standard Model.
Whenever possible, we will present results in the 
context of the EWCL, in order
to be as model-independent as possible.

As the EWCL is to be regarded as an effective theory, one generally includes
non-renormalizable interactions.  Such interactions have coupling constants
with dimensions of inverse mass, and are thus attributed to residual low
energy effects from high energy physics (which presumably are renormalizable
if one were to know the full high energy theory).  Thus, by observing such
effects one hopes to learn something about the scale at which these effects
become important, and the details of the full theory could be studied.
An example of such an operator is a flavor-changing neutral current (FCNC)
operator which connects the top quark, the charm quark, and the gluon.
In order to respect the ${\rm SU(3)}_C$ gauge symmetry, the 
lowest possible mass dimension of
operator is dimension 5 and may be written \cite{tcgmenehab},
\bea
   {\cal L}_{gtc} = \frac{g_S}{\Lambda_{gtc}} G^a_{\mu \nu} \left(
     \kappa_1 \, \bar{t} \, \sigma^{\mu \nu} \, \lambda^a \, c \;
   + \; \kappa_2 \, \bar{t} \, \sigma^{\mu \nu} \, \gamma_5 \, \lambda^a \, c 
   + H.c. \right) ,
\eea
where $\kappa_{1,2}$ parameterize the strength of the interaction in terms
of $g_S = g_3$ 
and $\Lambda_{gtc}$, which contains the mass dimension of the
coupling, may be thought of as
the scale at which the effective theory
breaks down.  The matrix $\sigma^{\mu \nu}$ is related to the
Dirac matrices by,
\bea
   \sigma^{\mu \nu} = \frac{i}{2} \left[ \gamma^\mu, \gamma^\nu \right]
                    = \frac{i}{2} \left( \gamma^\mu \, \gamma^\nu
                      - \gamma^\nu \, \gamma^\mu \right) .
\eea
In constructing this operator, we have followed the usual
EWCL procedure of defining the fermion fields such that they transform
under ${\rm SU(2)}_L \times {\rm U(1)}_Y$ the same way they transform
under ${\rm U(1)}_{\rm EM}$\footnote{This may be accomplished by including
an exponential of the Goldstone bosons in the definition of
the fermion field.  In the Unitary gauge, this corresponds with the usual
definition.}.
As we shall see in Chapter \ref{chapSingletop}, 
this operator may have important
implications for single top production.

\subsection {Final Remarks}

In presenting the SM, and in exposing its weaknesses, we have obtained
some sense of what a theory that hopes to improve our understanding of the
electroweak symmetry-breaking needs to accomplish.  Many theories propose
a wide variety of ways to accomplish this, and finding ways to prove or
disprove these theories is one of the current challenges for experimental
high energy physics.  The remainder of this dissertation is an exploration
of several classes of models, with an eye towards the question of how we
could discover whether or not these models represent a viable picture of 
reality.  As we have argued, the top quark is a likely place to find
new phenomena because of its huge mass.  Thus, we begin by studying the
process of single top production, which is expected to provide us with the
first real understanding of the top's weak interactions.  We will employ
a mixture of model-independent tools (such as the EWCL) as well as predictions
of specific models to show that Run II of the Fermilab Tevatron and
the CERN Large Hadron Collider
(LHC) represent a wealth of information about the top quark itself,
and thus most likely about the EWSB as well.

\chapter{Single Top Production}
\label{chapSingletop}

As we saw in Chapter~\ref{intro}, the SM suffers from a number of weaknesses
that are in one way or another related to the mechanism of the EWSB, both
in the generation of the gauge boson masses and the fermion masses.
The attractive idea that the top quark may play a special role in
the EWSB was introduced.  The definitive test of this hypothesis must come
from studying the properties of the top quark.  Careful measurement will
reveal if it is indeed a SM top, or something different.
Indeed, if signals of something beyond the SM
exist in top quark observables, careful study of them will provide
a means to determine what properties the more fundamental theory must possess
in order to explain the observed deviation.

The question of how to discover physics beyond the SM related to the top
quark
reduces to the question of how the top's properties may be determined
in a model-independent fashion, and without making strong assumptions that
will bias the interpretation of the measurements.  
Experiments at the Tevatron\footnote{Run~II of the Tevatron will involve
$p \, \bar{p}$ collisions with an expected center-of-mass energy of
$\sqrt{s} = 2$ TeV.}
Run~II 
and the LHC\footnote{We use LHC to denote a $p \, p$ collider with
center-of-mass energy $\sqrt{s} = 14$ TeV.}
will observe thousands of top quarks, and thus it is important
to examine various ways to probe top quark properties.  In this chapter we
present a detailed exposition of the available means to obtain information
about top, weighing the strengths and weaknesses of each.  In particular we
will see that single top production at a hadron collider represents a vital
means to study the weak interactions of the top quark, and thus test
the possibility of a relationship between top and EWSB.
In this entire chapter, we assume a top mass of $m_t = 175$ GeV,
unless explicitly noted otherwise.

\section{Top Quark Properties at a Hadron Collider}

\begin{figure}[t]
\epsfysize=1.4in
\centerline{\epsfbox{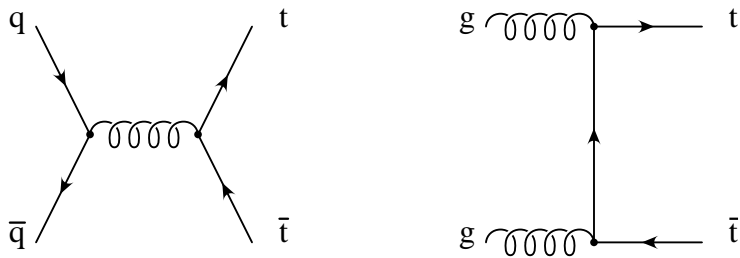}}
\caption{Representative Feynman diagrams showing QCD production of
$t \, \bar{t}$ pairs: $q \, \bar{q}, g \, g \ra t \, \bar{t}$.}
\label{ttbarfig}
\end{figure}

At hadron colliders, the dominant mechanism for producing top quarks is
to produce pairs of $t$ and $\bar{t}$
through the strong interaction \cite{ttbar},
as is shown in Figure~\ref{ttbarfig}.  As dictated by the QCD-improved
parton-model \cite{qcdparton}, the cross section for hadrons scattering into
$t \, \bar{t}$ pairs is computed by considering the partonic reactions
$q \, \bar{q} \ra t \, \bar{t}$ through a virtual gluon and through fusion of
two gluons, $g \, g \ra t \, \bar{t}$.  These partonic cross sections are then
convolved with universal parton distribution functions (PDF's)
\cite{pdf} which contain
non-perturbative information about the likelihood of finding a 
particular parton inside a parent hadron carrying
a given fraction of the parent hadron's momentum.
It is well known that the gluon distribution function is much larger at
very low momentum fraction than the 
corresponding valence quark distributions,
but falls much more rapidly as the momentum fraction increases.  For the
production of massive top quarks, this has the consequence that at a collider
with relatively low center of mass energy such as the Tevatron the dominant
subprocess will be from $q \, \bar{q}$ fusion, whereas a collider with
a much higher center of mass energy such as the LHC has a dominant
contribution from $g \, g$ fusion.  It is clear that the rate of 
$t \, \bar{t}$ production (coming from either subprocess)
represents a measure of the top's coupling to the gluons.

The large rate of $t \, \bar{t}$ production (about 7.55 pb 
at the Tevatron Run~II
and 760 pb at the LHC \cite{ttbar,ttbarrate})
insures that it is an important means to study the top quark.
As we have seen, it is an important measure of the top's strong interactions,
and could also be sensitive to some kind of new physics resonance in
$t \, \bar{t}$ production.
It also allows one to measure the top quark mass, $m_t$,
by reconstructing the
top mass from the top decay products.  From Run I of the Tevatron,
a combined CDF and D\O \, measurement of $m_t = 174.3 \pm 5.1$ GeV
based on direct observation of top has been
made, and it is hoped that improved statistics at Run~II of the
Tevatron will allow a more precise measurement of $\pm 2$ GeV
\cite{tevdata}.

\begin{figure}[t]
\epsfysize=1.4in
\centerline{\epsfbox{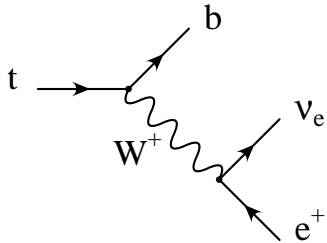}}
\caption{Feynman diagram showing the top decay into $W^+ \, b$,
including the leptonic decay $W^+ \ra e^+ \, \nu_e$.}
\label{tdecayfig}
\end{figure}

Top quarks are identified by their decay products.  In the SM, the top
decays into a $W^+$ boson and a down-type quark (predominantly bottom
because $V_{tb} \gg V_{ts}, V_{td}$) \cite{topdecay},
through Feynman diagrams such as that shown in Figure~\ref{tdecayfig}.
Its width can thus be computed in terms of the top mass, the gauge couplings,
and the CKM elements $V_{td}$, $V_{ts}$, and $V_{tb}$.  The SM prediction
is found to be about 1.5 GeV, much larger than for any other quark.
This large decay width indicates that the top
decays very quickly, before it has time to hadronize\footnote{A simple
heuristic way to understand this is to notice that the top width (1.5 GeV)
is very much larger than $\Lambda_{\rm QCD} \sim 200$ MeV, 
the scale at which
non-perturbative effects in the strong force become important.}
\cite{tophadron}.  This fact
means that even aside from the strong motivation to study top as a means to
understand the EWSB, there is also interest in top because it is the only
quark that we are able to study ``bare''.  Since the top decays into 
a $W^+$
and $b$ with a branching ratio (BR) close to one, top decays are
distinguished by the $W^+$ decay products.  The hadronic decays
($W^+ \ra q \, q^\prime$) are dominant (with BR $\sim$ 6 / 9), however
the leptonic decays ($W^+ \ra \ell^+ \, \nu_\ell$) generally provide a clean
signature at a hadron collider.

Clearly studying top decays provides some information about the top's weak 
interactions.  However, there is an important fact to keep in mind while
considering top decays; a study of decays can measure BR's
but since it does not actually measure the decay width itself, it is not
directly proportional to the coupling, and thus cannot measure the
magnitude of the $W$-$t$-$b$ coupling.  Thus, if the $W$-$t$-$b$
vertex is modified, but no new decay modes appear, the BR for
$t \ra W^+ \, b$ will remain close to 1, despite the fact that new physics
is affecting the structure of the interaction.  A further problem in
using top decays to search for new physics is that exotic top decays
may be unobservable or unrecognized as originating from top quarks,
and therefore could be missed.
Despite these weaknesses, as we shall see in Section~\ref{polarization},
top decays are an excellent opportunity to explore the Dirac structure
of the $W$-$t$-$b$ interaction, testing the left-handed nature of the
SM weak interactions of the top.

A powerful probe of the top's weak interactions is provided by single top
production, in which a top (or anti-top) quark is produced singly through
the weak interaction.  There are three important modes of single top
production in the SM: the $s$-channel $W^*$ mode \cite{schan}
in which a virtual
off-shell $W$ boson is produced which then decays into $t \, \bar{b}$;
the $t$-channel $W$-gluon fusion mode \cite{tchan}
in which a $W$ is exchanged between
a bottom quark and a light quark, resulting in a top and a jet;
and the $t \, W^-$ mode \cite{tw}
in which a bottom quark radiates an on-shell $W^-$
boson, resulting in a $t \, W^-$ 
final state.  The SM rates of these three processes at the Tevatron
and LHC, as a function of the top mass, are presented in 
Figure~\ref{smratefig}.

Single top production represents
a genuine opportunity to probe the magnitude of the
$W$-$t$-$b$ vertex because in this case
the size of the cross section is directly proportional to the $W$-$t$-$b$
coupling.
In the SM, this allows one to measure $V_{tb}$.  In a model
of new physics involving the top, this could lead to a discovery of the
new physics.  In the following sections we will discuss the three modes
of single top production in some detail, first in the context of the SM,
and then in regards to their sensitivity to new physics effects.
The issue of the top polarization will also be discussed, and 
it will be
demonstrated that not only can the polarization of 
the top be observed, but it
can provide interesting information about the structure of the interactions
of the top.

\begin{figure}[p]
\epsfysize=3.0in
\centerline{\epsfbox{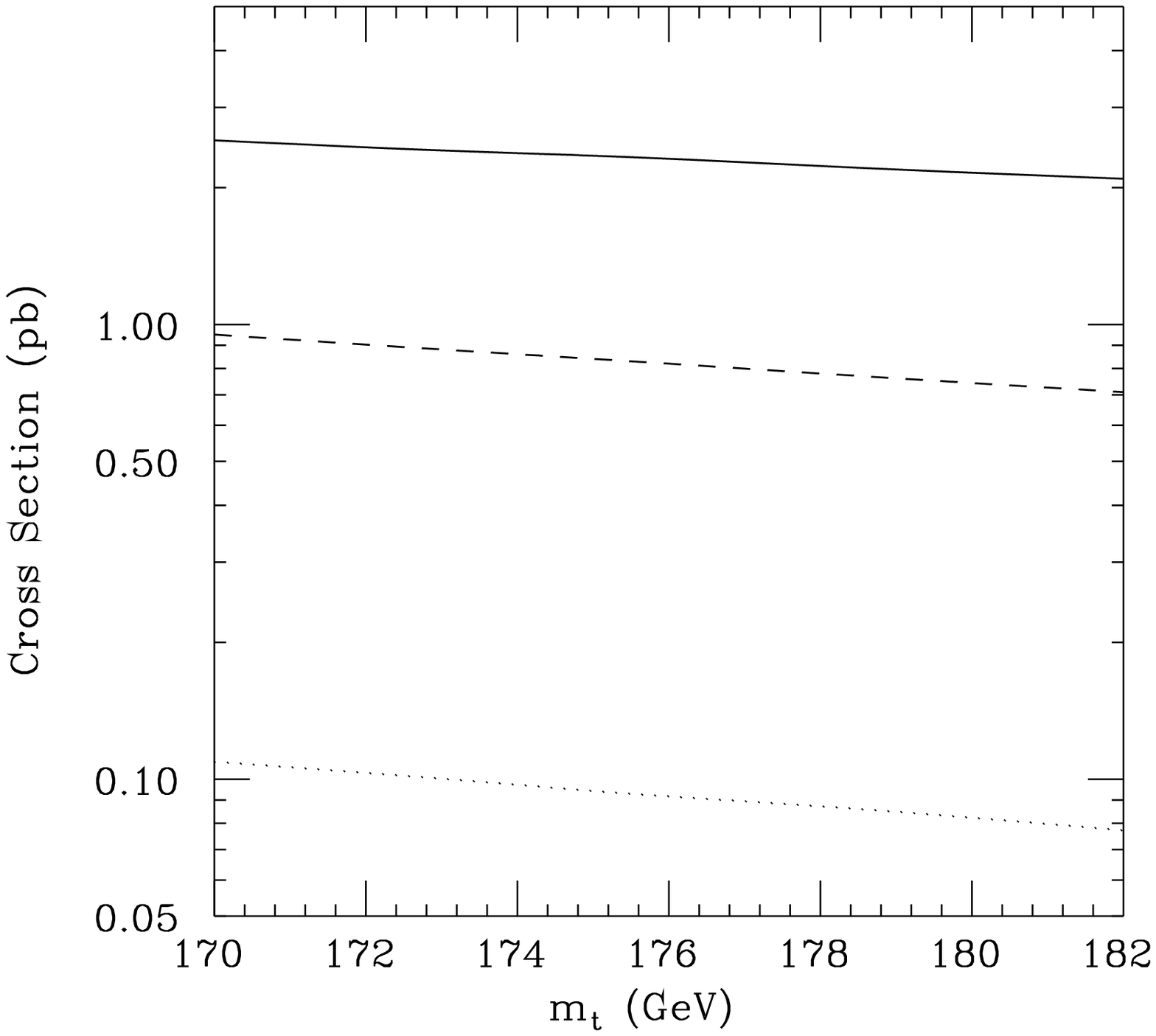}}
\epsfysize=3.0in
\centerline{\epsfbox{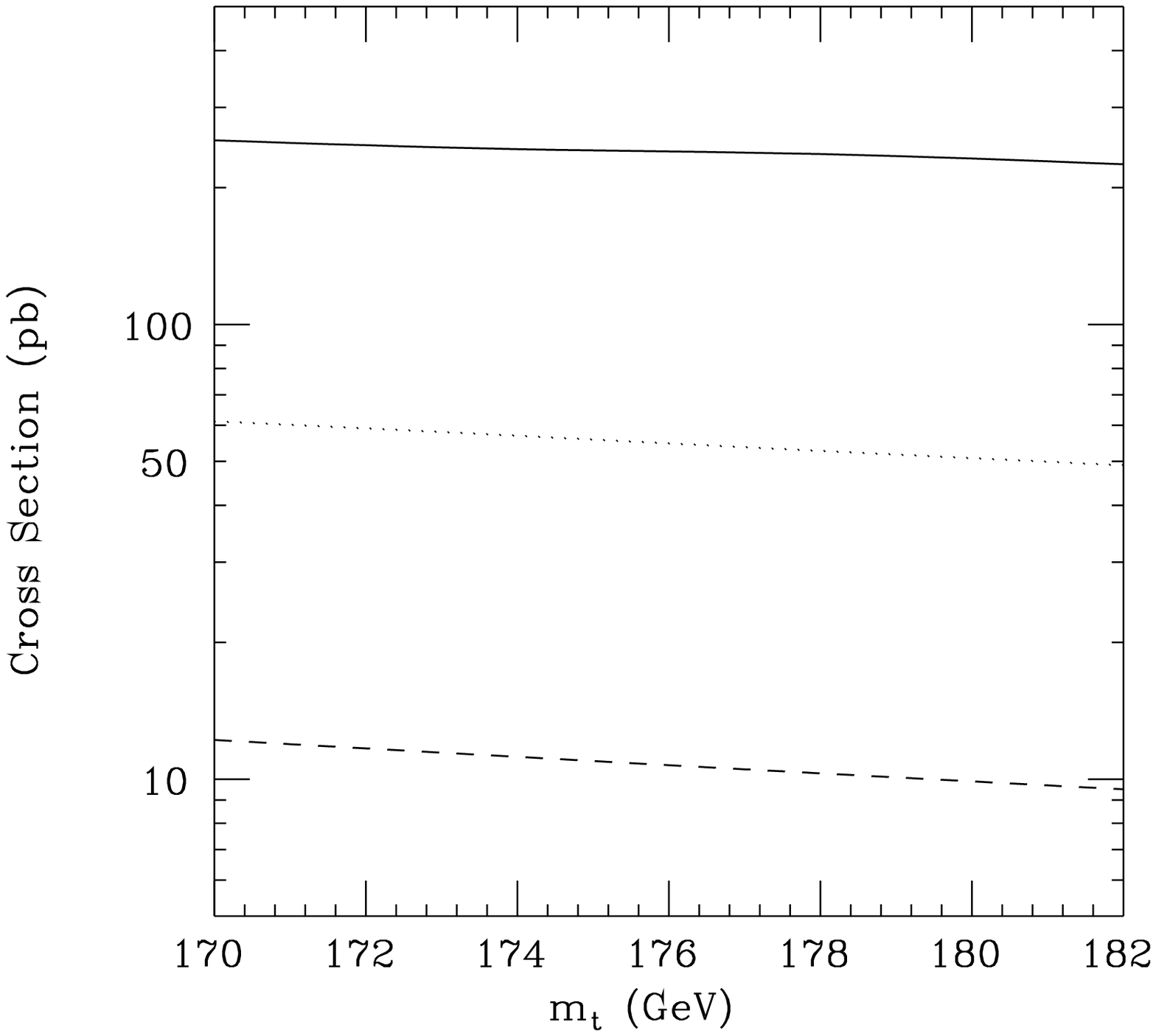}}
\caption{The SM rate of the three modes of single top production,
as a function of $m_t$ (summing the rates of $t$ and $\bar{t}$
production),
at the Tevatron (upper figure) and LHC (lower figure).  The solid
curve is the NLO $t$-channel rate, taken as the average of the 
results from CTEQ4M
and MRRS(R1) PDF's.  The dashed curve is the NLO $s$-channel rate,
taken as the average of the results from CTEQ4M and MRRS(R1) PDF's.
The dotted curve is the LO $t \, W^-$ rate, including large
$\log$ corrections, taken as the average of the CTEQ4L and MRRS(R1)
results.}
\label{smratefig}
\end{figure}

\clearpage

\section{Single Top Production in the SM}

The production mechanisms for the three modes of single top production are
quite different, and it is worth spending some time discussing the
particular physics aspects of each mode individually.  In this discussion,
we avoid detailed consideration of the particular kinematics and detection
strategy for each mode, as this has been considered elsewhere 
\cite{schan, tchan, doug}.  
Instead we concentrate on the inclusive rates
and the effects of nonstandard physics on each process, as our goal is
to understand how single top production serves as an important probe of new
physics effects.  We begin with the SM rates, and discuss the theoretical
issues involved in SM single top production.

In fact, it will be
shown that the three modes are separately susceptible to quite different
types of new physics \cite{bigst}, and can potentially
be observed independently from each other
\cite{doug}.
Thus, each mode is an independent source of information about the 
top quark.  One sometimes finds in the literature \cite{lumped1, lumped2} 
analyses
that treat all of the single top modes together as one signal.  This practice
of combining three signals with quite distinct kinematic signatures together
is not good physics; it wastes the information contained in each mode
separately.  As we will see, the three modes 
provide complimentary information about the top,
and thus are worth examining independently.  Further, because the
$W$-gluon fusion rate is generally much larger than the other two modes,
these ``combined analyses'' are really optimized to see that mode, with 
a small fraction of the other modes that manages to fake the characteristics
of a typical $t$-channel event included as well.  
Thus there is little practical difference between a combined analysis
and one focused on the $t$-channel process.
For these reasons, it is highly
preferable to avoid thinking of single top as one process, when it is
really three separate ones.

\subsection{$W^*$ Production}

\begin{figure}[t]
\epsfysize=1.4in
\centerline{\epsfbox{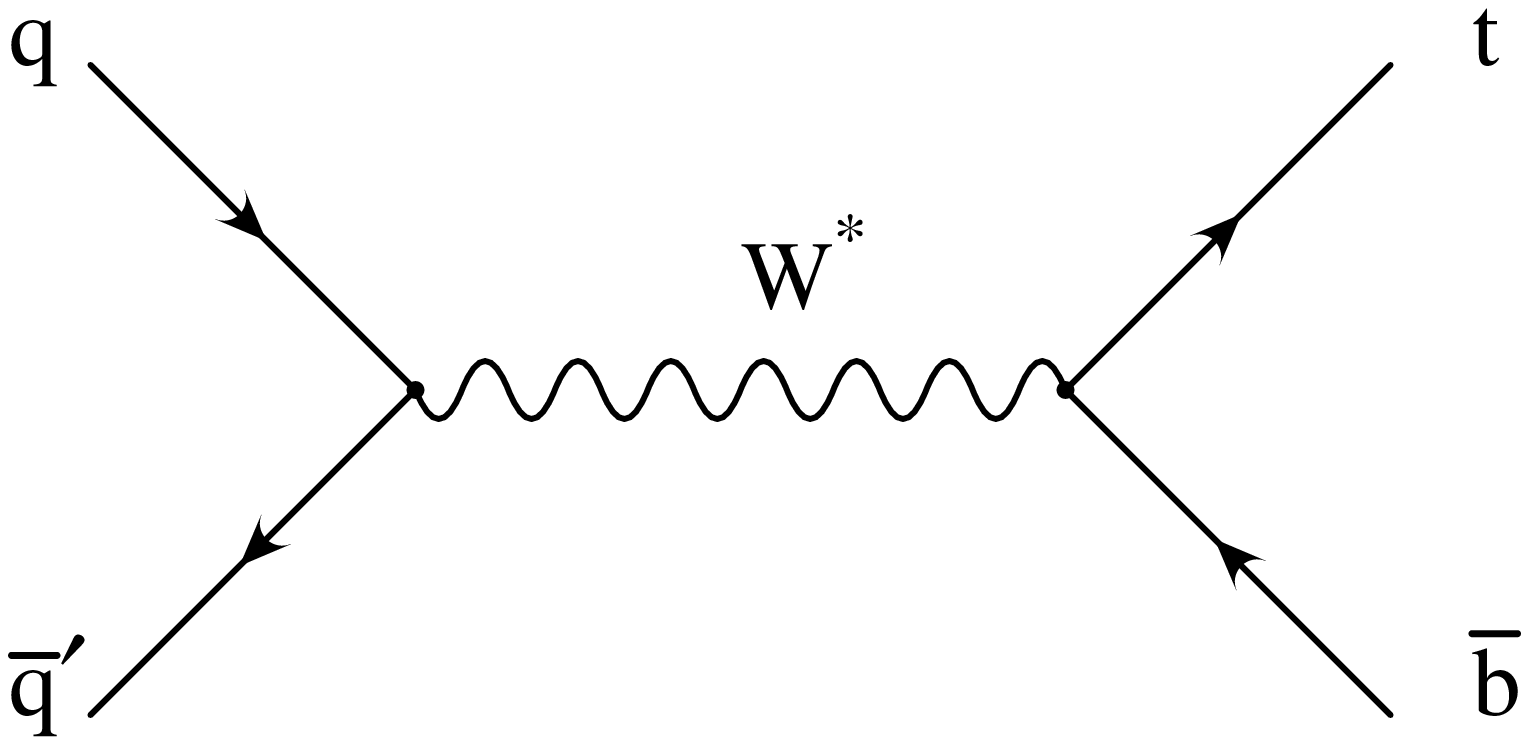}}
\caption{Feynman diagram for the $s$-channel mode of single top
production: $q \, \bar{q}^\prime \ra W^* \ra t \, \bar{b}$.}
\label{schanfig}
\end{figure}

The $W^*$ mode of single top production proceeds through an $s$-channel $W$
boson, as shown in Figure~\ref{schanfig}.  The final state consists of a
top quark and a central jet containing a $\bar{b}$.
Because the initial partons include
both a quark and an anti-quark, this process is relatively large at a
high energy $p \, \bar{p}$ collider such as the Tevatron, where valence
anti-quarks are present in the $\bar{p}$.
It has been computed
at NLO in QCD corrections \cite{nlos}, 
and it has been found that the corrections
coming from initial state radiation of soft and collinear
gluons are rather strong and
substantially increase the cross section.  The resulting 
NLO cross section
is $\sigma_s = 0.84$ pb at the Tevatron Run~II and 
$\sigma_s = 11.0$ pb
at the LHC.
The rather small increase in the cross section in going from
the Tevatron to LHC (compared to the other channels)
can be understood from the fact that the LHC is
a $p \, p$ collider, and thus has no valence anti-quarks.
In Tables~\ref{schantab1}, \ref{schantab2} and \ref{schantab3},
we present the NLO cross section,
as a function of the top quark mass for the Tevatron Run~II and LHC,
for several choices of the scale\footnote{In principle the factorization
scale and the renormalization scale may be chosen independently.  In practice,
we follow the usual procedure of choosing them to be equal to each other.}
and the CTEQ4M \cite{cteq4} and MRRS(R1) \cite{mrrs} PDF's.
The mean cross section is defined to be the average of the CTEQ4M 
and MRRS(R1) results evaluated at the canonical scale choice.
These rates are for the production process, 
$q \, \bar{q}^\prime \ra t \, \bar{b}$ only, and do not include 
the branching ratios for any particular top decay.  $V_{tb}$ has
been assumed to be one, and $V_{ts}$ and $V_{td}$ have been neglected,
as they
are so small that their effect on the cross section is much
less than $1\%$ of the $s$-channel rate.

The theoretical prediction for the
cross section shows a rather small dependence on the renormalization and
factorization scales of about
$\pm 5\%$, when the scale is varied from the default value
of $\mu^s_0 = \sqrt{s}$ by a factor of 2,
indicating that the uncomputed
higher order QCD corrections are probably small.
The distribution functions of
quarks and anti-quarks in the proton are relatively well-determined by
deeply inelastic scattering (DIS) data, and thus this important input to the
theoretical prediction is rather well understood.
The mass of the top quark is another important quantity that will affect the
predicted cross section.  The $W^*$ mode is particularly sensitive to this
quantity, because it not only determines the phase space of the produced
particles, but also controls how far off-shell the virtual 
$W^*$ boson must be.

\begin{table}[t]
\caption{The NLO rates of
$q \, \bar{q}^\prime \ra W^* \ra t \, \bar{b}$
(in pb) at the Tevatron Run~II. At the Tevatron,
the rate of $\bar{t}$ production is
equal to the $t$ production rate.}
\label{schantab1}
\begin{center} 
\begin{tabular}{lccccccc}
 & \multicolumn{3}{c}{CTEQ4M} & 
\multicolumn{3}{c}{MRRS(R1)} &  \\[0.1cm] 
 $m_{t}$ (GeV) & $\mu = \mu^s_0 / 2$ & $\mu = \mu^s_0$ &
$\mu = 2 \mu^s_0$ & $\mu = \mu^s_0 / 2$ & 
$\mu = \mu^s_0$ & $\mu = 2 \mu^s_0$ & $\sigma_s^{(mean)}$
\\[0.2cm] \hline \hline \\
170 & 0.53  & 0.49  & 0.46  & 0.495 & 0.46  & 0.425 & 0.475 \\
171 & 0.52  & 0.485 & 0.445 & 0.485 & 0.45  & 0.415 & 0.465 \\     
172 & 0.51  & 0.475 & 0.435 & 0.475 & 0.435 & 0.405 & 0.455 \\ 
173 & 0.495 & 0.46  & 0.425 & 0.46  & 0.425 & 0.395 & 0.44  \\ 
174 & 0.48  & 0.445 & 0.415 & 0.45  & 0.415 & 0.385 & 0.43  \\ 
175 & 0.465 & 0.43  & 0.405 & 0.44  & 0.405 & 0.375 & 0.42  \\
176 & 0.45  & 0.415 & 0.395 & 0.425 & 0.395 & 0.365 & 0.41  \\ 
177 & 0.445 & 0.405 & 0.385 & 0.415 & 0.385 & 0.36  & 0.405 \\ 
178 & 0.435 & 0.40  & 0.375 & 0.405 & 0.375 & 0.35  & 0.39  \\ 
179 & 0.43  & 0.395 & 0.365 & 0.395 & 0.365 & 0.34  & 0.38  \\ 
180 & 0.42  & 0.39  & 0.355 & 0.385 & 0.355 & 0.335 & 0.37  \\ 
181 & 0.41  & 0.38  & 0.345 & 0.375 & 0.35  & 0.325 & 0.36  \\ 
182 & 0.395 & 0.365 & 0.34  & 0.37  & 0.34  & 0.315 & 0.355 \\ \\
\hline \hline 
\end{tabular}
\end{center}
\vspace{0.5in}
\end{table}

\begin{table}[p!] 
\vskip 0.08in 
\caption{The NLO rates of 
$q \, \bar{q}^\prime \ra W^* \ra t \, \bar{b}$ (in pb) at the LHC.}
\label{schantab2}
\begin{center} 
\begin{tabular}{lccccccc}
 & \multicolumn{3}{c}{CTEQ4M} & 
\multicolumn{3}{c}{MRRS(R1)} & \\
$m_{t}$ (GeV) & $\mu = \mu^s_0 / 2$ & $\mu = \mu^s_0$ &
$\mu = 2 \mu^s_0$ & $\mu = \mu^s_0 / 2$ & 
$\mu = \mu^s_0$ & $\mu = 2 \mu^s_0$ & $\sigma_s^{(mean)}$
\\[0.2cm] \hline \hline \\
170 & 7.2 & 7.5 & 7.8 & 7.0  & 7.3  & 7.4  & 7.4  \\
171 & 7.1 & 7.4 & 7.6 & 6.8  & 7.1  & 7.25 & 7.25 \\     
172 & 6.9 & 7.2 & 7.5 & 6.7  & 6.9  & 7.1  & 7.05 \\ 
173 & 6.8 & 7.1 & 7.3 & 6.55 & 6.8  & 6.95 & 6.95 \\ 
174 & 6.7 & 6.9 & 7.1 & 6.4  & 6.65 & 6.8  & 6.78 \\ 
175 & 6.5 & 6.8 & 7.0 & 6.3  & 6.5  & 6.65 & 6.65 \\
176 & 6.4 & 6.6 & 6.9 & 6.2  & 6.4  & 6.5  & 6.5  \\ 
177 & 6.3 & 6.5 & 6.7 & 6.05 & 6.25 & 6.4  & 6.38 \\ 
178 & 6.1 & 6.4 & 6.6 & 5.9  & 6.1  & 6.25 & 6.25 \\ 
179 & 6.0 & 6.3 & 6.4 & 5.8  & 6.0  & 6.1  & 6.15 \\ 
180 & 5.9 & 6.1 & 6.3 & 5.7  & 5.9  & 6.0  & 6.0  \\ 
181 & 5.8 & 6.0 & 6.2 & 5.6  & 5.75 & 5.9  & 5.88 \\ 
182 & 5.7 & 5.9 & 6.1 & 5.5  & 5.65 & 5.8  & 5.78 \\ \\
\hline \hline 
\end{tabular}
\end{center} 
\end{table}

\begin{table}[p!] 
\vskip 0.08in 
\caption{The NLO rates of 
$q \, \bar{q}^\prime \ra W^* \ra b \, \bar{t}$ (in pb) at the LHC.}
\label{schantab3}
\begin{center} 
\begin{tabular}{lccccccc}
 & \multicolumn{3}{c}{CTEQ4M} & 
\multicolumn{3}{c}{MRRS(R1)} & \\
$m_{t}$ (GeV) & $\mu = \mu^s_0 / 2$ & $\mu = \mu^s_0$ &
$\mu = 2 \mu^s_0$ & $\mu = \mu^s_0 / 2$ & 
$\mu = \mu^s_0$ & $\mu = 2 \mu^s_0$ & $\sigma_s^{(mean)}$
\\[0.2cm] \hline \hline \\
170 & 4.5  & 4.7  & 4.8  & 4.2  & 4.4  & 4.5  & 4.55 \\
171 & 4.4  & 4.6  & 4.7  & 4.1  & 4.3  & 4.4  & 4.45 \\     
172 & 4.3  & 4.5  & 4.6  & 4.0  & 4.2  & 4.3  & 4.35 \\ 
173 & 4.2  & 4.4  & 4.5  & 3.9  & 4.1  & 4.2  & 4.25 \\ 
174 & 4.1  & 4.3  & 4.4  & 3.85 & 4.0  & 4.1  & 4.15 \\ 
175 & 4.0  & 4.2  & 4.3  & 3.8  & 3.9  & 4.0  & 4.05 \\
176 & 3.9  & 4.1  & 4.2  & 3.7  & 3.8  & 3.9  & 3.95 \\ 
177 & 3.8  & 4.0  & 4.1  & 3.6  & 3.7  & 3.8  & 3.85 \\ 
178 & 3.75 & 3.9  & 4.0  & 3.5  & 3.65 & 3.75 & 3.78 \\ 
179 & 3.7  & 3.8  & 3.9  & 3.45 & 3.6  & 3.7  & 3.7  \\ 
180 & 3.6  & 3.7  & 3.85 & 3.4  & 3.5  & 3.6  & 3.6  \\ 
181 & 3.5  & 3.65 & 3.8  & 3.3  & 3.4  & 3.5  & 3.53 \\ 
182 & 3.45 & 3.6  & 3.7  & 3.25 & 3.35 & 3.4  & 3.48 \\ \\
\hline \hline 
\end{tabular}
\end{center} 
\end{table}

\clearpage

\subsection{$W$-gluon Fusion}

\begin{figure}[t]
\epsfysize=1.8in
\centerline{\epsfbox{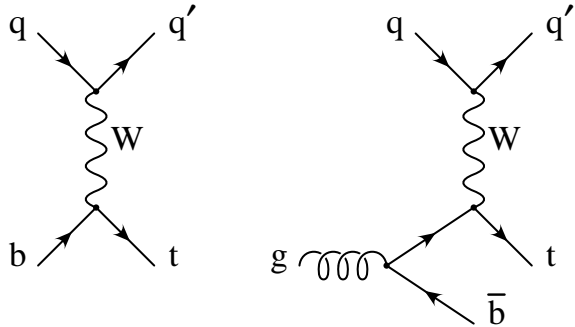}}
\caption{Feynman diagrams for the $t$-channel mode of single top
production: $b \, q \ra t \, q^\prime$.  A second process in which
the incoming light quark is switched with a light $\bar{q}$ is also
possible.}
\label{tchanfig}
\end{figure}

The $W$-gluon fusion production mode involves a $t$-channel $W$ exchange,
as shown in Figure~\ref{tchanfig}.  Thus, its final state consists of
a top quark and a jet that tends to be forward.
It relies on the possibility of
finding bottom quarks inside the hadrons involved in
a high energy collision in order to produce a single top quark.  The name
``$W$-gluon fusion'' can be understood in that the physical picture is
that the process actually involves a virtual gluon splitting into a
$b \, \bar{b}$ pair, with one of the bottom quarks participating in the
high energy scattering.  One could thus compute the inclusive cross section
starting from a quark-gluon initial state, but the result is not
perturbatively reliable because the kinematic region in which the 
$b \, \bar{b}$
pair from the gluon splitting is approximately collinear with the
initial gluon produces a contribution that is proportional to 
$\alpha_S \, \log m_t^2 / m_b^2$, which for $m_t \sim 175 \,$ GeV,
$m_b \sim 4.5 \,$ GeV, and $\alpha_S \sim 0.1$ is over-all
of order 1.  In fact,
the $n$th order correction always contains a collinear piece which has
the behavior $(\alpha_S \log m_t^2 / m_b^2)^n$,
which
spoils the perturbative description of this process.
A convergent perturbative expansion can be restored by resumming these large
logarithms into a bottom quark parton distribution function
\cite{bpdf}.  This PDF is different from the light quark PDF's in 
that it is actually perturbatively derived from the gluon 
distribution function.  In fact, this
two particle to two particle 
($2 \ra 2$)
description of the scattering represents the most important part of the
$W$-gluon fusion kinematics, because
the dominant kinematic configuration is
one in which the incoming bottom is collinear with the gluon
\cite{willendicus}.
It should be kept in mind that
the $b$ PDF has effectively
integrated out the $\bar{b}$ kinematics, so this formalism does not
accurately describe
the kinematic region in which the $\bar{b}$ has large transverse
momentum ($p_T$).  In this region, a description based on the two to
three scattering is more appropriate (and since this is precisely
the region in which the $b$ is not collinear with the incoming gluon,
it is well-defined in perturbation theory).  The resulting NLO cross
section is $\sigma_t = 2.53$ pb at the Tevatron and 241 pb at the LHC.

This strong dependence on the gluon PDF is a large source of uncertainty
in the prediction for the $W$-gluon fusion cross section.  The DIS
experiments are much less sensitive to the gluon density than to the
quark density, and thus the gluon density is much less well determined,
particularly in the high momentum fraction region relevant for single top
production.  Though it is not a quantitative measure of the uncertainty
from the PDF, this fact is reflected in the larger dependence of the
$W$-gluon fusion rate on the choice of PDF in the computation.

The NLO QCD corrections to $W$-gluon fusion are
slightly negative at both the Tevatron and LHC \cite{nlot}.
The NLO cross section varies by about 
$\pm 6\%$ at the Tevatron and $\pm 5\%$ at the
LHC when the natural scale choice of $\mu^t_0 = \sqrt{Q^2 + m_t^2}$ is
varied by a factor of 2, where $Q^2$ is related to the $W$ boson
momentum by $Q^2 = - p_W^2$.
This again indicates that the NLO 
inclusive rate is expected to be fairly insensitive to the uncomputed
higher order QCD corrections.  In Tables~\ref{tchantab1}, 
\ref{tchantab2}, and \ref{tchantab3}, we show
$\sigma_t$ for various top masses, PDF choices, and scales
at the Tevatron Run~II and LHC.  These rates are for the production
process $b \, q \ra t \, q^\prime$, with $V_{tb}=1$ 
and $V_{ts}, V_{td}= 0$.

\begin{table}[p] 
\caption{The NLO rates of $b \, q \ra t \, q^\prime$ 
(in pb) at the Tevatron Run~II.  At the Tevatron, the
rate of $\bar{t}$ production is equal to the rate of $t$
production.}
\label{tchantab1}
\begin{center}
\begin{tabular}{lccccccc}
 & \multicolumn{3}{c}{CTEQ4M} & 
\multicolumn{3}{c}{MRRS(R1)} & 
 \\
$m_{t}$ (GeV) & $\mu = \mu^t_0 / 2$ & $\mu = \mu^t_0$ &
$\mu = 2 \mu^t_0$ & $\mu = \mu^t_0 / 2$ & 
$\mu = \mu^t_0$ & $\mu = 2 \mu^t_0$ & $\sigma_t^{(mean)}$
\\[0.2cm] \hline \hline \\
170 & 1.255 & 1.31  & 1.365 & 1.18  & 1.22  & 1.26  & 1.265 \\
171 & 1.235 & 1.285 & 1.355 & 1.16  & 1.195 & 1.235 & 1.24  \\     
172 & 1.215 & 1.26  & 1.34  & 1.14  & 1.175 & 1.215 & 1.22  \\ 
173 & 1.195 & 1.24  & 1.32  & 1.12  & 1.155 & 1.195 & 1.20  \\ 
174 & 1.175 & 1.225 & 1.30  & 1.105 & 1.135 & 1.175 & 1.18  \\ 
175 & 1.155 & 1.205 & 1.275 & 1.085 & 1.12  & 1.155 & 1.165 \\
176 & 1.135 & 1.19  & 1.25  & 1.07  & 1.105 & 1.135 & 1.15  \\ 
177 & 1.115 & 1.17  & 1.225 & 1.05  & 1.085 & 1.12  & 1.13  \\ 
178 & 1.095 & 1.115 & 1.20  & 1.035 & 1.07  & 1.10  & 1.115 \\ 
179 & 1.075 & 1.14  & 1.175 & 1.02  & 1.055 & 1.08  & 1.10  \\ 
180 & 1.06  & 1.12  & 1.155 & 1.00  & 1.035 & 1.065 & 1.08  \\ 
181 & 1.045 & 1.10  & 1.14  & 0.985 & 1.015 & 1.045 & 1.06  \\ 
182 & 1.03  & 1.08  & 1.125 & 0.97  & 0.995 & 1.03  & 1.04  \\ \\
\hline \hline 
\end{tabular}
\end{center}
\vspace{0.5in}
\end{table}

\begin{table}[p] 
\caption{The NLO rates of 
$b \, q \ra t \, q^\prime$ (in pb) at the LHC.}
\label{tchantab2}
\begin{center} 
\begin{tabular}{lccccccc}
 & \multicolumn{3}{c}{CTEQ4M} & 
\multicolumn{3}{c}{MRRS(R1)} & 
 \\
$m_{t}$ (GeV) & $\mu = \mu^t_0 / 2$ & $\mu = \mu^t_0$ &
$\mu = 2 \mu^t_0$ & $\mu = \mu^t_0 / 2$ & 
$\mu = \mu^t_0$ & $\mu = 2 \mu^t_0$ & $\sigma_t^{(mean)}$
\\[0.2cm] \hline \hline \\
170 & 156 & 161 & 165 & 154 & 157 & 164 & 159   \\
171 & 154 & 160 & 164 & 153 & 155 & 162 & 157.5 \\     
172 & 152 & 159 & 163 & 152 & 153 & 161 & 156   \\ 
173 & 150 & 157 & 162 & 150 & 152 & 160 & 154.5 \\ 
174 & 149 & 156 & 160 & 149 & 151 & 158 & 153.5 \\ 
175 & 147 & 155 & 159 & 148 & 150 & 157 & 152.5 \\
176 & 146 & 154 & 158 & 147 & 149 & 155 & 151.5 \\ 
177 & 144 & 153 & 157 & 146 & 148 & 154 & 150.5 \\ 
178 & 142 & 152 & 155 & 145 & 147 & 152 & 149.5 \\ 
179 & 141 & 151 & 154 & 143 & 146 & 151 & 148.5 \\ 
180 & 140 & 150 & 153 & 142 & 145 & 149 & 147.5 \\ 
181 & 139 & 148 & 152 & 141 & 144 & 148 & 146   \\ 
182 & 138 & 147 & 151 & 140 & 143 & 147 & 145   \\ \\
\hline \hline 
\end{tabular}
\end{center}
\end{table}

\begin{table}[p] 
\caption{The NLO rates of 
$\bar{b} \, q \ra \bar{t} \, q^\prime$ (in pb) at the LHC.}
\label{tchantab3}
\begin{center} 
\begin{tabular}{lccccccc}
 & \multicolumn{3}{c}{CTEQ4M} & 
\multicolumn{3}{c}{MRRS(R1)} & 
 \\
$m_{t}$ (GeV) & $\mu = \mu^t_0 / 2$ & $\mu = \mu^t_0$ &
$\mu = 2 \mu^t_0$ & $\mu = \mu^t_0 / 2$ & 
$\mu = \mu^t_0$ & $\mu = 2 \mu^t_0$ & $\sigma_t^{(mean)}$
\\[0.2cm] \hline \hline \\
170 & 90   & 93   & 96   & 90   & 98 & 95 & 95.5 \\
171 & 89.5 & 91   & 95   & 89.5 & 96 & 95 & 93.5 \\     
172 & 89   & 89.5 & 94   & 89   & 94 & 93 & 91.8 \\ 
173 & 88.5 & 89   & 93.5 & 88.5 & 92 & 92 & 90.5 \\ 
174 & 88   & 88   & 93   & 88   & 90 & 91 & 89   \\ 
175 & 87.5 & 87.5 & 92   & 87   & 89 & 90 & 88.3 \\
176 & 87   & 87   & 91   & 86   & 88 & 89 & 87.5 \\ 
177 & 86.5 & 86.5 & 90   & 84   & 87 & 88 & 86.8 \\ 
178 & 86   & 86   & 89   & 83   & 86 & 87 & 86   \\ 
179 & 85   & 85.5 & 88   & 82   & 85 & 86 & 85.3 \\ 
180 & 84   & 85   & 86   & 81   & 84 & 85 & 84.5 \\ 
181 & 83   & 84   & 85   & 80   & 83 & 84 & 83.5 \\ 
182 & 82   & 83   & 84   & 78   & 82 & 82 & 82.5 \\ \\
\hline \hline 
\end{tabular}
\end{center}
\end{table}

\clearpage

\subsection{$t \, W^-$ Production}
\label{twsec}

\begin{figure}[t]
\epsfysize=1.7in
\centerline{\epsfbox{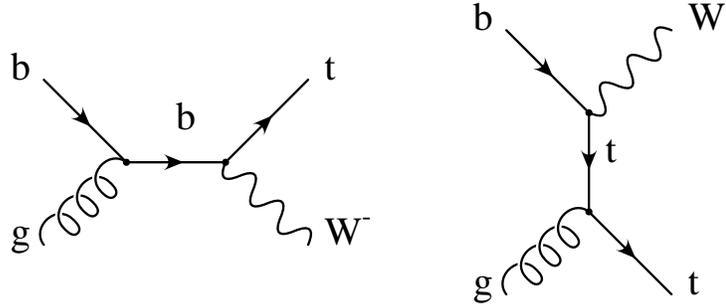}}
\caption{Feynman diagrams for the $t \, W^-$ mode of single top
production: $g \, b \ra t \, W^-$.}
\label{twfig}
\end{figure}

The $t \, W^-$ mode of single top production proceeds from Feynman
diagrams such as those presented in Figure~\ref{twfig}.  The final
state consists of an on-shell $W^-$ (which can decay either to quarks
or leptons) and a top quark.  It should be quite clear that the 
signature of this process is very distinct from the other two
modes, because of the extra $W^-$ decay products in the final state.
This process also involves finding
a bottom quark inside one of the colliding hadrons, and the
same issues related to this fact that were present in
the $W$-gluon fusion process discussed
above are also important here, in particular a rather strong
dependence on the gluon PDF used to obtain the prediction.

Though the complete NLO QCD corrections are not available,
one can improve the LO estimates by including the 
${\cal{O}}(1/ \log m_t^2 / m_b^2 )$ corrections coming from
Feynman diagrams such as those in Figure~\ref{twlogfig}
There are two subtle points that must
be carefully dealt with in carrying out this procedure.  The first
is that when the $b$ PDF was defined, the collinear contributions
from these diagrams was already resummed into what we called the
LO contribution.  Thus, in order to avoid double-counting this
collinear region one must subtract out the piece already
included in the LO contribution.  The full cross section 
for $A \, B \ra t \, W^-$ may thus be expressed as,
\bea
   \sigma_{tW} = \sigma^0(A \, B \ra t \, W^-)
   + \sigma^1(A \, B \ra t \, W^- \, \bar{b}) 
   - \sigma^{S}(A \, B \ra t \, W^- \, \bar{b}),
\eea
with the individual terms given by,
\bea
   \sigma^0(A \, B \ra t \, W^-) &=&
     \int  dx_1 \, dx_2 \left\{
     {f_{g / A}}(x_1, \mu) \, f_{b / B}(x_2, \mu) \,
     {\sigma}(b \, g \ra t \, W^-) \right. \\
   & & \left. + \; {f_{b / A}}(x_1, \mu) \, f_{g / B}(x_2, \mu)
     \, {\sigma}(g \, b \ra t \, W^-) \right\} \nonumber \\
   \sigma^1(A \, B \ra t \, W^- \, \bar{b}) &=&
     \int dx_1 \, dx_2 \;
     {f_{g / A}}(x_1, \mu) \, f_{g / B}(x_2, \mu) \,
     {\sigma}(g \, g \ra t \, W^- \, \bar{b}) \nonumber \\[0.1cm]
   \sigma^{S}(A \, B \ra t \, W^- \, \bar{b}) &=& 
     \int dx_1 dx_2 \left\{
     {\tilde{f}_{b / A}}(x_1, \mu) \, f_{g / B}(x_2, \mu) \,
     {\sigma}(b \, g \ra t \, W^-) \right. \nonumber \\
   & & \left. + \; {f_{g / A}}(x_1, \mu) \, {\tilde{f}_{b / B}}(x_2, \mu)
     \, {\sigma}(g \, b \ra t \, W^-) \right\} . \nonumber
\eea
The
``modified $b$ PDF'', ${\tilde{f}_{b / H}}$, 
contains the collinear logarithm and splitting
function $P_{b \leftarrow g}$ convoluted with the gluon PDF,
\bea
   {\tilde{f}_{b / H}}(x, \mu) = \frac{ \alpha_S(\mu)}{2 \, \pi}
   \log \left( \frac{ \mu^2 }{ m_b^2 } \right) \int^1_x
   \frac{dz}{z} \left[ \frac{ z^2 + (1 - z)^2 }{2} \right]
   f_{g/H}\left(\frac{x}{z}, \mu \right).
\eea
Having included this subtraction piece, the problem of double-counting
the collinear region is resolved.

\begin{figure}[t]
\epsfysize=2.2in
\centerline{\epsfbox{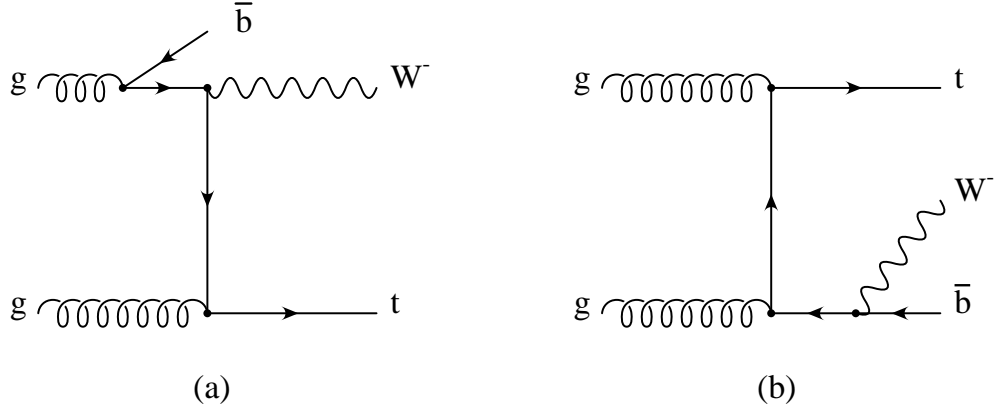}}
\caption{Representative Feynman diagrams for 
corrections to the $t \, W^-$ mode of single top
production corresponding to (a) large $\log$ corrections associated with
the $b$ PDF and (b) LO $t \, \bar{t}$ production followed by the LO
decay $\bar{t} \ra W^- \, \bar{b}$.}
\label{twlogfig}
\end{figure}

The second subtle point
in evaluating the large $\log$ contributions is that they contain
contributions such as those found in Figure~\ref{twlogfig}b
that correspond to LO $g \, g \ra t \, \bar{t}$ 
production followed by the LO
decay $\bar{t} \ra W^- \, \bar{b}$.  This expresses the fact that as
one considers higher orders in perturbation theory, the distinction
between $t \, \bar{t}$ production and various types of single top
production is blurred.  However, when considering quantities that are
properly defined, these corrections are small, and there is no problem
distinguishing these processes.  As a matter of book keeping,
the corrections to $t \, W^-$ production involving an on-shell top
are more intuitively
considered a part of the LO $t \, \bar{t}$ rate, and thus it
is important to subtract them out to avoid double counting in this
kinematic region. This may be done by noting that in the region
where the invariant mass of the $\bar{b} \, W^-$ system, $M_{Wb}$,
is close to the top mass, the behavior of the partonic cross
section ${\sigma}(g g \ra t \, W^- \bar{b})$ may be expressed,
\bea
   \frac{d{\sigma}}{dM_{Wb}}(g \, g \ra t \, W^- \, \bar{b}) &=&
   \sigma^{LO}(g \, g \ra t \bar{t}) \, 
   \frac{ m_t \, \Gamma^{LO}(\bar{t} \ra W^- \bar{b}) }
   {\pi \, [(M_{Wb}^2 - m_t^2)^2 + m_t^2 \, \Gamma_t^2]} \\
   &=& \sigma^{LO}(g \, g \ra t \bar{t}) \,
   \frac{ m_t \, \Gamma_t \, BR(\bar{t} \ra W^- \bar{b}) }
   {\pi \, [(M_{Wb}^2 - m_t^2)^2 + m_t^2 \, \Gamma_t^2]} \nonumber \\[0.3cm]
   &\ra& \sigma^{LO}(g \, g \ra t \bar{t}) \,
   BR(\bar{t} \ra W^- \bar{b}) \, \delta (M_{Wb}^2 - m_t^2) \nonumber
\eea
where $\sigma^{LO}(g \, g \ra t \bar{t})$ and 
$\Gamma^{LO}(\bar{t} \ra W^- \bar{b})$ are the LO cross section and
partial width, and $\Gamma_t$ is the inclusive top decay width.
The last distribution identity holds in the limit $\Gamma_t \ll m_t$.
Having identified this LO on-shell piece, it may now be simply subtracted
from ${\sigma}(g \, g \ra t \, W^- \, \bar{b})$.
The advantage to this formulation of the subtraction is that
by taking the narrow width limit, one removes all of the on-shell 
$\bar{t}$ contribution.
The interference terms between one of the on-shell $\bar{t}$ amplitudes 
and an amplitude without an on-shell $\bar{t}$ involve a Breit-Wigner
propagator of the form, $( M_{Wb}^2 - m_t^2 + i \, m_t \Gamma_t )^{-1}$,
which in the limit of small $\Gamma_t$, may be expressed as a 
principle valued integral in $M_{Wb}$.
Following this prescription,
and choosing a canonical scale choice
of $\mu_0 = \sqrt{s}$, leads to a large $\log$ correction to the
$t \, W^-$ rate of $-9.5\%$ at the LHC, which is consistent with
previous experience from the $W$-gluon fusion mode \cite{doug}.

This problem of the on-shell top was dealt with
another way in \cite{lumped2}, where
a cut was applied on $M_{Wb}$, to exclude the region of
$|M_{Wb} - m_t| \leq 3 \, \Gamma_t$.
Following this prescription, one finds a much larger correction
of about $+50\%$
to the $t \, W^-$ rate at the LHC.  
However, this is misleading because the large
corrections are mostly coming from the region where the $\bar{t}$ is
close to on-shell (though still at least 3 top widths away).  In other
words, the large positive correction comes from the tails of the Breit-Wigner
distribution for on-shell $\bar{t}$ production.  This can be simply understood
by taking the prescription in \cite{lumped2}
and varying the cut by increasing
the interval about the on-shell $\bar{t}$ region that is excluded.  
One finds that the correction computed in this way varies quite strongly with
the cut, and reproduces the subtraction method we have employed for
the cut $|M_{Wb} - m_t| \leq 12 \, \Gamma_t$.  A further theoretical
advantage of the subtraction method is that when one determines the
$t \, \bar{t}$ and $t \, W^-$ rates, one would like to actually fit the
data to the sum of the two rates, and thus the subtraction method
allows one to simply separate this sum into the two contributions
without introducing an arbitrary cut-off into the definition
of the separation.

Even if one were to use a cut-off to effect the separation, there is
a further problem in employing
the cut $|M_{Wb} - m_t| \leq 3 \, \Gamma_t$
to remove on-shell $t \, \bar{t}$ production.
This is that from a purely
practical point of view $3 \, \Gamma_t \sim 4.5 \,$ GeV, which is much smaller
than the expected jet resolution at the Tevatron or LHC.  Thus, it is not
experimentally possible to impose this definition of the separation between
$t \, W^-$ and $t \, \bar{t}$.  A more realistic resolution is
about 15 GeV \cite{lhcjetres}, which corresponds to a subtraction of
$|M_{Wb} - m_t| \leq 10 \, \Gamma_t$.  As we have seen above, this 
choice of the $M_{Wb}$ cut agrees rather well with
our subtraction method result.

The $t \, W^-$ process has been studied
much less intensively than the other two modes, mostly owing
to the fact that it has a small rate at the Tevatron Run~II
($\sigma_{tW} = 0.094$ pb)
that is probably unobservable.  On the other hand, the rate
is fairly considerable at the LHC 
($\sigma_{tW} = 55.7$ pb) and it
may be observable there.  However,
detailed simulations
studying means by which the signal may be extracted from the
large $t \bar{t}$ background are still underway.
For completeness, we include in Tables~\ref{twtab1} and
\ref{twtab2} the LO rate (including the large $\log$ corrections
described above) of $t \, W^-$ production at the Tevatron
and LHC, for various choices of $m_t$, PDF, and scale,
with the canonical scale choice set to $\mu_0 = \sqrt{s}$.  As
usual, the $t$ and $\bar{t}$ rates have been summed, 
$V_{tb} = 1$, $V_{ts}, V_{td} = 0$ and no
decay BR's are included.  From these results, we see that
varying the scale by a factor of two produces a variation in the
resulting cross section of about $\pm  25 \%$ at the Tevatron
and $\pm 15 \%$ at the LHC.
This large scale dependence signals
the utility in having a 
full NLO in $\alpha_S$
computation of this process in order to have
a more theoretically reliable estimate for the cross section.

\begin{table}[t] 
\caption{The LO (with ${\cal O}( 1 / \log m_t^2 / m_b^2)$
corrections)
rates of $b \, g \ra t \, W^-$ (in pb) at the Tevatron Run~II.
The rate of $\bar{t}$ production is equal to the rate of $t$
production.}
\label{twtab1}
\begin{center} 
\begin{tabular}{lccccccc}
 & \multicolumn{3}{c}{CTEQ4L} & 
\multicolumn{3}{c}{MRRS(R1)} & 
 \\
$m_{t}$ (GeV) & $\mu = \mu_0 / 2$ & $\mu = \mu_0$ &
$\mu = 2 \mu_0$ & $\mu = \mu_0 / 2$ & 
$\mu = \mu_0$ & $\mu = 2 \mu_0$ & $\sigma_{tW}^{(mean)}$
\\[0.2cm] \hline \hline \\
170 & 0.0645 & 0.0505 & 0.0405 & 0.076  & 0.058  & 0.046  & 0.0545 \\
171 & 0.063  & 0.049  & 0.0395 & 0.074  & 0.0565 & 0.0445 & 0.053  \\
172 & 0.061  & 0.048  & 0.0385 & 0.072  & 0.055  & 0.0435 & 0.0515 \\
173 & 0.0595 & 0.0465 & 0.0375 & 0.07   & 0.053  & 0.042  & 0.05   \\
174 & 0.0575 & 0.045  & 0.0365 & 0.068  & 0.0515 & 0.041  & 0.0485 \\
175 & 0.056  & 0.044  & 0.0355 & 0.066  & 0.05   & 0.0395 & 0.047  \\
176 & 0.0545 & 0.0425 & 0.0345 & 0.064  & 0.049  & 0.0385 & 0.046  \\
177 & 0.053  & 0.0415 & 0.0335 & 0.062  & 0.0475 & 0.0375 & 0.0445 \\
178 & 0.0515 & 0.0405 & 0.0325 & 0.06   & 0.046  & 0.0365 & 0.0435 \\
179 & 0.05   & 0.039  & 0.0315 & 0.0585 & 0.0445 & 0.0355 & 0.042  \\
180 & 0.0485 & 0.038  & 0.0305 & 0.057  & 0.0435 & 0.0345 & 0.041  \\
181 & 0.0475 & 0.037  & 0.03   & 0.0555 & 0.0425 & 0.0335 & 0.040  \\
182 & 0.046  & 0.036  & 0.029  & 0.054  & 0.041  & 0.0325 & 0.0385 \\
\hline \hline 
\end{tabular}
\end{center}
\vspace{0.5in}
\end{table}

\begin{table}[t]
\caption{The LO (with ${\cal O}( 1 / \log (m_t^2 / m_b^2)$
corrections) rates for
$b \, g \ra t \, W^-$ (in pb) at the LHC.
The rate of $\bar{t}$ production is equal to the rate of $t$
production.}
\label{twtab2}
\begin{center} 
\begin{tabular}{lccccccc}
 & \multicolumn{3}{c}{CTEQ4L} & 
\multicolumn{3}{c}{MRRS(R1)} & 
 \\
$m_{t}$ (GeV) & $\mu = \mu_0 / 2$ & $\mu = \mu_0$ &
$\mu = 2 \mu_0$ & $\mu = \mu_0 / 2$ & 
$\mu = \mu_0$ & $\mu = 2 \mu_0$ & $\sigma_{tW}^{(mean)}$
\\[0.2cm] \hline \hline \\
170 & 33.0 & 28.2 & 24.5 & 39.0 & 33.0 & 28.4 & 30.6 \\
171 & 32.2 & 27.5 & 24.0 & 38.3 & 32.5 & 27.9 & 30.0 \\
172 & 31.6 & 27.1 & 23.6 & 37.6 & 31.8 & 27.4 & 29.4 \\
173 & 31.1 & 26.6 & 23.1 & 38.0 & 31.3 & 26.9 & 28.9 \\
174 & 30.5 & 26.1 & 22.7 & 36.2 & 30.7 & 26.4 & 28.4 \\
175 & 29.9 & 25.6 & 22.2 & 35.4 & 30.1 & 26.0 & 27.9 \\
176 & 29.4 & 25.2 & 21.8 & 34.8 & 29.6 & 25.5 & 27.4 \\
177 & 28.9 & 24.7 & 21.5 & 34.2 & 28.9 & 25.0 & 26.8 \\
178 & 23.3 & 24.2 & 21.1 & 33.6 & 28.4 & 24.6 & 26.3 \\
179 & 27.8 & 23.7 & 20.7 & 33.0 & 27.9 & 24.1 & 25.8 \\
180 & 27.2 & 23.3 & 20.3 & 32.4 & 27.4 & 23.7 & 25.4 \\
181 & 26.8 & 22.9 & 20.0 & 31.8 & 26.9 & 23.2 & 24.9 \\
182 & 26.3 & 22.5 & 19.6 & 31.2 & 26.4 & 22.9 & 24.5 \\
\hline \hline 
\end{tabular}
\end{center}
\end{table}

\clearpage

\section{New Physics in Single Top Production}
\label{stnewphysics}

As we have argued above, single top production is an important place
to search for physics beyond the SM.  This is reflected in the
growing body of literature in which the effect of loops of
new particles
on the single top rate is examined \cite{genstnp}.
In this section, we analyze several
possible signals for new physics that could manifest themselves in single top
production.  These signals can be classified as to whether they involve
the effects of a new particle (either fundamental or composite) that couple
to the top quark, or the effect of a modification of the SM coupling between
the top and other known particles.  In fact these two classifications
can be seen to overlap in the limit in which the additional particles
are heavy and decouple from the low energy description.  In this case
the extra particles are best seen through their effects on the couplings
of the known particles.

\subsection{Additional Nonstandard Particles}
\label{extrap}

Many theories of physics beyond the SM predict the existence of particles
beyond those required by the SM itself.  Examples include both the fundamental
super-partners in a theory with SUSY, and the composite top-pions found in
top-condensation and top-color models.  In order for some kind of additional
particle to contribute to single top production at a hadron collider, the new
particle must somehow couple the top to one of the lighter SM particles.
Thus, the new particle may be either a boson (such as a
$W^\prime$ vector boson that couples to top and bottom) 
or a fermion (such as a $b^\prime$ quark that couples to the 
$W$ boson and top).

Additional fermionic particles can couple the top and either one of
the gauge bosons or the Higgs boson.  In order to respect the 
color symmetry, this requires that the extra fermion occurs in a 
color triplet, and thus it is sensible to think of it as some
type of quark.  In order to be invariant under the electromagnetic symmetry,
this new ``quark'' should have either electric charge ($Q$)
$+2/3$ or $-1/3$ in order
for one to be able to construct couplings between the extra quark, the
top quark, and the known bosons.  Generally, we can refer to a
$Q = +2/3$ extra quark as a $t^\prime$ and a $Q = -1/3$ extra quark
as a $b^\prime$, though this does not necessarily imply that the
extra quarks are in the same representation under 
${\rm SU(2)}_L \times {\rm U(1)}_Y$ as the SM top and bottom.
Additional fermions are not generally expected to be
a large source of new contributions to single top production,
because of strong constraints from other observables.  On the other
hand we will see that there are models with additional fermions
to which single top production is a sensitive probe.

``Extra'' bosons can contribute to single top production
either by coupling top to the down-type quarks, in which case the boson
must have electric charge $Q = \pm 1$ in order to maintain the 
electromagnetic symmetry, or by coupling top to the charm or up quarks, in
which case the boson should be electrically neutral.  One could 
also imagine a boson
carrying an odd combination of color and electric charge that would allow
it to couple to both top and a lepton field.  Such bosons carrying
both baryon- and lepton-
number (leptoquarks) could arise, for example, from
the generators of the part of the gauge group of a GUT that connects the
electroweak and strong sectors of the GUT.  In that case one would naturally
suppose that these bosons have mass of the order of $M_{GUT}$, and thus
may not play an important role in single top production at the weak scale.
This GUT picture has the leptoquark as part of the gauge interactions,
so the question as to whether or not top observables are
an interesting means to study leptoquarks becomes a question
as to whether or not the leptoquark has some reason to prefer to couple to
the top quark.  One could imagine that the grand unified interaction contains
a sector corresponding to a family symmetry that could somehow
cause this to be the case.  Another interesting picture of leptoquarks
is one in which the SM quarks and leptons are bound states of some more
fundamental set of particles (which we will refer to as preons).
In that case the question as to whether or not the top quark is a good place
to look for evidence of the preons depends on how the model arranges the
various types of preons to build quarks and leptons.
However, at a hadron collider the possible
light parton initial states available are not suitable for production of
a single leptoquark, and thus are not particularly 
interesting in the context of single top production\footnote{
It is interesting to note that a leptoquark with
$Q = +2 / 3$ could play an important role in top decays through a process
such as $t \ra \nu \, L \ra \nu \, b \, \ell^+$.  This leads to a final
state that is identical to a SM top decay, but with a
very distinct kinematic structure}.  For this reason,
we will not focus on leptoquarks in the discussion below.

\subsubsection{Extra Quarks}

A simple extension of the SM is to allow for an extra set of quarks.
Such objects exist in a wide variety of extensions to the SM.
Examples of such theories include
the top see-saw versions of the top-color \cite{tctopseesaw}
and top-flavor \cite{tftopseesaw} models, which rely on additional
fermions to participate in a see-saw mechanism to generate the top mass;
SUSY theories with gauge mediated SUSY breaking that must be communicated
from a hidden sector in which SUSY is broken to the visible sector
through the interactions of a set of fields with SM gauge 
quantum numbers \cite{gaugemediated}; 
and even models with a fourth generation of fermions.

Direct search limits on extra
quarks require that they be quite massive ($m_{q^\prime} \ge 46 - 128$
GeV at the $95\%$ C. L., depending on the decay mode \cite{pdg}),
and thus they cannot appear
as partons in the incident hadrons at either Tevatron or LHC.
This prevents them from significantly affecting the $W$-gluon fusion
and $t \, W^-$ rates.
Thus, they are best observed either through their mixings
with the third family (and thus their effect on
the top couplings), or through direct production.

As a particular example, a fourth generation of quarks could
mix with the third generation through a generalized CKM matrix,
and this could allow $V_{tb}$ to deviate considerably from unity.
In this case, all three modes of single top production would be
expected to have considerably lower cross sections than the SM 
predicts.  This already shows how the separate modes of single top
production can be used to learn about physics beyond the SM.  Other
types of new physics could scale the three rates
independently.  Thus, if all three modes are measured to have cross
sections that are the same fraction of the SM rates, it is an
indication that the new physics modifies the top's coupling to the
bottom and $W$ (and not another pair of light particles), and
further that the modification is the same regardless of the momentum
flowing through the vertex (as is the case with the $W$-$t$-$b$
interaction in the SM).

\begin{figure}[t]
\epsfysize=1.5in
\centerline{\epsfbox{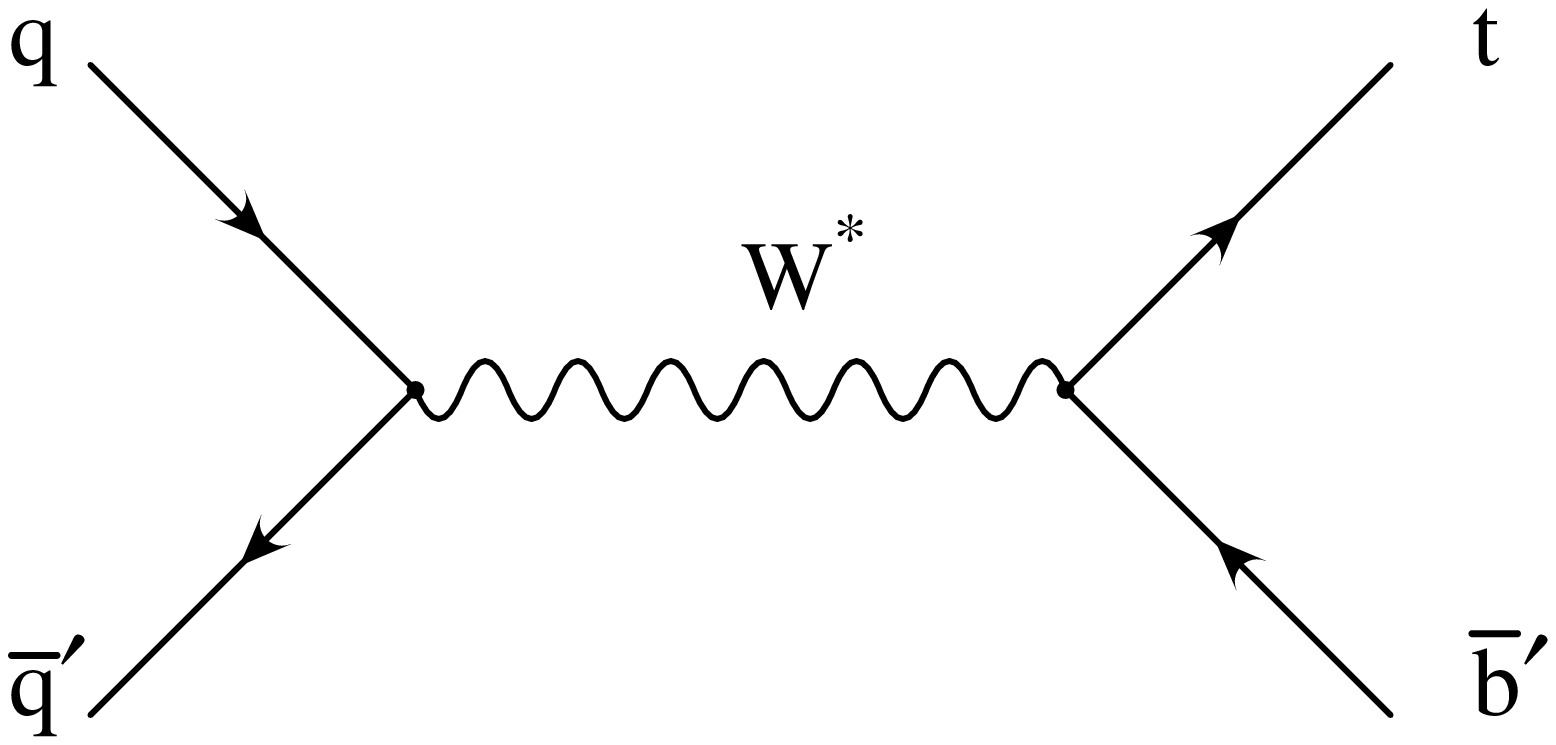}}
\caption{Feynman diagram for $s$-channel production of
a single top and a $b^\prime$:
$q \, \bar{q}^\prime \ra t \, \bar{b}^\prime$.}
\label{tbprimefig}
\end{figure}

In addition to mixing effects, one could also hope to observe direct
production of one of the fourth generation quarks, through reactions
such as $q \, \bar{q}^\prime \ra t \, \bar{b^\prime}$, shown in
Figure~\ref{tbprimefig}.  If the $b^\prime$ is somewhat heavier
than the top, and $V_{tb^\prime}$ is large, this process could
be more important than QCD production of $b^\prime \, \bar{b}^\prime$
because of the greater phase space available to the lighter top.
The production rates will depend on the
magnitude of the $W$-$t$-$b^\prime$ coupling ($|V_{tb^\prime}|^2$ in the
model with a fourth family) and the mass of the $b^\prime$.  In 
Figure~\ref{bprimeratefig} we present the NLO rate for $t \bar{b^\prime}$
production (as well as $\bar{t} b^\prime$ production) without any
decay BR's.
Since the
$|V_{tb^\prime}|^2$ dependence may be factored out, these rates
assume $V_{tb^\prime} = 1$.  
The collider signatures resulting from
such a process depend on the decay modes available to the $b^\prime$.
If $m_{b^\prime} > m_t + m_W$, it is likely to decay into a top quark and
a $W^-$, and the events will have a $t \, \bar{t}$ pair with an additional
$W^\pm$ boson.  If this decay mode is not open, loop induced decays
such as $b^\prime \ra b \, \gamma$ may become important, resulting in
a signature $t \, \bar{b}$ plus a hard photon whose invariant mass
with the $b$ quark will reconstruct the mass of the $b^\prime$.

\begin{figure}[p]
\epsfysize=5.0in
\centerline{\epsfbox{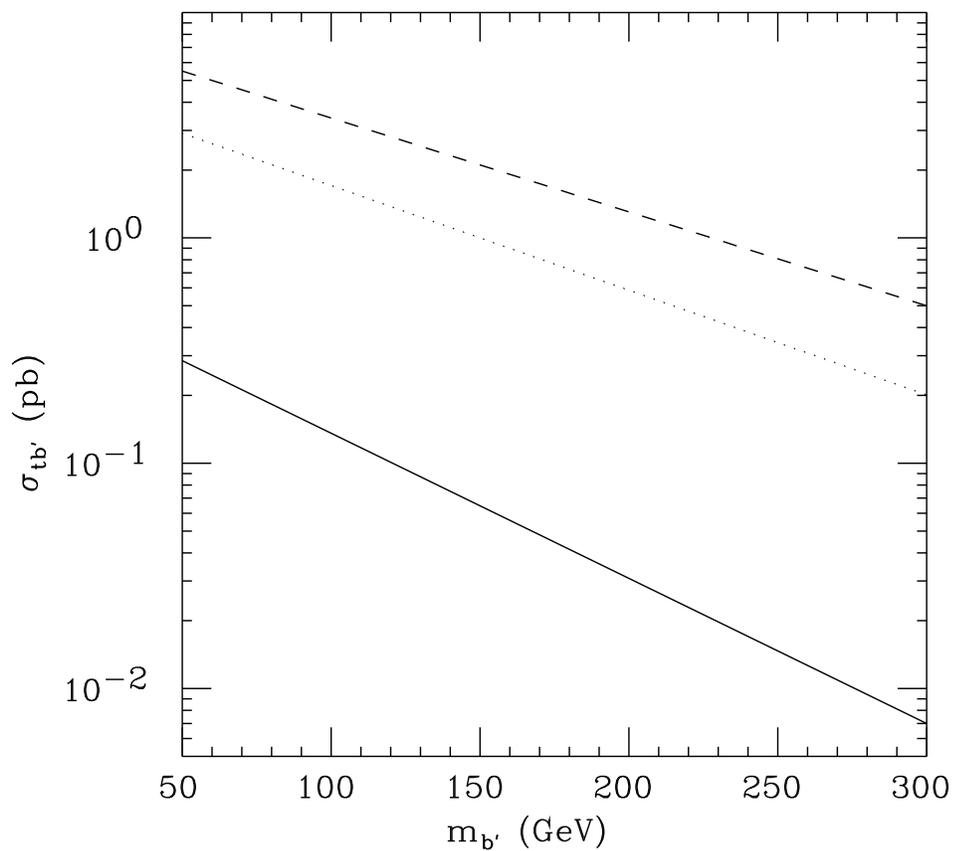}}
\caption{The NLO rates (in pb) for the process 
$q \, \bar{q}^\prime \ra W^* \ra t \, \bar{b}^\prime$ 
for various $b^\prime$
masses at the Tevatron (solid curve) and LHC (dashed curve),
assuming $V_{tb^\prime}= 1$.  At the Tevatron, the rates of
$q \, \bar{q}^\prime \ra W^* \ra \bar{t} \, b^\prime$ is equal to the
$t \, \bar{b}^\prime$ rate.  The $\bar{t} \, b^\prime$ rate at the
LHC is shown as the dotted curve.}
\label{bprimeratefig}
\end{figure}

\subsubsection{Extra Gauge Bosons}

Another simple extension of the SM is to postulate the existence of
a larger gauge group which somehow reduces to the SM gauge
group at low energies.  Such theories naturally have additional 
gauge bosons, some of which may prefer to couple to the top (or even
the entire third family).  Examples of such theories include 
the top-color \cite{topcolor} and top-flavor 
\cite{tftopseesaw, ehabsu2a} models,
which give special dynamics to the third family in order to explain the
large top mass.  As a specific example, we will consider the 
top-flavor
model with an extra ${\rm SU(2)}_h$ gauge symmetry that generates
a top mass through a see-saw effect \cite{tftopseesaw}.

This model has an
over-all gauge symmetry of
SU(2$)_{h} \times$ SU(2$)_{l} \times$ U(1$)_{Y}$, and thus
there are three additional weak
bosons ($W^{\prime \pm}$ and $Z^{\prime}$).
The first and second generation fermions 
and third family leptons transform under
SU(2$)_{l}$, while the third generation quarks transform
under SU(2$)_{h}$.  As was alluded to before,
in order to cancel the anomaly and provide
a see-saw mechanism to generate the top mass, an additional
doublet of heavy quarks whose left-handed components transform
under ${\rm SU(2)}_l$ and right-handed components transform
under ${\rm SU(2)}_h$ is also present.

A set of scalar fields transforming under
both SU(2$)_{l}$ and SU(2$)_{h}$ acquire a VEV, $u$,
and break the symmetry to SU(2$)_{l + h} \times$
U(1$)_{Y}$.  From here the usual electro-weak symmetry breaking
can be accomplished by introducing a scalar doublet which acquires
a VEV $v$, further breaking the gauge symmetry to U(1$)_{EM}$.
We write the
covariant derivatives for the fermions as,
\bea
D^{\mu} &=& \partial^{\mu} +
i g_{l} \; T^{a}_{l} \; {W^{a}}^{\mu}_{l} +
i g_{h} \; T^{a}_{h} \; {W^{a}}^{\mu}_{h} +
i g_1 \; \frac{Y}{2} \; B^{\mu} ,
\eea
where $T^{a}_{l (h)}$ are the generators for SU(2$)_{l (h)}$,
$Y$ is the hyper-charge generator, and ${W^{a}}^{\mu}_{l (h)}$
and $B^{\mu}$ are the gauge bosons for the SU(2$)_{l (h)}$ and
U(1$)_{Y}$ symmetries.  The gauge couplings may be written,
\bea
g_{l} = \frac{e}{\sin \theta_W \cos \phi} \, , \hspace{1cm}
g_{h} = \frac{e}{\sin \theta_W \sin \phi} \, , \hspace{1cm}
g_1 = \frac{e}{\cos \theta_W} \, ,
\eea
where $\phi$ is a new
parameter in the theory.  Thus this theory is determined
by two additional quantities $x = u / v$,
the ratio of the two VEV's, and $\sin^2 \phi$, which
characterizes the mixing between the heavy and light SU(2)
gauge couplings.

At leading order, the heavy bosons
are degenerate in mass,
\bea
{M^2}_{Z^{\prime}, W^{\prime}} =
{M_0}^2 \left( \frac{x}{\sin^2 \phi \cos^2 \phi}
+ \frac{\sin^2 \phi}{\cos^2 \phi} \right) ,
\eea
where 
${M_0}^2 = \frac{ e^2 v^2 }{4 \sin^2 \theta_W \cos^2 \theta_W}$.
We can thus parameterize the model by the
heavy boson mass, $M_{Z^{\prime}}$, and the
mixing parameter\footnote{As shown in \cite{ehabsu2a},
for $\sin^2 \phi \leq$ 0.04, the
third family fermion coupling to the heavy gauge bosons can
become non-perturbative.  Thus we restrict ourselves to considering
$0.95 \geq \sin^2 \phi \geq$ 0.05.},
$\sin^2 \phi$.
Low energy data requires that the mass
of these heavy bosons, $M_{Z^{\prime}}$,
be greater than about $900$ GeV \cite{ehabsu2b}.

\begin{figure}[t]
\epsfysize=1.5in
\centerline{\epsfbox{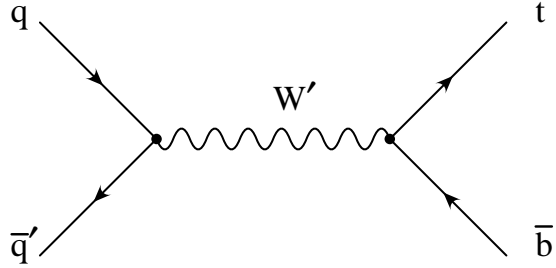}}
\caption{Feynman diagrams illustrating how a $W^\prime$ boson
can contribute to single top production through
$q \, \bar{q}^\prime \ra W^\prime \ra t \, \bar{b}$.}
\label{wpfig}
\end{figure}

The additional $W^\prime$ boson can contribute to the $s$-channel mode
of single top production through virtual exchange of a $W^\prime$
as shown in Figure~\ref{wpfig}
\cite{es}.  In particular, if enough energy is
available, the $W^\prime$ may be produced close to on-shell, and a
resonant enhancement of the signal may result.
Since the additional
diagrams involve a virtual $W^\prime$, they will
interfere with the SM $W$-exchange diagrams, and thus the net
rate of single top
production can be increased or decreased as a result, though the
particular model under study always results in an increased $s$-channel
single top rate.  In Figure~\ref{wpratefig} the resulting NLO $s$-channel rate
for $q \, \bar{q}^\prime \ra W, W^\prime \ra t \, \bar{b}$ at Tevatron
and LHC is shown, as a function of the $W^\prime$ mass, 
for a few values of $\sin^2 \phi$.  The rate for $\bar{t}$ production
through the same process is shown as well.
While the final state particles for this case
are the same as the SM $s$-channel mode, the distribution of the invariant
mass of the $t \, \bar{b}$ system could show a Breit-Wigner
resonance effect around $M_{W^\prime}$, which serves to identify this type
of new physics.  However, if the mass of the $W^\prime$ is large compared
to the collider energy, and its width broad, the resonance shape can
be washed out even at the parton level.
Jet energy smearing from detector
resolution effects will further make such a resonance 
difficult to identify.

\begin{figure}[p]
\epsfysize=5.0in
\centerline{\epsfbox{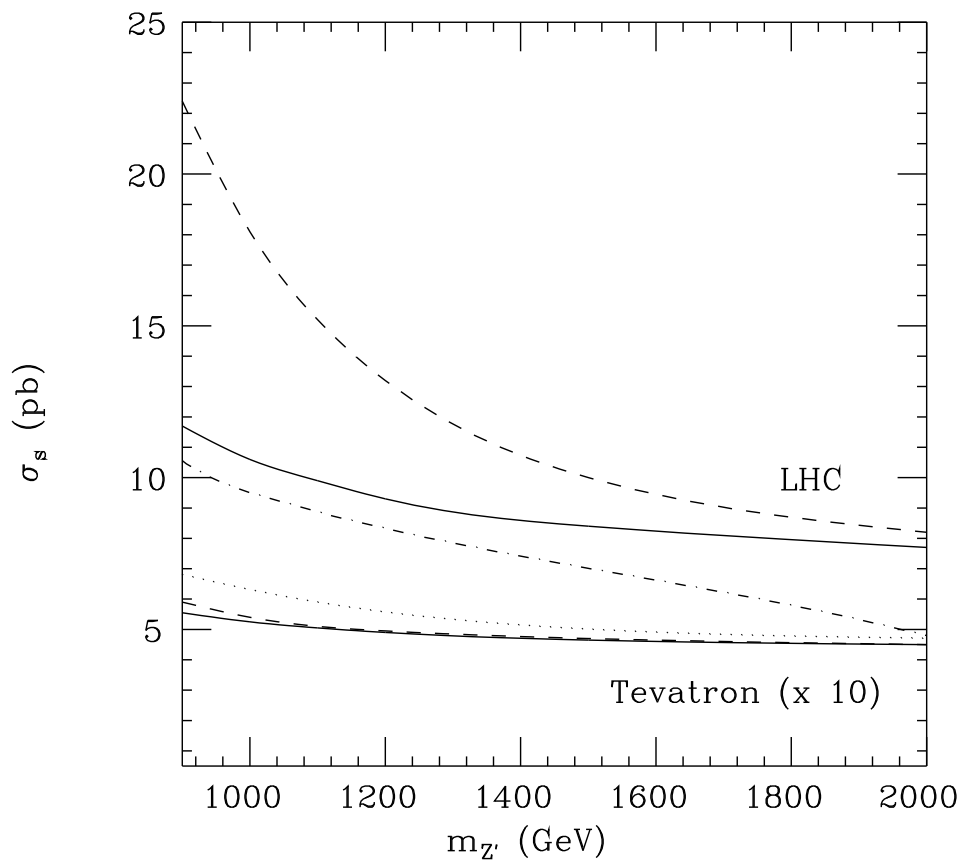}}
\caption{The NLO rate of 
$q \, \bar{q}^\prime \ra W, W^\prime \ra t \, \bar{b}$
(in pb)
at the Tevatron (lower curves)
and LHC (upper curves), for the top-flavor model with
$\sin^2 \phi = 0.05$ (solid curves) and 
$\sin^2 \phi = 0.25$ (dashed curves), as a function of
$M_{Z^\prime} = M_{W^\prime}$.
The Tevatron cross sections are multiplied by a factor of 10.
At the Tevatron, the $\bar{t}$ production rate is equal to the
$t$ rate.  At the LHC the $\bar{t}$ rates are shown for
$\sin^2 \phi = 0.05$ (dotted curve) and
$\sin^2 \phi = 0.25$ (dot-dashed curve).}
\label{wpratefig}
\end{figure}

A $t$-channel
exchange of the $W^\prime$ is also possible, but in that case a negligible
effect is expected because the boson must have a space-like momentum,
and thus the additional contributions are suppressed by $1 / M_{W^\prime}$,
and are not likely to be visible.  This argument applies quite
generally to any heavy particle's effect on single top production.  The 
$s$-channel rate is quite sensitive to a heavy particle because of the
possibility of resonant production, whereas the $t$-channel rate
is insensitive because the space-like exchange is suppressed by the
heavy particle mass.

Clearly, the existence of a $W^\prime$ will not influence the rate of
$t \, W^-$ production, but it could allow for exotic production modes
such as $b \, g \ra t \, {W^\prime}$.  If the ${W^\prime}$ has a strong
coupling with the third family, then one would expect that its dominant
decay should be into $b \, \bar{t}$, and thus a final state of
$t \, \bar{t} \, b$ would result with the $t \, \bar{b}$ invariant mass
reconstructing the $W^\prime$ mass.
Current limits on the $W^\prime$ mass in the
top-flavor model make this mode nonviable at the Tevatron
and unpromising at the LHC,
with a cross section of 1.14 pb
for $M_{W^\prime} = 900$ GeV
and $\sin^2 \phi = 0.05$ including the large $\log$ contributions
described in Section~\ref{twsec}. 
However, an observation of this signal would be a clear indication
of the nature of the new physics.  

\subsubsection{Extra Scalar Bosons}

Scalar particles appear in many theories,
usually associated with the spontaneous breaking of a symmetry.  
In the SM and the minimal
supersymmetric extension, fundamental scalar fields of both neutral
and charged character are present in the theory, and are expected to
have a strong coupling with the top because of the role they play in
generating fermion masses.  In dynamical models such as the top-condensate
and top-color assisted technicolor models, scalar particles exist as
bound states of top and bottom quarks
(as was seen in Chapter~\ref{intro} this is how these models deal
with the fine-tuning and naturalness problems of the SM).  
These composite scalars also
have a strong coupling to the top because of their role in the generation
of the top mass.  This illustrates the fact that the large top mass
naturally makes it a likely place to look for physics associated with the
EWSB.

\begin{figure}[t]
\epsfysize=1.5in
\centerline{\epsfbox{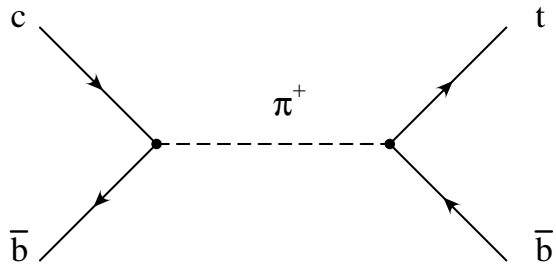}}
\caption{Feynman diagram illustrating how a charged top-pion
can contribute to single top production through 
$c \, \bar{b} \ra \pi^+ \ra t \, \bar{b}$.}
\label{toppionfig}
\end{figure}

An illustrative example is provided by the charged composite
top-pions ($\pi^\pm$)
of the top-color model,
which can be produced in the $s$-channel through
$c \, \bar{b}$ fusion \cite{toppion}, 
$c \, \bar{b} \ra \pi^+ \ra t \, \bar{b}$.
The leading order Feynman diagram is shown in Figure~\ref{toppionfig}.
In this case the strong $\pi^+$-$c$-$\bar{b}$ coupling comes from mixing
between the $t$ and $c$ quarks.  In order to avoid constraints from the CKM
matrix, this requires the mixing to occur between right-handed $t$ and $c$
quarks, and thus this interaction has a right-handed nature that will prove
interesting when we study top polarization below.

\begin{figure}[p]
\epsfysize=5.0in
\centerline{\epsfbox{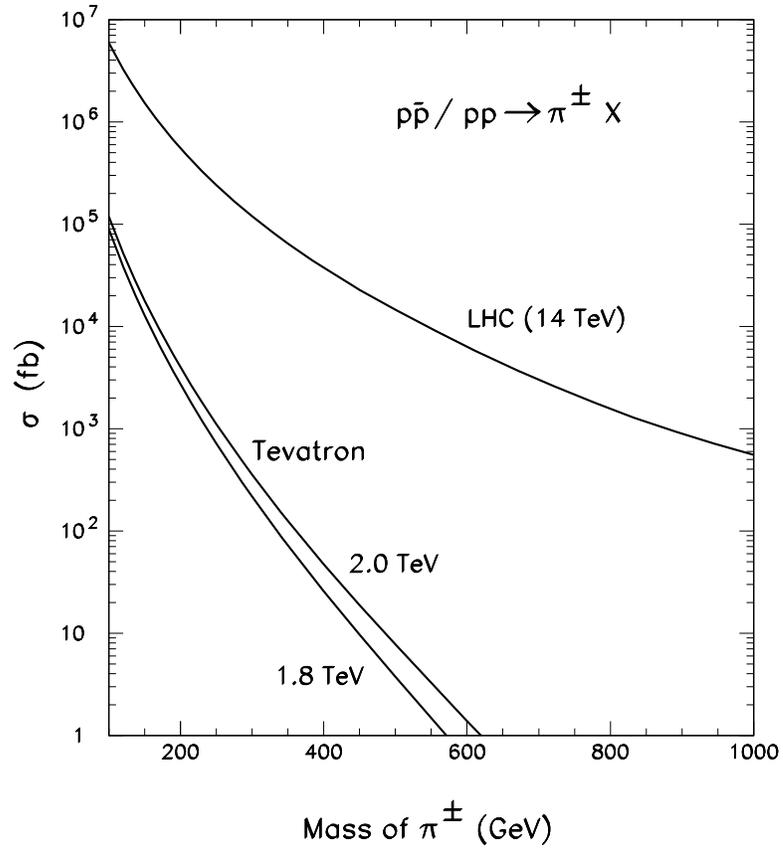}}
\caption{The LO rate of single top production through the
reaction $c \, \bar{b} \ra \pi^+ \ra t \, \bar{b}$
as a function of $M_{\pi^\pm}$, assuming a $t_R$-$c_R$ mixing
of $20\%$.  These rates include $t$ and $\bar{t}$
production, which are equal for both Tevatron and LHC.}
\label{toppionratefig}
\end{figure}

Like the $W^\prime$, the $\pi^\pm$ contributes to the $s$-channel topology
of single top production and can allow large resonant contributions.
However, unlike the $W^\prime$, the $\pi^+$ does not have a significant
interference with the SM amplitudes, because the SM contribution is mostly
from light quarks ($u$ and $\bar{d}$).  In Figure~\ref{toppionratefig}, we
present the NLO single top rate from the top-pion process \cite{nlotoppion},
for a variety of $\pi^\pm$ masses with the $t_R$-$c_R$ mixing set equal to
$20\%$.  
The two other modes of single top production are once again relatively
insensitive to the $\pi^\pm$.  The $t$-channel process has additional
contributions suppressed by $1/ M^2_{\pi^\pm}$ and the fact that the $\pi^\pm$
does not couple to light quarks.  The $t \, W^-$ mode is insensitive because
presumably the $\pi^\pm$ is generally distinguishable from a 
$W^\pm$ boson, and so $g \, b \ra \pi^- \, t \ra \bar{t} \, b \, t$
will not be mistaken for $t \, W^-$ production.

Different types of scalar particles that couple top and bottom can
be analyzed in a similar fashion.  The $s$-channel mode allows for
resonant production, which can show a large effect, where-as the
$t$-channel mode is suppressed by the space-like momentum 
(and large mass) of the
exchanged massive particle.  The $t \, W^-$ mode is insensitive because
in that case the $W$ is actually observed in the final state.
The technipions in a technicolor model can contribute
to single top production in this way \cite{tcst}.  
Another example is provided by SUSY models
with broken $R$-parity, in which
the scalar partners of the leptons (sleptons)
can couple with the top and bottom quarks, and will contribute to
single top production \cite{rpviol}.

\begin{figure}[t]
\epsfysize=1.5in
\centerline{\epsfbox{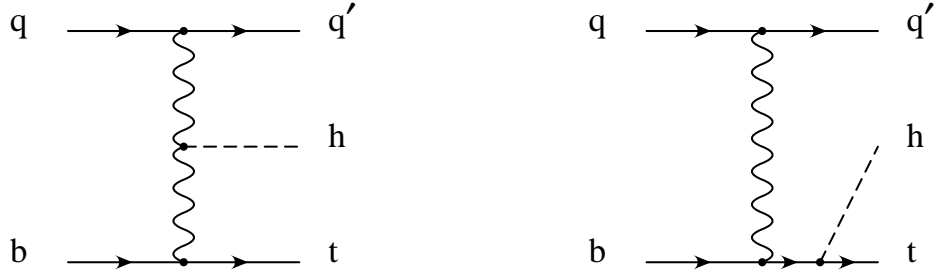}}
\caption{Feynman diagrams for associated production of a
neutral scalar and single top quark:
$q \, b \ra q^\prime \, t \, h$.}
\label{hstfig}
\end{figure}

As a final note, there is the interesting process in which a neutral
scalar (like the Higgs of the SM) is produced in association with a
single top quark \cite{hsinglet}.
Feynman diagrams are shown in Figure~\ref{hstfig}.
This process is of interest because while the magnitude of the $h$-$W$-$W$
and $h$-$t$-$t$ couplings can be measured independently by studying 
$q \, \bar{q}^\prime \ra W^* \ra W \, h$ and 
$q \, \bar{q} \: (g \, g) \ra h \, t \, \bar{t}$, the relative phase between
the couplings can be found from the process $q \, b \ra q^\prime \, t \, h$,
as that phase information is contained in the interference between the
two diagrams shown in Figure~\ref{hstfig}.  This process is extremely small
compared to the other two mentioned (with a SM cross section of 
$6 \times 10^{-5}$ pb at the
Tevatron and 0.1 pb at the LHC for $m_h = 100$ GeV and
including both $t$ and $\bar{t}$ production),
and thus it is not promising a discovery
mode.  The small SM rate results from the fact that the interference
term provides a strong cancellation of the rate, reducing it by a factor
of about 5.  This indicates that this process is very strongly
sensitive to any physics that modifies the relative phase between the
$h$-$t$-$t$ and $h$-$W$-$W$ couplings from the SM relation.
Thus, it contains important information not available in the other two
processes.

\subsection{Modified Top Quark Interactions}

Another interesting set of properties of the top that can be 
studied in single top production are the top couplings to light particles.
As was shown in Section~\ref{ewcl}, the electroweak chiral Lagrangian provides
a powerful way to study such effects model-independently.
Following the EWCL approach, we write an effective Lagrangian to describe
low energy physics as,
\bea
 {\cal L}_{eff} &=& {\cal L}_{SM} + {\cal L}_4 + {\cal L}_5,
\eea
where ${\cal L}_{SM}$ refers to the usual SM Lagrangian described in
Chapter~\ref{intro}, and ${\cal L}_4$ and ${\cal L}_5$ are the Lagrangians
containing deviations from the SM in terms of operators of
mass dimension 4 and 5, respectively.

Terms which
have the potential to modify single top production include
mass dimension 4 operators \cite{ehabewcl},
\bea
  \label{ewcleq1}
  {\cal L}_4 &=& \frac{ e }
  {\sqrt{2} \, \sin \, \theta_W} 
  {W^+}_\mu \left( 
    \kappa_{Wtb}^L \, e^{i \, \phi_{Wtb}^L} \,
    \bar{b} \, \gamma^\mu \, P_L \, t +
    \kappa_{Wtb}^R \, e^{i \, \phi_{Wtb}^R} \,
    \bar{b} \, \gamma^\mu \, P_R \, t \right) \\[0.3cm] & & \,
    + \frac{e \, \kappa_{Ztc}}
    {2 \, \sin \theta_W \, \cos \theta_W} 
    {Z}_\mu \left( \;
    e^{i \, \phi_{Ztc}^L} \,
    \sin \theta_{Ztc} \, \bar{c} \, \gamma^\mu \, P_L \, t \:+
    \right. \nonumber \\[0.5cm] & & \, \left. \,
    e^{i \, \phi_{Ztc}^R} \,
    \cos \theta_{Ztc} \, \bar{c} \, \gamma^\mu \, P_R \, t \; 
    \right) + H.c. , \nonumber
\eea
which can be classified as two operators which modify the SM top weak
interactions, as well as two flavor-changing neutral current operators
involving the $Z$ boson, $t$, and $c$ quarks.
Additional dimension 4 FCNC operators with the $c$ quark replaced by the
$u$ quark are also possible.
We have included the $CP$ violating
phases $\phi_{Wtb(Ztc)}^{L(R)}$ in the interactions
for generality, though they are not always considered in the literature.
In addition there are dimension 5 operators that involve interactions
between new sets of particles and the top\footnote{There are also dimension
five operators involving the sets of particles that already appear in
Equation~\ref{ewcleq1} \cite{dim5}.  Since naive dimensional analysis
\cite{nda} suggests that at low energies
these operators are less significant than their
dimension four counterparts, we limit ${\cal L}_5$ to the dimension
5 operators which involve only new sets of fields.}
and can contribute to single top production.  These include
the FCNC operators,
\newpage
\bea
  \label{ewcleq2}
  {\cal L}_5 &=& \frac{ g_S \, \sqrt{2} \, G^a_{\mu \nu} }
    { \Lambda_{gtc} }  \left( \:
    e^{i \, \phi_{gtc}^L} \;
    \sin \theta_{gtc}^L \: \bar{c} \: T^a \;
    \sigma^{\mu \nu} \, P_L \, t
    \right. \\[0.4cm] & & \; \left.
  + \; e^{i \, \phi_{gtc}^R} \;
    \sin \theta_{gtc}^R \: \bar{c} \: T^a \; 
    \sigma^{\mu \nu}
    \, P_R \, t
    \right) \nonumber \\[0.3cm] & & \,
  + \frac{2\, \sqrt{2} \, e \, F_{\mu \nu} }{3 \, \Lambda_{\gamma tc} }
    \left( \; e^{i \, \phi_{\gamma tc}^L} \;
    \sin \theta_{\gamma tc}^L \, \bar{c} \:
    \sigma^{\mu \nu} \, P_L \, t
  + e^{i \, \phi_{\gamma tc}^R} \;
    \cos \theta_{\gamma tc}^R \, \bar{c} \:
    \sigma^{\mu \nu}
    \, P_R \, t
    \right) \nonumber + H.c.,
\eea
which couple the charm quark to the top and gluon or photon
fields.  Once again, we have included $CP$ violating phases
$\phi_{gtc (\gamma tc)}^{L(R)}$ which are not generally considered
in the literature.
Additional operators with the charm replaced by the up
quark are also possible.
As dimension 5 operators, these terms have couplings with
dimension of inverse mass that have been written in the form of
$1 / \Lambda_{gtc}$ and $1 / \Lambda_{\gamma tc}$.
If the underlying theory is strongly coupled, these mass scales
may be thought of as the energy scale in which the SM breaks down
and must be replaced with the underlying theory.  However, 
it should be kept in mind that if the
underlying theory is weakly coupled, this interpretation is somewhat
obscured by the fact that the energy scales $\Lambda$ will also include
small factors of the fundamental interaction strength and loop suppression
factors.  Even in this case, an experimental constraint on $\Lambda$
is very useful because it will provide constraints on the parameters
of an underlying model that makes a prediction for it.

The dimension 4 terms which modify the $W$-$t$-$b$ vertex will clearly
have a large impact on single top production \cite{doug}.  However,
$\kappa_{Wtb}^R$ is already very strongly constrained 
by low energy
$b \ra s \, \gamma$ data \cite{cleo}, which requires \cite{kappaR},
\bea
  -.0035 \geq ( \; \kappa_{Wtb}^R \cos \phi_{Wtb}^{R} 
  + 20 \, {\kappa_{Wtb}^R}^2 \; ) \leq 0.0039 ,
\eea
provided that $\kappa_{Wtb}^L$ is somewhat smaller than 1.
Given this strong constraint, it is
unlikely that further information about $\kappa_{Wtb}^R$
can be gleaned from single top production, so
we will assume $\kappa_{Wtb}^R = 0$ in the discussion below.
On the other hand, all three modes are sensitive to $\kappa_{Wtb}^L$,
and will be proportional to 
$(1 + {\kappa_{Wtb}^L}^2 + 2 \, \kappa_{Wtb}^L \, \cos \phi_{Wtb}^L)$ 
much the same way that
they will all be sensitive to $V_{tb}$ in the SM\footnote{This is because 
the dimension 4 term that is proportional to $\kappa_{Wtb}^L$ in 
${\cal L}_4$ does not depend on the momenta of the interacting particles,
as is the case for the SM $W$-$t$-$b$ interaction.  
For higher dimension $W$-$t$-$b$ operators,
which may depend on the momenta, each single top mode will respond
differently to the new interaction, and thus could be used to distinguish
one operator from another.}.

\begin{figure}[t]
\epsfxsize=6.0in
\centerline{\epsfbox{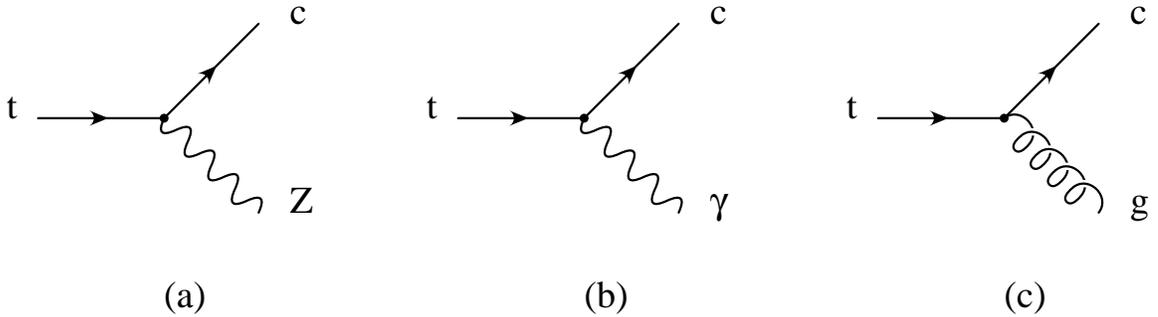}}
\caption{Feynman diagrams showing FCNC top decays through
(a)~$t \ra Z \, c$, (b)~$t \ra \gamma \, c$, and (c)~$t \ra g \, c$.}
\label{tncdecayfig}
\end{figure}

The flavor-changing neutral current terms in ${\cal L}_4$ and ${\cal L}_5$
will also contribute to single top production, and since they involve particles
lighter than the top mass, will also contribute to top decays
through Feynman diagrams such as those shown in Figure~\ref{tncdecayfig},
which illustrate FCNC $t$ decays to $c$.  The FCNC interactions between
$t$ and $u$ will allow for exotic decays of the same type, but with
the $c$ quark exchanged with a $u$ quark.
One
could hope to learn about these anomalous FCNC couplings both by studying
single top production and top decays.  However, this brings us back to the
problem with using top decays to determine the magnitude of a coupling -
the decay can provide information about the relative branching fraction of
the exotic decay compared to the SM top decay $t \ra W^+ \, b$, but since
it does not allow one to measure the top decay width, it cannot provide
a limit on the size of the exotic operator without 
first making an assumption
concerning the nature of the $W$-$t$-$b$ interaction.
In fact, one might think that single top would suffer from the same
difficulty in distinguishing the magnitude of
new physics in the $W$-$t$-$b$ interaction from new physics in a
FCNC interaction.  However, as we shall see, one can use the three
modes of single top production separately to disentangle the FCNC
new physics from the possibility of $W$-$t$-$b$ new physics.

The three FCNC operators have a similar structure of a light $c$ (or $u$)
quark interacting with a top and a neutral vector boson.  Thus, we can
discuss their impact on the three single top processes rather generally
by considering the specific example of the $Z$-$t$-$c$ operator.
In examining the FCNC operators in Equations~\ref{ewcleq1} and
\ref{ewcleq2}, we note that they can have left-handed and
right-handed interactions with different interaction coefficients
(and even different phases).
For now we will restrict our discussion to the case where
all of the phases are zero, and discuss only the magnitude of the
interactions, set by $\Lambda_{gtc}$, $\Lambda_{\gamma tc}$, 
and $\kappa_{Ztc}$.  We will return to the subject of exploring their
chiral structure when we consider top polarization in 
Section~\ref{polarization}.

\begin{figure}[t]
\epsfysize=1.4in
\centerline{\epsfbox{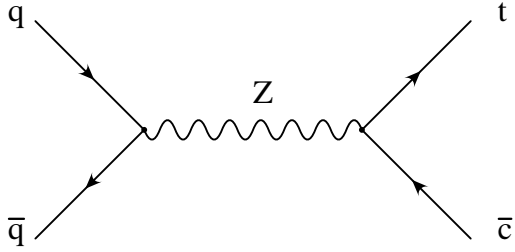}}
\caption{Feynman diagram showing how a FCNC $Z$-$t$-$c$ interaction
contributes to the $s$-channel mode of single top production
through $q \, \bar{q} \ra Z^* \ra t \, \bar{c}$.}
\label{schanztcfig}
\end{figure}

The $Z$-$t$-$c$ 
operator will allow for additional contribution to the $s$-channel
mode of single top production through reactions such as 
$q \, \bar{q} \ra Z^* \ra t \, \bar{c}$,
shown in Figure~\ref{schanztcfig}.  This reaction has different
initial and final state from the SM $s$-channel mode, and thus there is
no opportunity for interference between SM and new physics contributions.
The fact that the new physics process has a $\bar{c}$ instead of a
$\bar{b}$ in the final state has a drastic practical consequence that the
new physics production mechanism probably cannot be experimentally extracted
at all, because in order to separate the $s$-channel mode from the large
$t \, \bar{t}$ and $W$-gluon fusion backgrounds, it is necessary to tag
the $\bar{b}$ produced in association with the top in the $s$-channel
mode, in addition to the $b$ from the top decay.  Thus, while a FCNC operator
could contribute to $s$-channel production of a single top, it will not be
counted as such\footnote{It could be possible to search for $s$-channel
production via a FCNC with a specialized strategy differing from the
usual one employed to extract the $W^*$ process, but such a search
will suffer from large backgrounds from $t \, \bar{t}$ and
$W$-gluon fusion single top processes.}.

\begin{figure}[t]
\epsfysize=1.8in
\centerline{\epsfbox{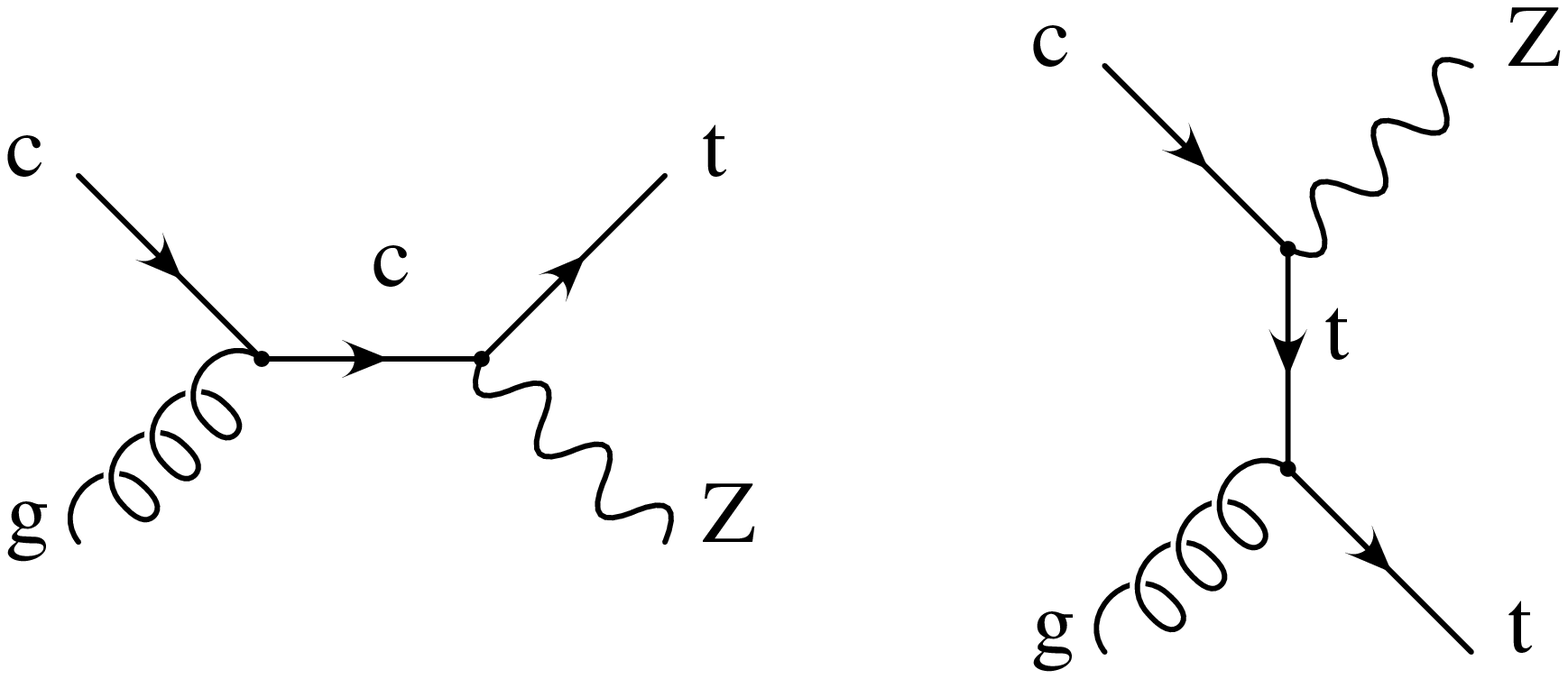}}
\caption{Feynman diagrams showing how a FCNC $Z$-$t$-$c$ interaction
contributes to the exotic mode of single top production
$g \, c \ra t \, Z$.}
\label{gctZfig}
\end{figure}

The $t \, W^-$ mode cannot receive a contribution from a FCNC, though
a FCNC will generally allow for new exotic production mechanisms such
as $g \, c \ra t \, Z$ shown in Figure~\ref{gctZfig}.  From this
consideration, along with the analysis of the $t \, W^-$ mode in
Section~\ref{extrap}, we see that the $t \, W^-$ mode has a special
quality because both the top and the $W$ are in the final state 
(and thus identifiable).  Thus, it is sensitive to new physics which
modifies the $W$-$t$-$b$ interaction\footnote{
Of course it is also sensitive to the $W$-$t$-$s$ and $W$-$t$-$d$
interactions, but these have been measured to be small 
\cite{pdg}.}, but it is not sensitive to
nonstandard physics involving new particles or FCNC's.  Thus, the
$t \, W^-$ mode represents a chance to study the $W$-$t$-$b$
vertex without contamination from other types of new physics.

\begin{figure}[t]
\epsfysize=1.8in
\centerline{\epsfbox{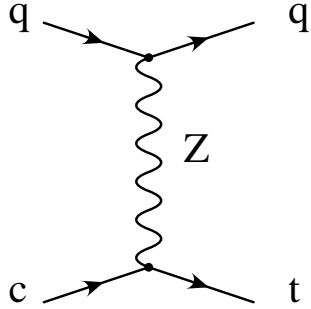}}
\caption{Feynman diagram showing how a FCNC $Z$-$t$-$c$ interaction
contributes to the $t$-channel mode of single top production
through $c \, q \ra t \, q$.}
\label{tchanztcfig}
\end{figure}

The $W$-gluon fusion mode of single top production is quite sensitive to
a FCNC involving the top and one of the light partons, through
processes such as $c \, q \ra t \, q$, from Feynman diagrams
such as those shown in Figure~\ref{tchanztcfig}.
The FCNC operators involve a different set of spectator quarks
in the reaction, and thus they do not interfere with the SM
$t$-channel process.  In fact,
because the $W$-gluon fusion mode
requires finding a $b$ inside a hadron, which has less probability
than finding a lighter parton, the FCNC's involving $u$ or
$c$ quarks already receive an enhancement relative to the SM $t$-channel
rate purely from the parton densities.  This can somewhat compensate
for a (presumably) smaller FCNC coupling.  This shows the sense in which
the $t$-channel single top mode is sensitive to the top quark's decay
properties.  The same type of new physics which opens up new top decay
modes (and thus modifies the top's total width) will also modify the
$t$-channel rate of single top production, because the same light partons
into which the top may decay are also responsible for producing
single tops in the $t$-channel process.  Thus, one can think of
the $t$-channel process as a kind of measure of the 
inclusive top width.

\begin{figure}[p]
\epsfysize=4.0in
\centerline{\epsfbox{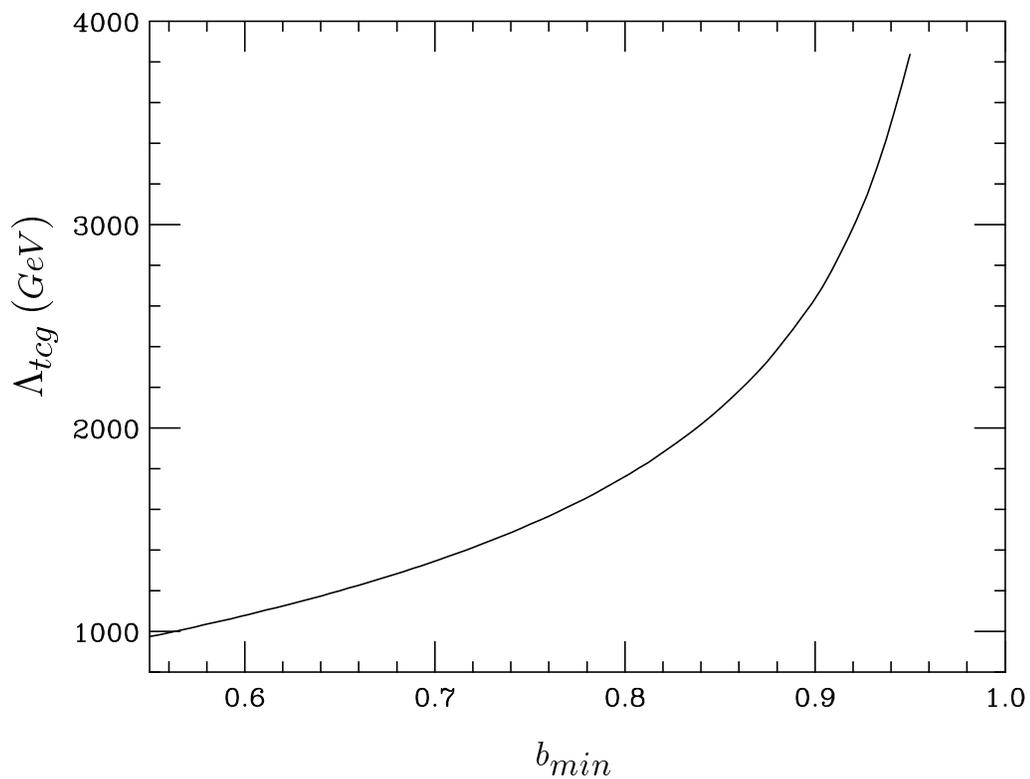}}
\caption{The correlation between the
maximum cross section of $q \, \bar{q} \ra t \, \bar{c}$,
$\sigma_{tc}$,
and the 
minimum BR($t \ra W \, b$) assuming the $t$-$c$-$g$ operator
is the only source of nonstandard physics in top decays, $b_{min}$.}
\label{tcgcorrfig}
\end{figure}

Because of the strong motivation to use single top production to
study FCNC operators involving the top quark, detailed
simulations of the effect of the $g$-$t$-$c$ operator on single top
production were performed \cite{tcgmenehab}, and found that this operator
could be constrained by the process $q \, \bar{q} \ra t \, \bar{c}$ to
$\Lambda_{gtc} \geq 4.5$ 
TeV at Run~II of the Tevatron if no new physics
signal were to be found.  Further refinements on this idea
\cite{tcgother} showed that it could be improved by including other
reactions such as $g \, c \ra t$, $g \, c \ra g \, t$,
$q \, c \ra t \, q$, and $ g \, g \ra t \, \bar{c}$
to $\Lambda_{gtc} \geq 10.9$ TeV at the Tevatron Run~II.
In \cite{tcgmencp} it
was pointed out that since the same $g$-$t$-$c$ operator contributes
to both single top production and the top decay $t \ra c \, g$, one
can look for the correlation in the BR($t \ra c \, g$) and the
anomalous rate of single top production to verify that this particular
operator is responsible for a given observation of a 
new physics effect.  In Figure~\ref{tcgcorrfig}
we present the correlation between the BR for the FCNC decay 
(assuming no other new physics is present)
and the
process $q \, \bar{q} \ra t \, \bar{c}$ expected from the $g$-$t$-$c$
operator.  Observation of 
this correlation (or one relating a different single top production
process with the $t \ra g \, c$ decay)
could be the smoking gun in identifying
the $g$-$t$-$c$ operator as being responsible for a deviation in
single top production.

Detailed simulations of the $Z$-$t$-$c$ and $\gamma$-$t$-$c$ operators
have so far been confined to studies of top decays \cite{tcz,exoticdecays}.
The quantity $|\kappa_{Ztc} \sin \theta_{Ztc}|$ is constrained by
low energy data on flavor-mixing processes
to be less than the order of magnitude of 0.05 \cite{tcz}.
These studies indicate that from Run~II at the Tevatron
top decays should provide constraints
of $\Lambda_{gtc} \geq 7.9$ TeV,
$\kappa_{Ztc} \leq 0.29$, and will not improve the bounds on
$\Lambda_{\gamma tc}$ from the current $b \ra s \, \gamma$ limit of
about 5 TeV.  Of course, as we have argued before, it was necessary to
assume a SM $W$-$t$-$b$ interaction in order to use decays to say
anything at all about these operators.  The effect of the $Z$-$t$-$c$
operator to the inclusive $t$-channel production rate is
to contribute an additional
0.13 pb at the Tevatron Run~II and 12.6 pb at the LHC, assuming
$\kappa_{Ztc} = 0.29$, and including the NLO QCD corrections
for both $t$ and $\bar{t}$ production.
Low energy constraints indicate that
$\kappa_{Ztc} = 0.29$ requires $|\sin \theta_{Ztc}| \leq 0.17$,
and the inclusive cross sections are insensitive when $\theta_{Ztc}$
is varied in this range.
The $\gamma$-$t$-$c$
operator can be studied at a hadron collider
through the reaction $\gamma \, c \ra t$ (where the photon is treated
as a parton inside the proton) \cite{yuanphoton}, though this 
exotic production
mechanism suffers from potentially large SM backgrounds.

\clearpage

\section{Top Polarization}
\label{polarization}

The polarization of top quarks represents another way to probe the
properties of top interactions.  
In the SM, the $W$-$t$-$b$ vertex
is entirely left-handed, which means that the top polarization
information is passed on to the $W$ boson and $b$ quark into which the
top decays.  Since the $W$ interaction with the light fermions
into which it decays is also left-handed, the $W$ polarization information
is thus also reflected in the kinematics of its decay products.
The same weak interaction is also responsible for single top production,
which has the consequence that single tops also show a large degree of
polarization.  The discussion below is based on the SM amplitudes for
top production and decay presented in \cite{doug}.

\subsection{The $W^+$ Polarization: The $W$-$t$-$b$ Interaction}

In order to probe the chiral structure of the $W$-$t$-$b$ interaction,
it is enough to consider the $W$ polarization of top decays.
As was shown in \cite{doug}, the left-handed nature of the SM interaction
demands that the produced $W$ bosons be either left-handed or longitudinally
polarized, and predicts the specific ratio of
\bea
   \frac{N_0}{N_{-}} = \frac{ m_t^2}{2 \, M_W^2 + m_t^2} \simeq 70\%.
\eea
The degree of $W$ polarization from top decays can be reconstructed by
studying the angle between the $W$ momentum and the charged lepton momentum,
in the $W$ rest frame.

It is desirable to employ top decays in order to probe the $W$-$t$-$b$
interaction, because in the case of a top decay, the $W$ and $b$ are
observed, and thus one can be sure that it is this interaction that
is responsible for the effect one is seeing, which may not be the case
if there is new physics in single top production.  Further, once one has
probed the chiral structure of the $W$-$t$-$b$ interaction, one can
then employ this information to unfold the top decay and reconstruct the
polarization of the top itself, as will be explained below.

\subsection{The Top Polarization}

Once the chiral structure of the $W$-$t$-$b$ interaction 
has been probed through top decays, and the
SM left-handed structure verified, the top decay products can be
used in order to study the polarization of the produced top quarks
themselves.  As we will see, this can be very useful in determining
what sort of new physics is responsible for an observed deviation in
single top production.  Currently, there are two important bases for
describing the top polarization.  The usual helicity basis measures
the component of top spin along its axis of motion (in the center of
mass frame - because the top mass is large its helicity is not a Lorentz
invariant quantity).  The so-called ``optimized basis'' 
\cite{optpol} relies on the SM
dynamics responsible for single top production in order to find
a direction (either along the direction of one of the incoming hadrons
or produced jets) which results in a larger degree of polarization for
the top quark.  In the discussion below, we will describe the
modes of single top production in both bases, and analyze the
particular strengths and weaknesses of each.

\begin{figure}[t]
\epsfxsize=6.0in
\centerline{\epsfbox{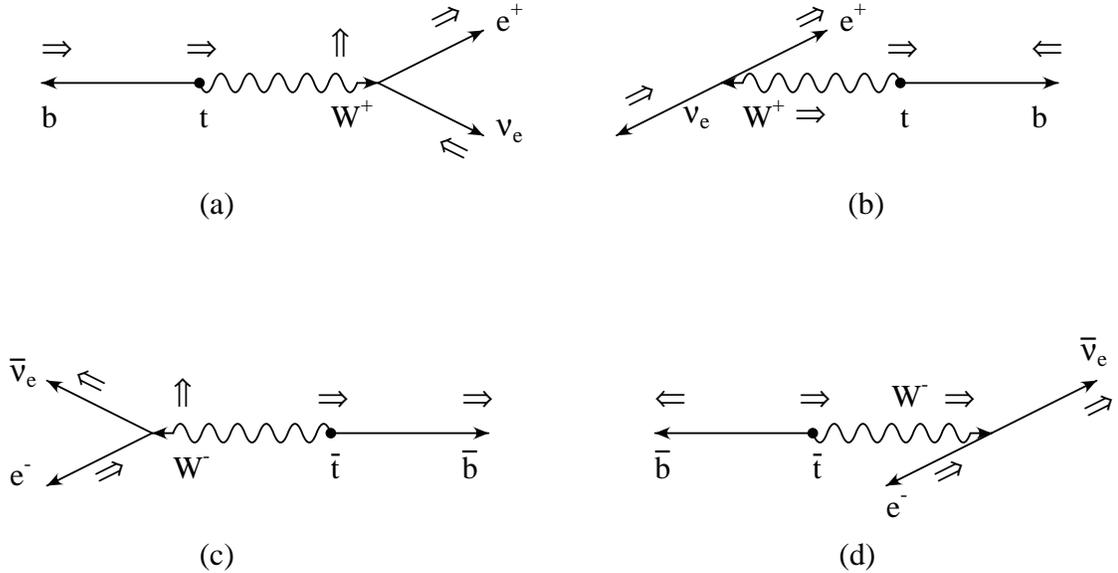}}
\caption{A diagram indicating schematically the correlation between the
charged lepton ($e^\pm$) from a top decay, and the top spin, in the top
rest frame.  The arrows on the lines indicate the preferred direction
of the momentum in the top rest frame, while the large arrows 
alongside the lines indicate
the preferred direction of polarization.  As shown, the $e^+$ ($e^-$)
from a $t$ ($\bar{t}$) decay prefers to travel along (against) the
direction of the $t$ ($\bar{t}$) polarization.}
\label{toppolfig}
\end{figure}

Before looking at a particular process or 
basis, it is worth describing how one can determine the top polarization
from its decay products \cite{doug}.  
A simple heuristic argument based on the
left-handed nature of the $W$ interactions and the conservation of
angular momentum can be made, and is displayed
diagrammatically in Figure~\ref{toppolfig}.  
The analysis is carried out
in the rest frame of the top quark, and is
slightly different for a left-handed or a longitudinally
polarized $W$ boson participating in the decay.
In the left-handed $W$ case, the fact that the
$b$ quark must be left-handed forces it to move along the direction
of the top polarization.  The $W$ thus moves against this direction.
When the $W$ decays, the charged lepton ($\ell^+$)
must be right-handed, so it prefers
to move against the $W$ direction, in the same direction as the top
polarization.  When the $W$ boson is longitudinally polarized, it prefers
to move in the same direction as the top spin.  Its decay products
prefer to align along the $W$ polarization, and since the $W$ is boosted
in the direction of the top polarization, the charged lepton again
prefers to move along the top spin axis.  As shown in 
Figure~\ref{toppolfig}, a similar argument can be made for the 
$\bar{t}$ spin, but in this case the charged lepton prefers to
move against the $\bar{t}$ spin axis.  From this point onward,
we restrict our discussion to top quarks, but it should be clear
how they apply to $\bar{t}$ as well.
The simple angular momentum argument is
reflected in a more detailed computation of the distribution,
\bea
 \frac{1}{\Gamma} \frac{d \Gamma}{d \cos \theta}(t \ra W^+ \, b 
 \ra \ell^+ \, \nu_\ell \, b)
 &=& \frac{1}{2} 
 \left( 1 + \cos \theta \right),
\eea
where $\theta$ is the angle between the top polarization and the
direction of the charged lepton, in the top rest frame, and
$\Gamma$ is the partial width for a semi-leptonic top decay in the SM.
In principle, one has only to decide on a scheme for relating the top
polarization to some axis, and one can fit the distribution,
\bea
  F(\cos \theta) &=& \frac{A}{2} \left( 1 + \cos \theta \right)
  + \frac{1 - A}{2} \left( 1 - \cos \theta \right) ,
\eea
to determine the degree of polarization ($A$) along this axis.
In practice, there are complications arising from the fact that the
endpoints of the distribution tend to be distorted by the cuts
required to isolate the signal from the background, and the fact that
in reconstructing the top rest frame, the component of the unobserved
neutrino momentum along the beam axis ($p^{z}_\nu$) is unknown.
One may determine this quantity up to a two-fold ambiguity by
requiring the top decay products to have an invariant mass that
is close to $m_t$.  However, the ambiguity in this procedure will
also have some effect on the distribution, and so careful study is
required.  One can also use the asymmetry between events with
$\cos \theta > 0$ and $\cos \theta < 0$ to characterize the
degree of polarization of the top, which may be helpful if
the data set is limited by poor statistics.

\subsubsection{$W^*$ Production}

The degree of top polarization in the $W^*$ process is straight-forward
to compute in the helicity basis \cite{doug}.
Using the CTEQ4M PDF's, we find that about 
$75\%$ of the top quarks produced through the
$s$-channel process at the Tevatron are
left-handed, and $76\%$ of them are left-handed at the LHC.

The optimized basis improves the helicity basis result 
at the Tevatron by noting that
in the SM, the $W^*$ process produces top quarks whose polarization is
always along the direction of the initial anti-quark involved in the
scattering.  At the Tevatron, the vast majority ($\sim 97\%$) of these 
anti-quarks come from the $\bar{p}$ (which has valence anti-quarks).
Thus, one expects that by choosing to measure the top polarization along
the $\bar{p}$ direction in the top rest frame, one can raise the degree
of polarization from $75\%$ to $97\%$.  This represents a large improvement
for Tevatron polarization studies of the $W^*$ process.  However, at the
LHC there are no valence anti-quarks, and thus no optimized basis to
analyze the $W^*$ top polarization (though as we have seen, at the LHC
the helicity basis results in a fair degree of left-handed top production
anyway).

\subsubsection{$W$-gluon Fusion}

The discussion of polarization in the $W$-gluon fusion process is somewhat
tricky, mostly owing to the fact that as we have seen above, the 
detailed kinematics
of this process are sensitive to higher orders of perturbation theory.
It is clear that the kinematic region described by the process
$q \, b \ra q^\prime \, t$ is the dominant one, but a precise calculation
of the interplay between the $2 \ra 2$ scattering contribution and the
$2 \ra 3$ scattering contribution is still lacking.  Thus, one must be careful
in claiming what
degree of polarization results from a particular basis.

In the helicity basis, the $2 \ra 2$ description has the top
quarks $100\%$ left-handed when
produced from the $u \, b \ra d \, t$ sub-process.  In fact,
at both Tevatron and LHC the $\bar{d} \, b \ra \bar{u} \, t$
sub-process is quite small, and thus the over-all degree of polarization
is about $97\%$.  On the other hand, the $2 \ra 3$ description
shows a degree of polarization that is much lower, and depends 
on the choice of the bottom mass used in the computation.  This is an
indication that this method of computation is not perturbatively stable.
Thus, it is fair to say that the degree of polarization in the helicity
basis is high, but at the moment no reliable determination is available.

The optimized basis once again makes use of the fact that the top
polarization is $100\%$ along the direction of the spectator
anti-quark in the reaction.  At both Tevatron and LHC, this is
dominantly the spectator jet in the final state.  This basis thus
results in a top which is about $96\%$ polarized along the direction
of the spectator jet.  In \cite{optpol}, it was shown that this
basis is also not sensitive to the value of the bottom mass, and thus
is perturbatively reliable.

\subsection{New Physics and Top Polarization}

As we have seen, new physics may alter the structure of single top
production.  It may be that the new physics effects will reveal
themselves, and tell us something about their nature by causing a
large deviation in one or more of the single top production cross
sections.  In that case one can study the distribution of the top
polarization in order to learn something further about the nature of
the nonstandard production mechanism.

In Section~\ref{extrap}, it was demonstrated that either a
charged scalar top-pion or $W^\prime$ gauge boson can have
a substantial effect on single top production in the $s$-channel
mode.  Assuming for the moment that such a deviation has been
observed, one can then use the top polarization in order to narrow
down the class of models responsible for such an effect.  The $W^\prime$
boson couples to the left-handed top and bottom quarks, and thus an
analysis of the resulting top polarization will be the same as the SM
prediction.  Namely, the helicity basis will show $75\%$ of the tops
to be left-handed ($76\%$ at the LHC) and the optimized basis will
show $97\%$ at the Tevatron.  However, the $\pi^\pm$ has a right-handed
interaction, completely at odds to the SM.  In fact, there is another
difference between the $W^\prime$ and the $\pi^\pm$ that is also very
important.  Like the SM $W$ boson, the $W^\prime$ is a vector particle,
and thus carries angular momentum information between the initial state
and final state in the $s$-channel process.  However, the $\pi^\pm$,
as a scalar particle, does not carry such information.  
Thus, the optimized
basis, which relies on the correlation between 
top spin and the initial $\bar{d}$
momentum fails to apply to a scalar production mechanism, and if one
were to use it to analyze the polarization of the top coming from this
type of new physics effect, one would come to the wrong conclusion that
the produced tops were unpolarized.  On the other hand, in the
helicity basis the top quarks produced from the $\pi^\pm$ show
very close to $100\%$ right-handed polarization.  This demonstrates the
utility of using {\em both} bases.  If there is new physics in single top
production, not only is it unclear at the outset which basis 
will show a larger
degree of polarization, but we can use them together to distinguish
a vector from a scalar exchange, thus learning about the nature of
the new particle without directly observing it.

Study of polarization can also be useful in disentangling the operators
in the effective Lagrangian in Equations~\ref{ewcleq1} 
and \ref{ewcleq2}.  As we saw, those operators have left-handed and
right-handed versions, and thus the distribution of top polarizations
will depend on the relative strength of the two.  Thus, by studying top
polarization, one could begin to disentangle the chiral structure of
the operator responsible for a deviation in single top production,
giving further insight into the nature of the full theory that 
accurately describes higher energies.

\section{Top Quark Properties}

Having gone over in detail the physics one can probe with single top
production, it is worth summarizing what we have learned and examining
how one can use the different top quark observables to extract information
about the top that maximizes the available information.  In the preceeding
sections we have seen that single top production allows one to measure the
magnitude of the top's weak interactions (unlike top decays).  The three
modes of single top production are sensitive to different types of
new physics.  All three modes are sensitive to modification of the
$W$-$t$-$b$
interaction, with the $t \, W^-$ mode distinguished by the fact
that it is rather insensitive to any other types of new physics.
The $s$-channel mode is sensitive to certain types of additional
particles.  And the $t$-channel mode is sensitive to physics which modifies
the top decay properties, in particular to FCNC interactions.
In this light, it is rather unfortunate that the $t \, W^-$ mode is
so small at the Tevatron that it is not likely to be useful there,
as it can allow one to measure the strength of the $W$-$t$-$b$ vertex,
which would be a good first step in disentangling the information from
the $s$- and $t$-channel modes.

\begin{figure}[p]
\epsfxsize=6.0in
\centerline{\epsfbox{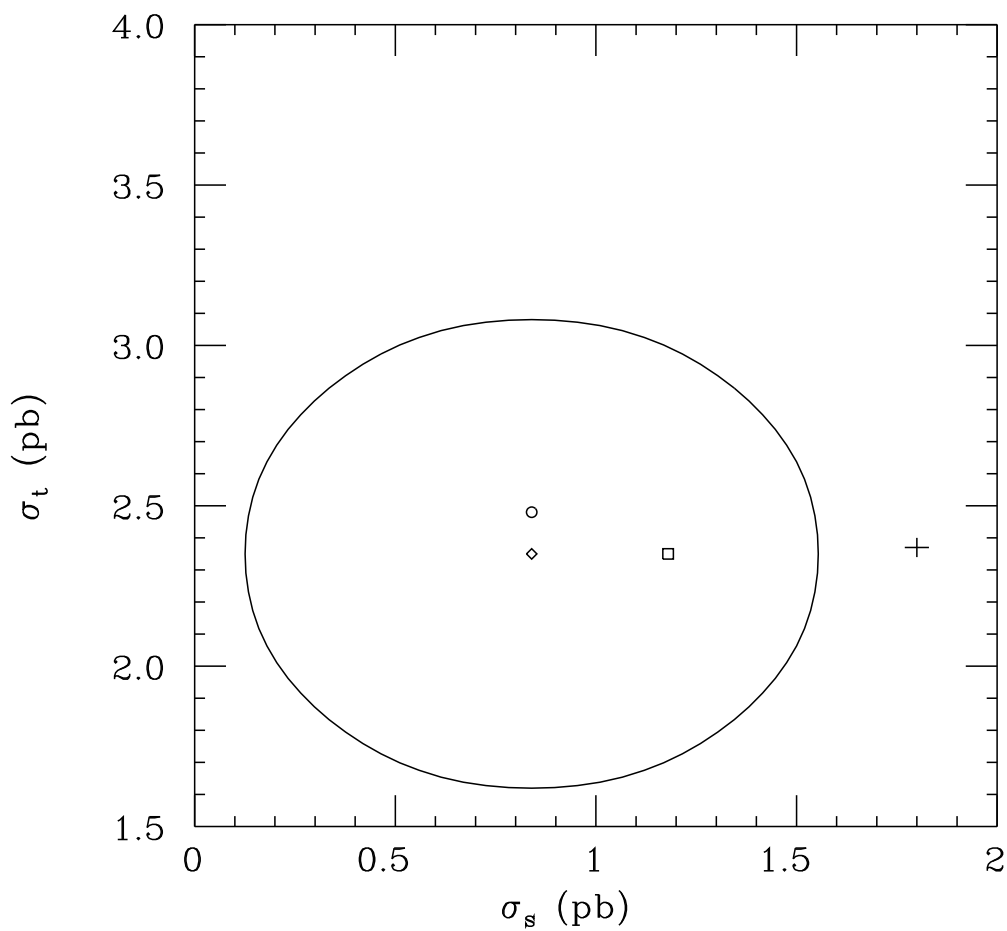}}
\caption{The location of the Tevatron SM point
(the diamond) in the $\sigma_s$-$\sigma_t$ plane,
and the $3 \sigma$ deviation
curve.  Also shown are the points for
the top-flavor model (with $M_Z^{\prime} = 900$ GeV
and $\sin^2 \phi = 0.05$) as the square, the
FCNC $Z$-$t$-$c$ vertex
($\kappa^Z_{tc} = 0.29$) as the circle,
and a model with a charged top-pion ($m_{\pi^{\pm}} =$
250 GeV and $t_R$-$c_R$ mixing of $\sim 20\%$) as the cross.
All cross sections sum the $t$ and $\bar{t}$ rates.}
\label{sigmaplanefig}
\end{figure}

Without the $t \, W^-$ mode, one will most likely have to study the
correlation of the $s$- and $t$- channel rates in the plane of 
$\sigma_s-\sigma_t$ in order to attempt to understand if a new
physics effect is present, and how one should interpret it if it is
observed.  In Figure~\ref{sigmaplanefig} we show this plane,
including the SM point (with the contour of $3 \, \sigma$
deviation around it)
and the points from the the top-flavor model
(with $M_{Z^\prime} = 900$ GeV and $\sin^2 \phi = 0.05$), 
the top-color model with a charged top-pion (with mass
$m_\pi^\pm = 250$ GeV and $t_R$-$c_R$ mixing of $20\%$)
and a FCNC
$Z$-$t$-$C$ operator (with $\kappa_{Ztc} = 0.29$, 
$\sin \theta_{Ztc} = 0.2$, and $\phi_{Ztc}^R = \phi_{Ztc}^L = 0$).
This illustrates how to use the knowledge we have about the sensitivity
of the $W^*$ and $W$-gluon fusion modes to find a likely explanation for
a new physics effect.  A deviation in $\sigma_s$ that is not
also reflected in $\sigma_t$ is most likely due to the effect of
nonstandard particles.  A deviation in $\sigma_t$ that
is not also seen in $\sigma_s$ is likely from a FCNC.  A deviation
that is comparable in both rates is most likely from a modification
of the $W$-$t$-$b$ interaction.  In the very least, if the SM is a
sufficient description of single top production, the fact that the
two rates are consistent will allow one to use them to extract
$V_{tb}$ with confidence that new physics is not distorting the
measurement.

Additional information is provided by polarization information.
By studying the $W$ polarization from top decays, one learns
about the nature of the $W$-$t$-$b$ interaction.  By studying the
top polarization, in both the helicity and optimized bases,
one can learn more about the chiral structure of nonstandard top
interactions, either by probing the chiral structure of the
interactions, or even the scalar/vector nature of a virtual
particle participating in single top production.

\chapter{Higgs with Enhanced Yukawa Coupling to Bottom}

\section{Introduction}

As we have seen, the mystery of the EWSB is one of the
primary challenges for modern particle physics.  The large
top mass, of the same order as the
EWSB scale, suggests that top may play a special
role in the generation of mass.
This occurs in models with dynamical 
top-condensate or top-color scenarios 
\cite{topcondensate,topcolor} as well as 
in SUSY theories \cite{mssm}.
Since the
bottom quark is the iso-spin partner of 
the top quark, its Yukawa 
coupling with a Higgs boson can be closely related to that of the
top quark. In \cite{ushbb}, we demonstrated that because of 
the small mass of bottom ($m_b \sim 4.5$ GeV) relative to 
top ($m_t\sim 175$\,GeV), studying the $b$ Yukawa coupling can
effectively probe new physics beyond the SM.

In this Chapter, we study the detection of a Higgs
boson ($\phi$) at hadron colliders in the context of
models where the 
bottom has an enhanced Yukawa coupling ($y_b$)
to the scalar Higgs.
We begin with a model-independent analysis for
Higgs production associated with $b \bar{b}$ jets,
through the reactions
$p \, \bar{p} \ra \phi \, b \, \bar{b} \ra \bbbb$, and 
$p \, {p} \ra \phi \, b \, \bar{b} \ra \bbbb$
at the Tevatron Run~II and LHC, to determine their
ability to probe models of dynamical EWSB and SUSY theories
through this process.

\section{Signal and Background}

We are interested in studying production of $\hbb \to \bbbb$
at the Run II of the Tevatron and the LHC.  
The signal events result from QCD production of a 
primary $b \bar{b}$ pair, with a Higgs boson ($\phi$) radiated
from one of the bottom quark lines as shown in Figure~\ref{sigfeynfig}.
The Higgs boson then decays into
a secondary $b \bar{b}$ pair to form a $\bbbb$ final state.
Because our detection strategy relies upon observing the primary $b$
quarks in the final state (and thus demands that they have large
transverse momentum), our calculation 
of the $\phibb$ signal rate from diagrams such as those
shown in Figure~\ref{sigfeynfig} is expected to be reliable.  This is in
contrast to the {\it inclusive} rate of $\phi$ production 
at a hadron collider, in which one does not require a final
state topology with four distinct jets.  In this case a calculation
based upon Feynman diagrams such as those shown in 
Figure~\ref{sigfeynfig} may not be reliable.
It would be better to consider the Higgs boson production via 
bottom quark fusion, such as
$b \bar{b} \to \phi$ and $g b \to \phi b$, with cares to avoid
double counting its production rate \cite{dicus}.
(This calculation would resum some large logarithms which are included 
in the definition of the 
bottom parton distribution function within the proton, much
as was true for single top production in Chapter~\ref{chapSingletop}.)
 We have chosen to search in the four jet
final topology because the QCD background for 3 jets is much
larger than that for 4 jets, and thus it would be more
difficult to extract a 3 jet signal.
Since the signal consists of four $b$ (including $\bar b$) jets,
the dominant backgrounds at a hadron collider come from 
production of $\zbb \to \bbbb$, seen in Figure~\ref{zbbfeynfig},
purely QCD production of $\bbbb$, seen in Figure~\ref{bbbbfeynfig},
and
$\bbjj$, where $j$ indicates a light quark or a gluon, shown in
Figure~\ref{bbjjfeynfig} which can occasionally fake a $b$-jet
signature in the detector. 

\begin{figure}
\epsfysize=2.0in
\centerline{\epsfbox{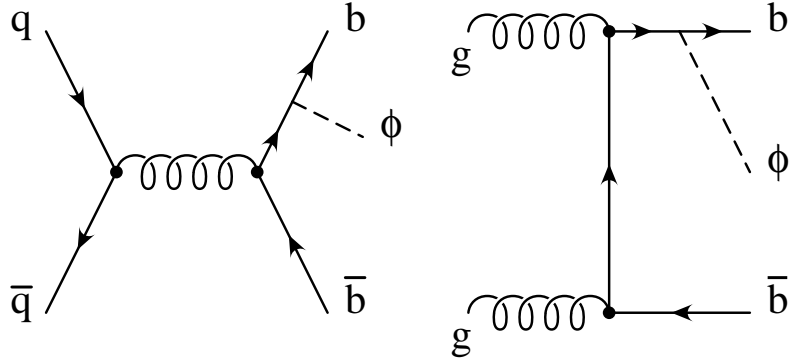}}
\caption{Representative leading order Feynman diagrams for 
$\hbb$ production
at a hadron collider.  The decay $\phi \to b \bar{b}$ is not shown.}
\label{sigfeynfig}
\end{figure}

\begin{figure}
\epsfysize=2.0in
\centerline{\epsfbox{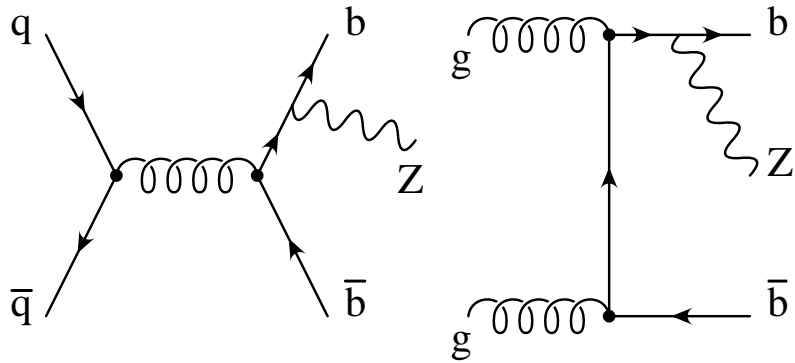}}
\caption{Representative Feynman diagrams for leading 
order $\zbb$ production
at a hadron collider.  The decay $Z \to b \bar{b}$ is not shown.}
\label{zbbfeynfig}
\end{figure}

\begin{figure}
\epsfysize=2.3in
\centerline{\epsfbox{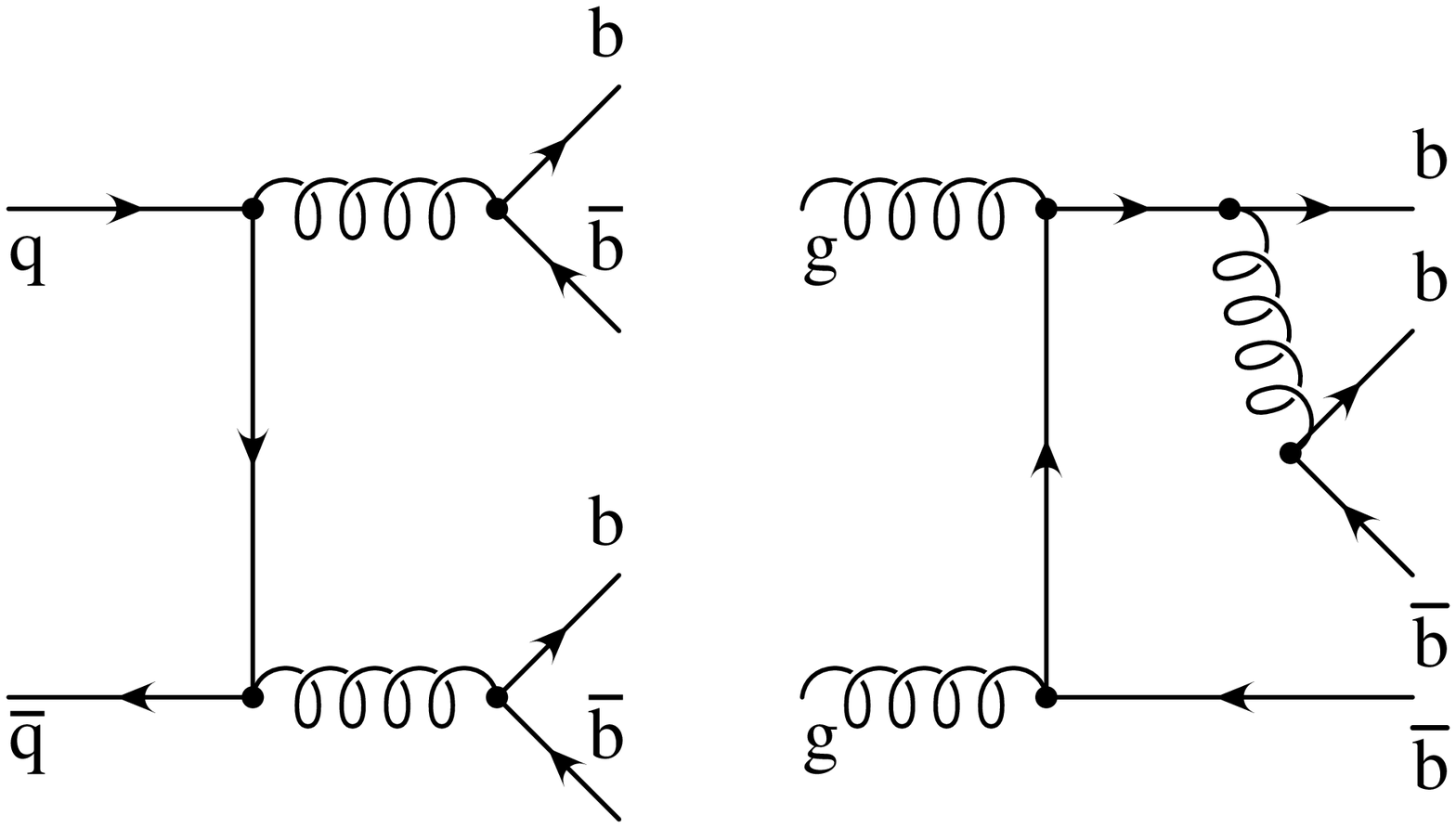}}
\caption{Representative leading order Feynman diagrams for QCD
$\bbbb$ production at a hadron collider.}
\label{bbbbfeynfig}
\end{figure}

\begin{figure}
\epsfysize=4.0in
\centerline{\epsfbox{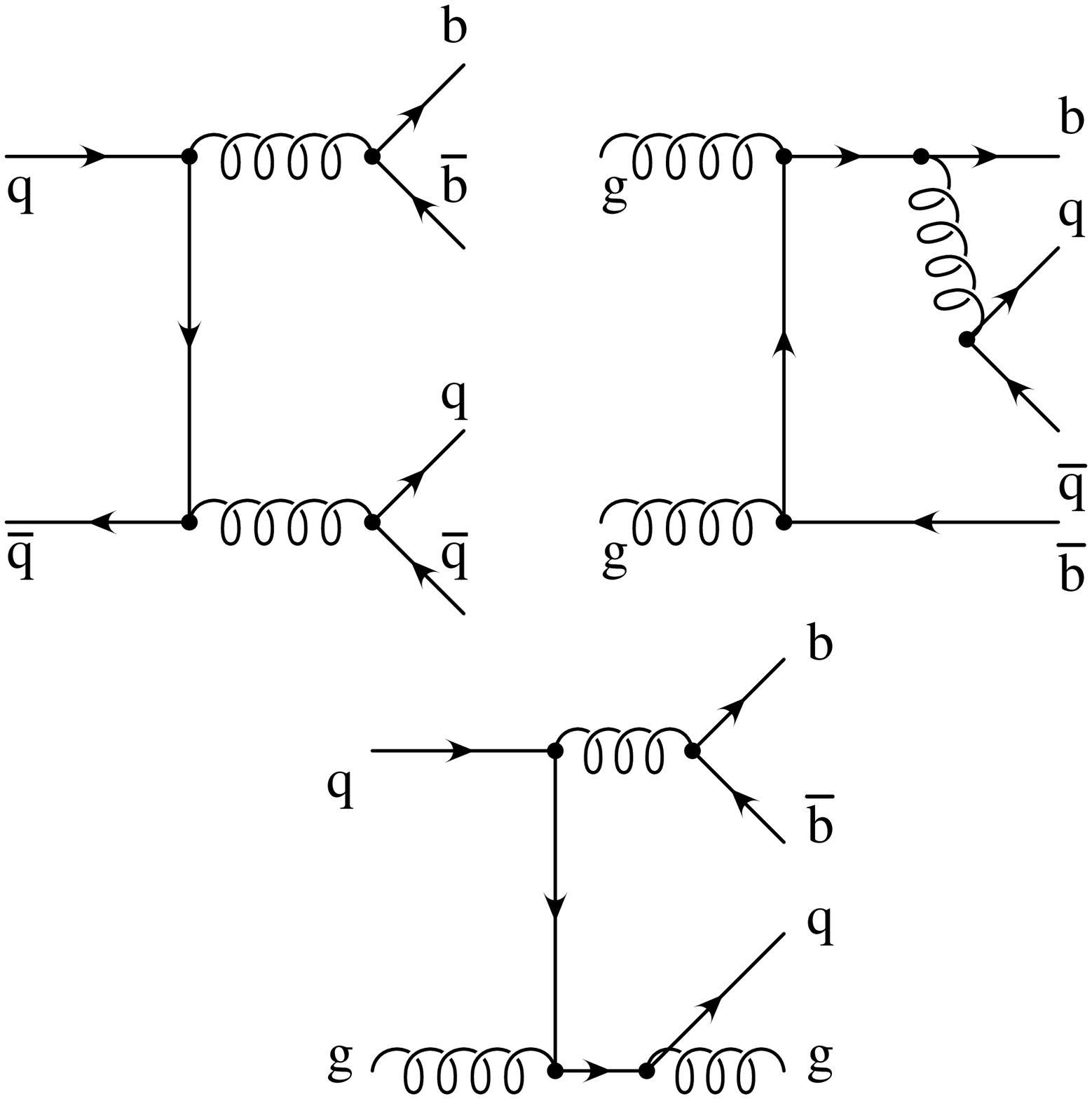}}
\caption{Representative leading order Feynman diagrams for 
QCD $\bbjj$ production at a hadron collider.}
\label{bbjjfeynfig}
\end{figure}

In order to derive model-independent bounds on 
the couplings of the scalar particles with
the bottom quark, we consider $K$, the square-root
of the enhancement factor for the production of 
$\phibb \to \bbbb$ over the SM prediction.
By definition,
\bea
K &=& \frac{y_b}{(y_b)_{\rm SM}}  ,
\eea
in which $(y_b)_{\rm SM} = \sqrt{2} \, m_b / v$ 
is the SM bottom Yukawa coupling 
and $y_b$ is the bottom Yukawa coupling in the new physics model under
the consideration.
The decay branching ratio of $\phi$ to $b\bar b$ is model-dependent, and
is not included in the calculations of this section. 
(Namely, $BR(\phi\to b\bar b)$ 
is set to one).
When analyzing the specific models
in the following sections, we include the appropriate $BR$
for that model.

We compute the signal and the backgrounds at the parton level, using 
leading order (LO)
results from the MADGRAPH package \cite{madgraph}
for the signal and the backgrounds,
including the sub-processes initiated by $q \bar q$ and $g g$
(and in the case of $\bbjj$, $q g$ and $\bar q g$).
While the complete next-to-leading order (NLO) calculations are 
not currently
available for the signal or background cross sections, we draw upon
existing results for high $p_T$ 
$b \bar b$ production at hadron colliders \cite{kfac} and
thus estimate the NLO effects by including a
$k$-factor of 2 for all of the signal and background
rates.  
We will estimate the theoretical uncertainty in the signal and
background cross sections below.
We use the CTEQ4L \cite{cteq4} parton distribution functions
(PDF's) and set the factorization scale, $\mu_0$, to the average of
the transverse masses of the primary $b$ quarks, and the boson 
($\phi$ or $Z$) transverse mass\footnote{
The transverse mass of particle $i$ is given by 
$m^{(i)}_{T} \equiv \sqrt{ m_i^2 + {p^{(i)}_T}^2}$.}
for the $\phibb$ and $\zbb$ processes, and use a
factorization scale of $\mu_0 = \sqrt{\hat s}$,
where $\hat s$ is the square of the partonic center of mass energy,
for the $\bbbb$ and $\bbjj$ background processes.
It is expected that a large part of the total
QCD $\bbbb$ and $\bbjj$ rates
at the Tevatron or LHC energies will come from fragmentation effects,
which we have neglected in our LO matrix element calculation.  
However as we shall see below, due to the
strong $p_{T}$ and isolation cuts 
which are necessary to improve the
signal-to-background ratio,
we expect that these effects will be
suppressed, and thus will only have a small effect on our results.
Similarly, we expect that after imposing the necessary kinematic cuts,
the signal and the background rates are less 
sensitive to the choice 
of the factorization scale.
In this section, unless otherwise noted,
we will restrict our discussion of numerical results
to a signal rate corresponding to a scalar mass of 
$m_{\phi} = 100$ GeV, and an enhancement factor of 
$K = m_t / m_b \approx 40$.  
We will consider the experimental limits which may be placed on $K$ as
a function $m_{\phi}$ below.

In order to simulate the detector acceptance, we require the $p_{T}$
of all four of the final state jets to be  $p_{T} \geq 15$\,
GeV, and that they lie in the central region of the detector,
with rapidity $|\eta | \leq 2$.  
We also demand that the jets are resolvable as
separate objects, requiring a cone separation 
of $\Delta R \geq 0.4$, where 
$\Delta R \equiv \sqrt{ {\Delta \varphi}^{2} + {\Delta \eta}^{2} }$.
($\Delta \varphi$ is the separation in the azimuthal angles.)
In the second column of Table~\ref{cutstab} we present the number of
events in the signal and background processes at the Tevatron Run~II
which satisfy these
acceptance cuts, assuming 2 \ifb of integrated luminosity.
As can be seen, the large background makes it
difficult to observe a signal in the absence of a carefully
tuned search strategy to enhance the signal-to-background ratio.
In presenting these numbers, we have assumed that it will 
be possible to trigger on events containing high $p_{T}$ jets 
(and thus retain all of the signal and background events).
This capability is essential for our analysis.

\begin{table}[t]
\caption{The signal and background events for 2~\ifb
of Tevatron data, assuming
$m_\phi = 100$\,GeV, $2 \Delta m_\phi = 26$\,GeV, and $K = 40$
after imposing the acceptance cuts, $p_T$ cuts,
and reconstructed $m_\phi$ cuts described in the text.
(A $k$-factor of 2 is included in both the signal and the background 
rates.)}
\label{cutstab}
\begin{center}
\begin{tabular}{ccccc}
~~Process~~ & ~~Acceptance Cuts~~ & ~~$p_T$ Cuts~~ 
& ~~$\Delta R$ Cut~~ & ~~$\Delta M$ Cut~~ \\ \hline \hline \\
$\phibb$  & 4923            & 1936       & 1389       & 1389       \\
$\zbb$  & 1432              & 580        & 357        & 357        \\
$\bbbb$ & $5.1 \times 10^4$ & 3760       & 1368       & 1284       \\
$\bbjj$ & $1.2 \times 10^7$ & $1.5 \times 10^6$ & 
$6.3 \times 10^5$ & $5.9 \times 10^5$ \\[0.2cm] \hline \hline
\end{tabular}
\end{center} 
\end{table}

The typical topology of the bottom quarks in the
signal events is a ``lop-sided'' structure in which one of the
bottom quarks from the Higgs decay has a rather high $p_{T}$ of
about $m_{\phi} / 2$, whereas the other three are typically much softer.
Thus, the signal
events typically have one bottom quark which is much more energetic
than the other three.  On the other hand, 
the QCD $\bbbb$ (or $\bbjj$) background is typically much more 
symmetrical, with pairs of bottom
quarks (or fake $b$'s) with comparable $p_T$.
In order to exploit this, we order the
$b$ quarks by their transverse momentum, 
\be
p^{(1)}_{T} \geq p^{(2)}_{T} \geq p^{(3)}_{T} \geq p^{(4)}_{T},
\ee
and require that the bottom quark with highest transverse momentum
have $p^{(1)}_{T} \geq 50$ GeV, and that $p^{(2)}_{T} \geq 30$ GeV
and $p^{(3,4)}_{T} \geq 20$ GeV.
In the third column of Table~\ref{cutstab} we show the effect
of these cuts on the signal and backgrounds.  As can be seen, these
cuts reduce the signal by about $60\%$, while drastically
reducing the QCD $\bbbb$ background by about $90\%$.

Since the $p_T$ spectrum of the leading jets is determined by the mass
of the scalar boson produced, the leading $p_T$ cuts can be optimized
to search for a particular $m_{\phi}$.  
From the discussion above, the optimal cut on $p^{(1)}_T$ 
can be seen to be close to 
$m_{\phi} / 2$ whereas the optimal cut on $p^{(2)}_T$ is somewhat lower
(generally closer to $m_{\phi} / 3$).  
We adopt these optimized $p_T$ cuts
for each mass considered, when estimating the search reach of the
Tevatron or LHC.

Another effective method for reducing the QCD background is to tighten
the isolation cut on the bottom quarks.  
In the QCD $\bbbb$ background, one of the $b \, \bar b$ pairs is 
preferentially produced from gluon splitting. Because of the collinear
enhancement, the invariant mass of this $b\bar b$ 
pair tends to be small,
and the $\Delta R$ separation of these two $b$'s prefers to be as small as
possible. On the contrary, in the signal events, the invariant mass of
the $b\bar b$ pair from
the $\phi$-decay is on the order of $m_\phi$, and the
$\Delta R$ separation is large because the angular distribution of $b$
in the rest frame of the scalar $\phi$ is flat.  Thus, by increasing
the cut on $\Delta R$ to $\Delta R \geq 0.9$ we can improve the
significance of the signal.  As shown in column four of 
Table~\ref{cutstab}, this cut further decreases the signal 
by about $30\%$, and the QCD $\bbbb$ 
background by about $65\%$. 
In the end, their event rates are about the same.

One can further improve the significance of the signal by attempting to
reconstruct the mass of the scalar resonance.  This can be difficult in
principle, because one does not know {\it a priori} what this mass is,
or which bottom quarks resulted from the $\phi$ decay in a given event.
It may be possible to locate the peak in the invariant mass distribution
of the secondary $b$ quarks resulting from the $\phi$ decay, though with
limited statistics and a poor mass resolution this may prove
impractical.  However, one can also scan through a set of masses, and
provide $95\%$ C.L. limits on the presence of a Higgs boson 
(with a given enhancement to the cross section, $K$) in the $\bbbb$
data sample for each value of $m_{\phi}$ in the set.
In order to do this, we assume a Higgs mass, and find the
pair of $b$ quarks with invariant mass which best reconstructs this
assumed mass.
We reject the event if this ``best reconstructed'' mass
is more than $2 \Delta m_{\phi}$ away from our assumed mass, where
$2 \Delta m_{\phi}$ is the maximum of either twice
the natural width of the scalar
under study ($\Gamma_{\phi}$) or the twice 
experimental mass resolution.
We estimate the experimental mass resolution for 
an object of mass $m_{\phi}$
to be,
\bea
  \Delta m_{\phi} &=& 0.13 \, m_{\phi} \, 
  \sqrt{ 100 \,{\rm GeV} / \, m_{\phi}} .
\eea
Under this assumption, the natural width of the bosons in the
specific models of new physics considered below are
usually smaller than this experimental mass resolution.
As shown in the fifth column of
Table~\ref{cutstab}, this cut has virtually no effect on the signal or
$\zbb$ background (for a 100 GeV Higgs)
while removing about another $10\%$ of the $\bbbb$ background.

As will be discussed below, the natural width of
the Higgs bosons in both the MSSM and the models of strong EWSB that we
wish to probe in this paper
are generally much smaller than our estimated
experimental mass resolution, and thus one might think that an improved
experimental mass resolution could considerably improve the limits
one may place on a scalar particle with a strong $b$ interaction.
However,  the models in which we are interested generally have
one or more nearly mass-degenerate bosons with similarly enhanced bottom
Yukawa couplings.  If the extra scalars are much closer in mass than
the experimental mass resolution
(and the natural width of the bosons), the signal can thus include
separate signals from more than one of them.  Thus there is potentially
a trade-off in the $\Delta M$ cut between 
reduction of the background
and acceptance of the signal from more than one scalar resonance.  In
order to estimate the potential improvement for discovering a single
Higgs boson, we have examined the effect on the significance one obtains
if the cut on the invariant mass which best reconstructs $m_{\phi}$ is
reduced to $\Delta m_{\phi}$ as opposed to 
$2 \Delta m_{\phi}$ as was considered
above.  We find that this improved mass resolution further reduces the
QCD $\bbbb$ background by about another $40\%$.  Assuming four $b$
tags (as discussed below), this improved mass resolution increases the
significance of the signal from about 12.2 to 14.6, which will
improve the model-independent lower bound on $K$ by about $10\%$.
Thus, an improved mass resolution would most likely be helpful in
this analysis.

Another method to further suppress background rate is to observe that
in the background events,  the
$b$ quarks whose invariant mass best reconstructs $m_{\phi}$ 
come from the
same gluon.  This is because, after imposing all the kinematical cuts 
discussed above, the matrix
elements are dominated by Feynman diagrams in which one very far
off-shell gluon decays into a $b \bar{b}$ pair, as opposed to
interference of many production diagrams, which dominates the lower
invariant mass region.  
Thus, for $m_{\phi}$ greater than about 100 GeV,
the background event produces $b$ quarks with the characteristic
angular distribution of a vector decaying into fermions,
$1 + \cos^2 \theta$, in the rest frame of the $b \bar{b}$ system.
This is distinct from the signal distribution, which comes from
a scalar decay, and is flat in $\cos \theta$.  Thus, for masses above
$100$ GeV, we further require $|\cos \theta| \leq 0.7$
after boosting back to the rest frame of the $b \bar{b}$ 
pair which we have identified as coming from the scalar boson $\phi$.

In order to deal with the large QCD $\bbjj$ background, it is important
to be able to distinguish jets initiated by $b$ quarks from those
resulting from light quarks or gluons. We estimate the probability to
successfully identify a $b$ quark passing the acceptance cuts outlined
above to be $60\%$, with a probability of $0.5\%$ to misidentify a jet
coming from a light quark or gluon as a $b$ jet \cite{tev2000}.
In Table~\ref{btagtab} we show the resulting number of 
signal and background events passing our optimized cuts at the
Tevatron, assuming 2 \ifb of integrated luminosity, after
demanding that two or more, three or more, or 
four $b$-tags be present in the
events, and the resulting significance of the signal (computed as
the number of signal events divided by the square root of the number
of background events).  We find that requiring 3 or more 
$b$-tags results in about the same significance of
$12.2 \sigma$ as requiring 4
$b$-tags.  However,
we see that for the chosen parameters ($m_{\phi} = 100$ \, GeV and 
$K = m_t / m_b \approx 40$), even with only 2 or more $b$-tags, one
arrives at a significance of about $3 \sigma$, and thus has some ability
to probe a limited region of parameters.
From the large significance, we see that the 
Tevatron may be used to place
strong constraints on Higgs particles with enhanced bottom quark Yukawa
couplings, and that the ability to tag 3 or more of the bottom quarks
present in the signal can probe a larger class of models (or parameter
space of the models)
as compared to what is
possible if only 2 or more of the bottom quarks are tagged.
In the analysis below, to allow for the possibility that the $\bbjj$
background may be somewhat larger than our estimates, we
require 4 $b$-tags, though as we have demonstrated above, we do not
expect a large change in the results if 3 or 4 $b$-tags were
required instead.

\begin{table}[t]
\caption{The signal and background events for 2~\ifb
of Tevatron data, assuming
$m_\phi = 100$\,GeV, $2 \Delta m_\phi = 26$\,GeV, and $K = 40$ 
for two or more,
three or more, or four $b$-tags, 
and the resulting significance of the signal. }
\label{btagtab}
\vspace{0.5cm}
\begin{center}
\begin{tabular}{cccc}
~~Process~~~  & ~~~~~2 or more $b$-tags~~~~~ & 
~~~~~3 or more $b$-tags~~~~~ & ~~~~~4 $b$-tags~~~~~  \\ \hline \hline \\
$\phibb$     & 1139              & 660           & 180              \\
$\zbb$       & 293               & 170           & 46               \\
$\bbbb$      & 1054              & 610           & 166              \\
$\bbjj$      & $1.2 \times 10^5$ & 2141          & 4                \\ 
             &                   &               &                  \\
Significance & 3.3               & 12.21         & 12.25            
\\[0.2cm] \hline \hline \\
\end{tabular}
\end{center} 
\end{table}

This analysis can be repeated for any value of $m_\phi$, using
the corresponding $p_T$ for that particular mass described above.
It is interesting to note that the signal composition in terms of the
$g g$ or $q \bar{q}$ initial state depends on the collider type and 
the mass of the produced boson, which controls the
type of PDF and the typical region of $x \sim m^2_{\phi} / S$ 
at which it is evaluated.  
At the Tevatron, for
$m_\phi = 100\,$ GeV, the signal is $99\%$ $g g$ initial state before
cuts, and $87\%$ after cuts, while for $m_\phi = 200\,$ GeV,
it is  $99\%$ $g g$ initial state before cuts, and $85\%$ after cuts.
Thus, at the Tevatron, one ignores about $15\%$ of the signal if one
relies on a calculation employing only the $g g$ initial state.
At the LHC, for $m_\phi = 100$, the signal is 
very close to $100\%$ $g g$ initial state
before cuts and $99\%$ after cuts, and for $m_\phi = 500\,$ GeV,
it is $99\%$ $g g$ initial state before cuts, and $99\%$ after cuts.
This indicates that at the LHC, very accurate results are possible from a
calculation considering only the $g g$ initial state.
The resulting numbers of signal and (total) background events after
cuts for various boson masses are shown in Table~\ref{eventnumtab}.

\begin{table}[t] 
\caption{Event numbers of signal ($N_S$), for one Higgs boson, and 
background ($N_B$) 
for a 2~\ifb of Tevatron data and a 100~\ifb of LHC data, for various
values of $m_\phi$, after applying the cuts described in the
text, and requiring 4 $b$-tags. An enhancement of 
$K = 40$ is assumed for the signal, though the numbers may be simply
scaled for any $K_{\rm new}$ by multiplying by $(K_{\rm new}/40)^2$.}
\label{eventnumtab}
\begin{center}
\begin{tabular}{ccccc}
               & \multicolumn{2}{c}{Tevatron}  & \multicolumn{2}{c}{LHC} \\
~~~$m_\phi$~(GeV)~~~ & ~~~$N_S$~~~ & ~~~$N_B$~~~ & ~~~$N_S$~~~ & 
~~~$N_B$~~~     \\ \hline \hline \\
 75            & 583          & 640       & 3.4 $\times 10^6$ &
 4.8 $\times 10^6$  \\
 100           & 180          & 216       & 2.0 $\times 10^6$ &
 3.0 $\times 10^6$  \\
 150           & 58           & 92        & 9.2 $\times 10^5$ &
 1.2 $\times 10^6$  \\
 200           & 17           & 31        & 4.2 $\times 10^5$  &
 5.6 $\times 10^5$  \\
250            & 4.8          & 8.8       & 1.9 $\times 10^5$  &
 2.0 $\times 10^5$  \\
 300           & 1.3          & 2.1       & 83000       & 70000     \\
 500           &              &           & 12000       & 5700      \\
 800           &              &           & 1500        & 406       \\
 1000          &              &           & 407         & 70        
\\[0.2cm] \hline \hline \\
\end{tabular}
\end{center} 
\end{table}

From these results, one may derive the minimum value of $K$, $K_{\min}$,
for a scalar boson with mass $m_\phi$ to 
be discovered at the Tevatron or
the LHC via the production mode $b\bar{b}\phi (\to b\bar{b})$.
Similarly, if signal is not found, one can exclude models which 
predict the enhancement factor $K$ to be larger than $K_{\min}$.
To give a model-independent result, we
assume that the width of the $\phi$ is much less 
than the estimated experimental
mass resolution defined above, which is the case for the models studied
in this paper.  We determine $K_{\min}$ by noting that in
the presence of a Higgs boson with enhanced bottom Yukawa couplings, the
number of expected signal events passing our selection criterion
is given by $N_S = K^2 \, N_S^{(SM)}$,
where $N_S^{(SM)}$ is the number of signal events expected for a scalar
of mass $m_{\phi}$ with SM coupling to the $b$ quark (assuming
Br$(\phi\to b\bar{b})=1$), whereas the number of
background events expected to pass our cuts, 
$N_B$, is independent of $K$.
Thus, requiring that
no $95\%$ C.L. deviation is observed in the $\bbbb$ data sample
(and assuming Gaussian statistics) determines 
\be
K_{\min} = \sqrt{ \frac{1.96 \, \sqrt{N_B} }{N_S^{(SM)} } },
\label{kmindef}
\ee
where $1.96 \sigma$ is the $95\%$ C.L. in Gaussian statistics.
In Figure~\ref{kminfig}, we show the
resulting $95\%$ C.L. limits one may impose 
on $K_{\min}$ as a function of 
$m_{\phi}$ from the Tevatron with 2, 10, 
and 30 \ifb and from the LHC with
100 \ifb, as well as the discovery reach of the LHC at the 
$5 \sigma$ level.
Our conclusions concerning the LHC's ability to probe a Higgs
boson with an enhanced $b$ Yukawa coupling are very similar to those
drawn in \cite{dai}, but are considerably more
optimistic than those in \cite{froidevaux}, where the conclusion was
that the $\bbjj$ background is considerably larger than our estimate
(though there are elements of the search strategy which differ between
those of \cite{froidevaux} and ours as well, and their simulation of the
ATLAS detector is certainly more sophisticated).
In \cite{froidevaux} the backgrounds were simulated 
using PYTHIA \cite{pythia}
to generate two to two hard scatterings and then generating the
additional jets from a parton showering algorithm.  As noted above, in
the light of the strong (ordering of) $p_T$ 
and isolation cuts applied to select the
signal events, we feel that a genuine four body matrix element
calculation such as was used in our analysis provides a more reliable
estimate of this background.

\begin{figure}
\begin{center}
\epsfysize=3.5in
\centerline{\epsfbox{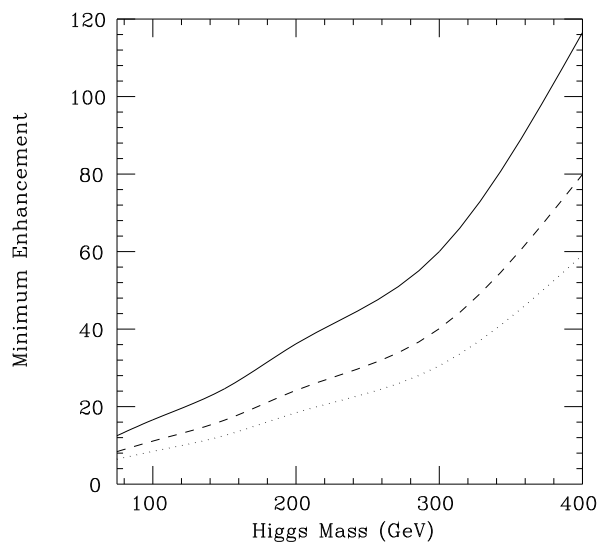}}
\epsfysize=3.5in
\centerline{\epsfbox{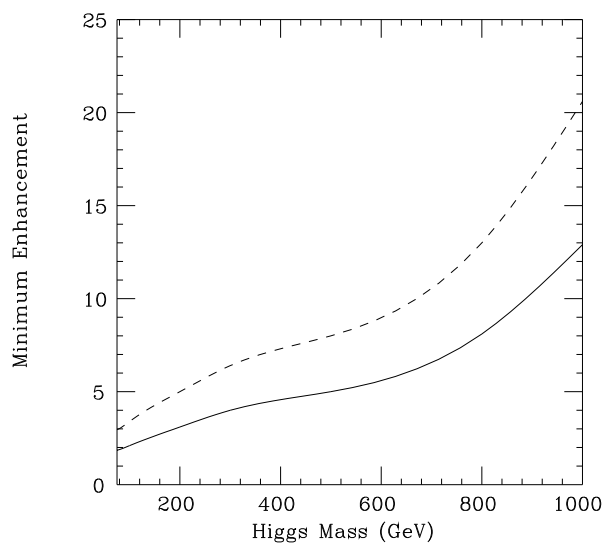}}
\end{center}
\caption{In the upper figure is
the model-independent minimum enhancement factor, $K_{\min}$, 
excluded at $95\%$ C.L. as
a function of scalar mass ($m_\phi$) for the Tevatron Run II with 2 \ifb
(solid curve), 10 \ifb (dashed curve) 
and 30 \ifb (dotted curve).  The lower figure shows
the same factor, $K_{\min}$, 
excluded at $95\%$ C.L. (solid curve) and discovered at
$5\sigma$ (dashed curve)
as a function of $m_\phi$ for the LHC with 100 \ifb.}
\vspace{0.5cm}
\label{kminfig}
\end{figure}

We have examined the scale and PDF dependence of our calculation for the
signal and background rates at the Tevatron, 
and find that in varying the scale between one half and twice its
default choice (defined above),
$\mu = \mu_0 / 2$ and $\mu = 2 \mu_0$, the $\phibb$
signal and $\zbb$ background rates both vary from the result at
$\mu = \mu_0$ by about $30\%$, while the
$\bbbb$ and $\bbjj$ backgrounds vary by about $45\%$.  This strong scale
dependence is indicative of the possibility of large higher order 
corrections to the leading order rate.  Thus, in order to better
understand the true signal and background rates, it would be useful to
pursue these calculations to NLO.  We have also compared the difference
in the results from the MRRS(R1) PDF \cite{mrrs} and the CTEQ4L 
PDF, and find a variation of about $10\%$ in the resulting 
signal and background rates.  Since these separate sources of
uncertainty (from PDF and scale dependence)
are non-Gaussianly distributed, there is no way to rigorously
combine them.  Thus, we 
conservatively choose to add them linearly, finding a total 
uncertainty of about $40\%$ in the signal rate ($N^{(SM)}_S$),
and $50\%$ in the background rate ($N_B$).  
From the derivation of $K_{\min}$ above, we see
that these uncertainties in signal and background rate 
(which we assume to be uncorrelated)
combine to give a fractional uncertainty in
$K_{\min}$,
\be
\frac{\delta K_{\min}}{K_{\min}} = \sqrt{ 
{\left( \frac{\delta N_S^{(SM)}}{2 \, N_S^{(SM)}}\right)}^2 + 
{\left( \frac{\delta N_B}{4 \, N_B}\right)}^2},
\label{kminerr}
\ee
where $\delta N_S^{(SM)}$ and $\delta N_B$ are the absolute
uncertainties in $N_S^{(SM)}$ and $N_B$, respectively.  From this
result, we see that in terms of a more precise theoretical
determination of $K_{\min}$, one gains much
more from a better understanding
of the signal rate than a better determination of the backgrounds.
Applying our estimate of the uncertainty from PDF and
scale dependence to Eq.~(\ref{kminerr}),
we find an over-all theoretical uncertainty in $K_{\min}$ of about
$25\%$.

\section{Implications for Models of Dynamical EWSB}

Examples of the strongly interacting EWSB sector  
with composite Higgs bosons are 
top-condensate and top-color models \cite{topcondensate, topcolor},
in which new strong dynamics associated with the top quark
play a crucial role for top and ${W,Z}$ mass generation. 
A generic feature of these models is a naturally large
Yukawa coupling of the bottom quark,  
of the same order as that of top ($y_t\sim 1$), due to the
infrared quasi-fixed-point structure \cite{RGE} and particular
boundary conditions for $(y_b,y_t)$ at the compositeness scale.

\subsection{The Two Higgs Doublet Extension of the BHL Model}

The effective theory of the top-condensate model 
is the SM without its elementary Higgs boson,
but with 4-Fermi interaction terms induced from 
(unspecified) strong dynamics at a high scale $\Lambda$ instead. 
The minimal Bardeen-Hill-Lindner
(BHL) top-condensate model with three 
families \cite{topcondensate},
contains only one type of 4-Fermi vertex for 
$< \tbar t >$ condensation
which generates
the masses for the top, and the $W$, and $Z$ bosons.
However, the top mass required to obtain the correct
boson masses is too large
to reconcile with experiment. 
Thus, we consider the two Higgs doublet
extension (2HDE) \cite{tt-2HDM} as an example
(which, with some improvements \cite{topcondensate,tctopseesaw}, 
is expected to produce an acceptable $m_t$), 
and examine its prediction for the $\hbb$ rate.
The 4-Fermi interactions of the 2HDE model 
produce condensates in both the $t\bar{t}$ 
and $b\bar{b}$ channels, 
which generate the EWSB and 
induce two composite Higgs doublets $\Phi_t$ and $\Phi_b$.
The
Yukawa interactions take the form,
\bea
  y_t \left( \Psibar_L \, {\Phi}_t \, t_R + H.c. \right) +
  y_b \left( \Psibar_L \, {\Phi}_b \, b_R +  H.c. \right) .
\eea
In the above equation,
$\Psibar_L$ is the left-handed third family quark doublet and 
$t_R$ is the right-handed top, and so forth.
This model predicts 
$~y_t(\Lambda )=y_b(\Lambda ) \gg 1~$ at the scale
$\Lambda$ \cite{topcondensate,tt-2HDM}.
In fact, one finds that 
$y_t(\mu ) \approx y_b(\mu)$~ for any $\mu <\Lambda$,
because the renormalization group equations governing the running of
$y_t$ and $y_b$ are identical except for the small difference in the
the $t$ and $b$ hypercharges \cite{topcondensate,RGE}. 
Due to the dynamical
$< \tbar t >$ and $< \bbar b >$ condensation, 
the two composite Higgs doublets develop VEV's,
\bea
 < \Phi_t > &=& \left(v_t, 0\right)^T/\sqrt{2} \\
 < \Phi_b > &=& \left(0, v_b\right)^T/\sqrt{2} .\nonumber
\eea
The bottom mass is given by $m_b = y_b \, v_b / \sqrt{2}$,
and must match the experimental value
at scale $\mu = m_b \sim 4.5$ GeV.
Assuming the Yukawa couplings $y_t \sim y_b \sim 1$, this
requires the two VEV's to have ratio,
\bea
   \frac{v_t}{v_b} \simeq 39 = \tan \beta. 
\eea
We thus see that in this model, the bottom quark has a
Yukawa coupling of the same order as the top quark, which
implies that $\tan \beta = v_t / v_b$ is naturally large.

The 2HDE has three neutral scalars, the lightest with enhanced
bottom coupling being the pseudoscalar,
\bea
   P &= & \sqrt{2} (  \sin \beta \; {\rm Im} \, \Phi_b^0
                    + \cos \beta \; {\rm Im} \, \Phi_t^0 ),
\eea
whose mass ($M_P$) is less than about $233$\,GeV
for $\Lambda =10^{15}$\,GeV \cite{tt-2HDM}.
Given $y_b$ and $M_P$, one can calculate the production rate of
$P \, b \, \bar{b}(\rightarrow b \, {\bar b} \, b \, {\bar b})$
at hadron colliders, and thus
for a given $M_P$
one can determine the minimal $y_b$ value needed 
for the Tevatron and LHC to observe the signal.
As shown in Figure~\ref{tcfig}a,  the Tevatron Run~II data 
with $2\,{\rm fb}^{-1}$
will exclude such a model 
with $M_P \sim 200$ GeV at $95\%$C.L.

\begin{figure}[p!]
\epsfxsize=6.0in
\epsfbox{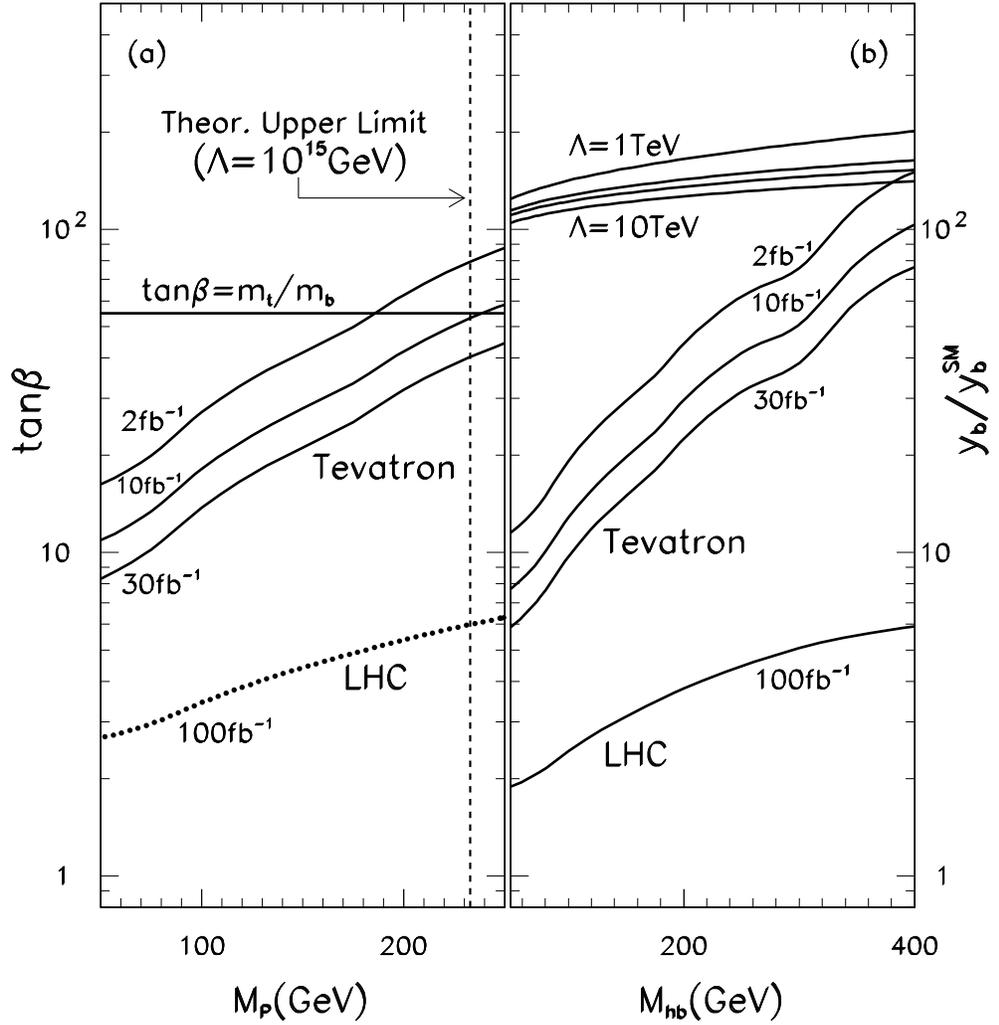}
\caption{The reach of the Tevatron and LHC 
for the models of (a) 2HDE and (b) TCATC.
Regions below the curves can be excluded at $95\%$C.L.  
In (b), the straight lines indicate $y_t(\mu =m_t)$
for typical values of the top-color breaking scale, $\Lambda$.
$y_b$ is predicted to be very close to $y_t$. } 
\label{tcfig}
\end{figure}

\subsection{Top-color Assisted Technicolor}

The top-color-assisted 
technicolor models (TCATC) \cite{topcolor} postulate the gauge structure
${\cal G}=
{\rm SU(3)}_1 \times {\rm SU(3)}_2 \times {\rm U(1)}_1 \times {\rm U(1)}_2 
\times {\rm SU(2)}_L$ 
at the scale above $\Lambda$ to explain the dynamic origin of the 
4-Fermi couplings described above.
At $\Lambda \sim 1$ TeV, ${\cal G}$ spontaneously breaks down to 
${\rm SU(3)}_C \times {\rm U(1)}_Y \times {\rm SU(2)}_L$, and
additional massive gauge bosons are produced in
color octet ($B^a$) and singlet ($Z^\prime$) states. 
Below the scale $\Lambda$, the effective 
4-Fermi interactions are generated in the form,
\bea
   {\cal L}_{4F} = \frac{4 \, \pi}{\Lambda^2}\left[
   \left(\kappa +\frac{2\kappa_1}{9 \, N_c}\right)
   \Psibar_L \, t_R \, \tbar_R \, \Psi_L +
   \left(\kappa -\f{\kappa_1}{9 \, N_c}\right)
   \Psibar_L \, b_R \, \bbar_R \, \Psi_L \right],
\eea
where $\kappa$ and $\kappa_1$ originate from the strong $SU(3)_1$ 
and $U(1)_1$ dynamics, respectively. 
In the low energy effective theory at the EWSB scale,
two composite Higgs doublets are induced with 
the Yukawa couplings 
\bea
 y_t =\sqrt{4 \pi \left( \kappa + 2 \frac{\kappa_1}{9 N_c}\right)} \\
 y_b =\sqrt{4 \pi \left( \kappa -   \frac{\kappa_1}{9 N_c}\right)}, \nonumber
\eea
It is clear that, unless $\kappa_1$
is unnaturally larger than $\kappa$,  $y_b$ is expected to be only slightly
below $y_t$.  The $U(1)_1$ force is attractive in the 
$<\tbar t>$ channel but repulsive in the 
$<\bbar b\>$
channel, thus $t$, but not $b$, acquires a dynamical mass,
provided
$y_b(\Lambda) < y_{\rm crit} = \sqrt{{8\pi^2}/{3}} < y_t(\Lambda )$.
(In this model, $b$ acquires a mass 
mainly from a top-color instanton effect \cite{topcolor}.)
Furthermore, the composite Higgs doublet $\Phi_t$, but not 
$\Phi_b$, develops a VEV, i.e., $v_t\neq 0$ and $v_b=0$. 

In TCATC, the top-color interaction generates $m_t$, but
is not responsible for the entire EWSB.
Thus, $\Lambda$ can be as low as
$O(1-10)$\,TeV (which avoids the severe fine-tuning needed in the
minimal models \cite{topcondensate,tt-2HDM}),
and correspondingly, $v_t=64-88$\,GeV for $\Lambda =1-5$\, TeV
by the Pagels-Stokar formula. 
The smaller value of $v_t$ predicted in the TCATC model,
compared to $v=246$\,GeV
makes the top coupling to $\Phi_t$ stronger,
i.e., $~y_t=2.8-3.9~$ at $\mu =m_t$,
than in the SM ($y_t \sim 1$).
As explained above, this results in a large bottom Yukawa
interaction, $y_b$, as well.
Thus, the neutral scalars
$h_b$ and $A_b$ in the doublet $\Phi_b$, which are about
degenerate in mass, have an enhanced coupling to the $b$-quark.

In Figure~\ref{tcfig}b, 
we show the minimal value of $y_b/(y_b)_{SM}$ needed
to observe the TCATC model signal as a function of $M_{h_b}$. 
As shown, if $M_{h_b}$ is less than about 400\,GeV, the Tevatron
Run II data can effectively probe the scale of the top-color breaking 
dynamics, assuming the TCATC model signal is observed. 
If the signal is not found, the LHC can further explore this
model up to large $M_{h_b}$.  For example, for
$M_{h_b}=800$\,GeV, the required minimal value of $y_b/(y_b)_{SM}$
is about 9.0 at $95\%$C.L.
Similar conclusions can be drawn for 
a recent left-right symmetric extension \cite{tt-lindner} of the 
top-condensate scenario, which
also predicts a large $b$-quark Yukawa coupling.

\section{Implications for Supersymmetric Models}

The EWSB sector of the MSSM model includes two Higgs doublets with
a mass
spectrum including two neutral CP-even scalars $h^0$ and $H^0$,
one CP-odd pseudoscalar $A^0$ and a charged pair $H^\pm$.
The Higgs sector is completely determined at tree level by 
fixing two parameters,
conventionally chosen to be the ratio of the VEV's, $\tan \beta$,
and the pseudoscalar mass, $m_A$ \cite{himssm}.
At loop level, large radiative corrections to the Higgs boson 
mass spectrum are dominated by the contributions of top quarks and 
squarks in loops \cite{himrc}.
In this study we employ the full one loop results \cite{effpot} to
generate the Higgs mass spectrum assuming the sfermion masses, $\mu$,
scalar tri-linear parameters, and SU(2$)_L$ gaugino masses
at the electro-weak scale are those chosen in the LEP Scan A2 set.
There is some sensitivity to this choice of parameters, coming from
the Higgs mass spectrum and coupling to bottom quarks \cite{ushbb, sens}.

The parameter
$\tan\beta$ is free in the MSSM, and the Higgs mass is constrained
by $m_h, m_A >75$\,GeV for $\tan\beta > 1$ \cite{lepdata}.
Since the couplings of $h^0$-$b$-$\bar b$, $H^0$-$b$-$\bar b$ and 
$A^0$-$b$-$\bar b$ are proportional to $\sin\alpha/\cos\beta$, 
$\cos\alpha/\cos\beta$ and $\tan\beta$, respectively, they
can receive large enhancing factor when $\tan\beta$ is large. 
This can lead to detectable $\bbbb$ signal events at the LHC,
as was previously studied in \cite{dai}. 
We calculate the enhancement factor $K$
predicted by the MSSM for given values of
$\tan\beta$ and $m_A$. 
In Figure~\ref{mssmfig} 
we present the discovery reach of the Tevatron and 
the LHC, assuming the LEP Scan A2 soft-breaking parameters,
and that all the superparticles 
are so heavy that Higgs bosons 
will not decay into them at tree level.
For comparison, the region that will be covered by LEP II is
also shown.

The BR for $\phi \to b\bar b$ is close to one for most of the 
parameter space above the discovery curves. 
Moreover, for $\tan\beta \gg 1$, the $h^0$ is nearly
mass-degenerate with the
$A^0$ (if $m_A$ is less than $\sim$120\,GeV)
and otherwise with $H^0$.
We thus include both scalars in the signal rate provided their
masses differ by less than $\Delta m_h$.
The MSSM can also produce additional $\bbbb$
events through production of $h^0 Z \to \bbbb$ and 
$h^0 A^0 \to \bbbb$, however these rates are expected to be 
relatively small
when the Higgs-bottom coupling is enhanced, and the resulting
kinematics are different from the $\hbb$ signal.  Thus we
conservatively do not include these processes in our signal rate.

From Figure~\ref{mssmfig} we deduce that if a signal is not found,
the MSSM with $\tan\beta > 45~(30,~20)$
can be excluded for $m_A$ up to 200 GeV
at the $95\%$ C.L. by Tevatron data with a luminosity of 
2~(10,~30)~fb$^{-1}$; while the LHC can exclude a much larger
$m_A$ (for $m_A=800$\,GeV, 
the minimal value of $\tan \beta$ is about 5). 
These Tevatron bounds thus improve a recent result obtained
by studying the $b\bar{b}\tau\tau$ channel \cite{dressetal}.
We note that studying the $\hbb$ mode can probe an important
region of the $\tan \beta$-$m_A$ plane which is not easily 
covered by other production modes at hadron colliders, such as 
$pp\to t\bar t + \phi(\to \gamma \gamma)+X$ and
$pp\to \phi (\to ZZ^*)+X$ \cite{gunorr}.
Also, in this region of parameter space the
SUSY Higgs boson $h^0$ is clearly distinguishable from a SM one.

\begin{figure}[p!]
\epsfxsize=6.0in
\epsfysize=6.0in
\epsfbox{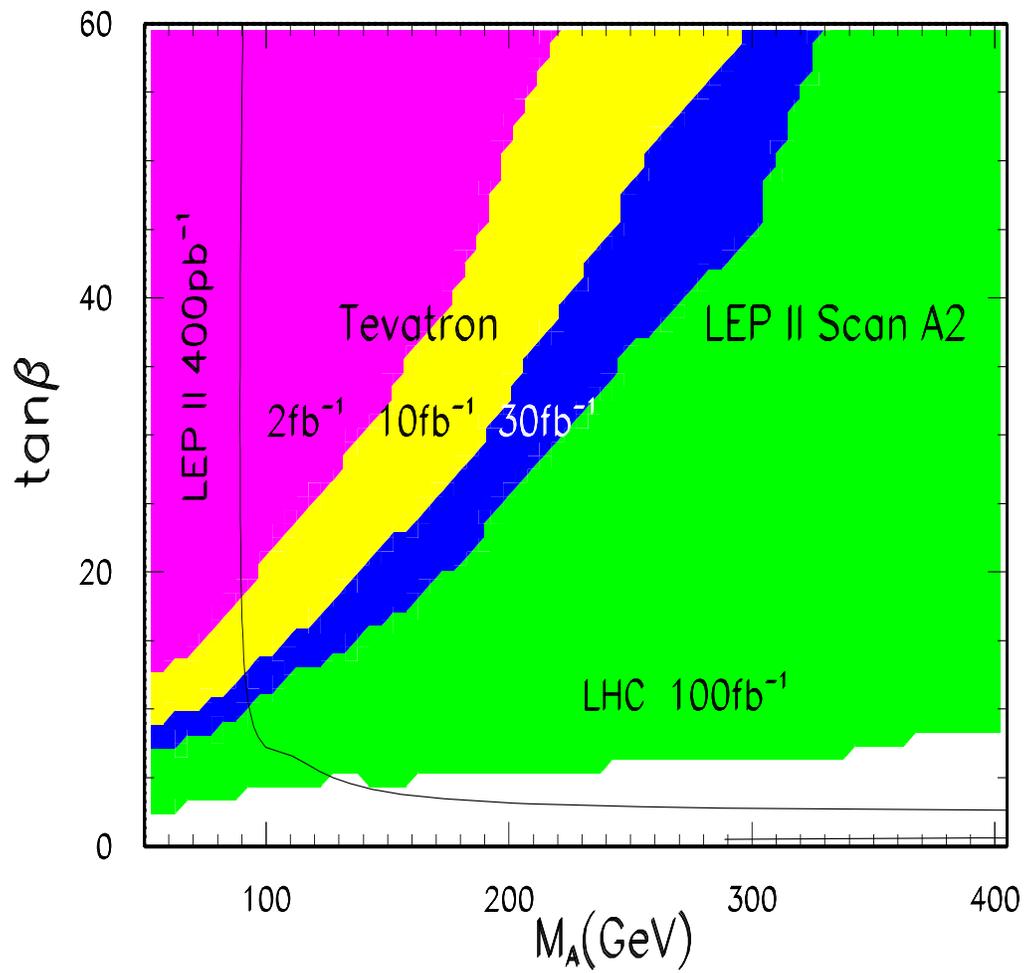}
\caption{
 The regions above the curves in the $\tan \beta$-$m_A$ plane can
be probed at the Tevatron and LHC with a $95\%$ C.L.. The soft breaking
parameters correspond to the LEP Scan A2 set. The 
region below the solid line
will be covered by LEP~II.} 
\label{mssmfig}
\end{figure}

The above results provide a general test for many SUSY models, 
for which the MSSM is the low energy effective theory. 
In the MSSM, the effect of SUSY breaking is parametrized by 
a large set of soft-breaking (SB) terms ($\sim$$O(100)$), 
which in principle should be derived from an underlying model.
We discuss, as examples, the 
Supergravity and Gauge-mediated (GM) models
with large $\tan \beta$.
In the supergravity-inspired model 
\cite{sugrarev} the SUSY breaking occurs in a hidden sector 
at a very large scale, of $O(10^{10-11})$\,GeV, 
and is communicated to the MSSM through gravitational 
interactions.  In the simplest model of this kind, 
all the SB parameters are expressed in terms of 5 universal inputs.
The case of large $\tan\beta$, of $O(10)$, has been examined
within this context \cite{sugratba}, and it was found that
in such a case $m_A \sim 100$~GeV.
Hence, these models can be cleanly confirmed or excluded
by measuring the $\bbbb$ mode at the Tevatron and LHC.

The GM models assume that
the SUSY-breaking scale is much lower, of $O(10^{4-5})$\, GeV, 
and the SUSY breaking is communicated to the MSSM superpartners 
by ordinary gauge interaction \cite{gaugemediated}.
This scenario can predict large $\tan\beta$ 
($\sim 30$).  However, some
models favor $m_A \gae 400$\,GeV \cite{bcerom},
which would be difficult to test at the Tevatron, though
quite easy at the LHC.
Nevertheless, in some other models,
a lighter pseudoscalar is possible
(for instance, $\tan\beta=45$ and $m_A=100$) \cite{gmpred},
and the $\bbbb$ mode at hadron colliders can easily explore
such a SUSY model.

\section{Conclusions}

In conclusion, the large QCD production rate at a hadron collider 
warrants the detection of a light scalar with large
$\phi$-$b$-$\bar b$ coupling.  This process can provide
useful information concerning dynamical models of EWSB and
on the MSSM, either through discovery or by limiting the viable
region of parameters in the model.

At
LEP-II and future $e^+ e^-$ linear colliders, because of the large
phase space suppression factor for producing a direct 3-body final state
as compared to first producing a 2-body resonant state, 
the $b {\bar b} A^0$ and $b {\bar b} h^0$ rates predicted by the MSSM are 
dominated by the production of $A^0 \, h^0$ and $h^0 \, Z$ pairs via 
electroweak interactions. Hence, the $e^+ e^-$ collider is less 
able to directly probe the $\phi$-$b$-$\bar b$ coupling.
This has the effect that our process is complimentary to the the LEP
studies, in that it is sensitive to a different region of SUSY
parameter space.

 \def\lp{\left. }
 \def\rp{\right. }
 \def\lr{\left( }
 \def\rr{\right) }
 \def\le{\left[ }
 \def\re{\right] }
 \def\lg{\left\{ }
 \def\rg{\right\} }
 \def\lb{\left| }
 \def\rb{\right| }

 \def\li{\mbox{Li}_2}

 \newcommand{\ms}{m_{\tilde{q}}}
 \newcommand{\mc}{m_{\tilde{\chi}}}
 \newcommand{\mg}{m_{\tilde{g}}}

\def\bea{\begin{eqnarray}}
\def\eea{\end{eqnarray}}
\def\ppbar{{\rm p} \bar{{\rm p}}}
\def\pp{ {\rm p} {\rm p} }
\def\ifb{ {\rm fb}^{-1} }
\def\del{\partial }
\def\ra{\rightarrow}
\def\Ra{\Rightarrow}
\def\dis{\displaystyle}
\def\f{\frac}

\def\chiplus{{\tilde{\chi}}^{+}}
\def\chiminus{{\tilde{\chi}}^{-}}
\def\chizero{{\tilde{\chi}}^{0}}
\def\chargino{{\tilde{\chi}}^{\pm}}
\def\neutralino{{\tilde{\chi}}^{0}}
\def\gaugino{{\tilde{\chi}}}
\def\gluino{\tilde{g}}
\def\bino{\tilde{B}}
\def\w3ino{\tilde{W_3}}
\def\h1ino{\tilde{H_1}}
\def\hino2{\tilde{H_2}}
\def\wpino{\tilde{W}^{+}}
\def\hpino{\tilde{H}^{+}}
\def\bbar{\bar{b}}
\def\tbar{\bar{t}}

\def\ifb{${\rm fb}^{-1}$}
\def\alphas{\alpha_{S}}
\def\alphash{\hat{\alpha}_{S}}
\def\MS{$\overline{\rm MS}$\,\,}

\def\u6d{u_{6 \Delta}}
\def\ud7{u_{7 \Delta}}
\def\s4d{s_{4 \Delta}}
\def\sd3{s_{3 \Delta}}
\def\delu{\Delta_{u}}
\def\delt{\Delta_{t}}

\newcommand{\msqu}[1]{m_{\tilde{q}_{#1}}}
\newcommand{\ih}[2]{\,\hat{I}\left( \frac{#1}{#2} \right)}

\chapter{Associated Production of Gauginos with Gluinos at NLO}

As we saw in Chapter~\ref{intro}, one of the attractive solutions to
the problems with the Higgs sector of the SM is to introduce weak scale
supersymmetry, which removes the instability of the Higgs mass
under quantum corrections, and deals with the triviality problem.
Thus, the discovery of SUSY would constitute a major development
in understanding the EWSB.  In this chapter we demonstrate how the NLO
SUSY-QCD corrections to the production of a gaugino($\gaugino$) in
association with a gluino ($\gluino$) are moderately sizable, and
significantly improve the theoretical stability of the cross section
\cite{ussusy}, which is important in interpreting experimental data
in terms of a SUSY discovery or exclusion.

Supersymmetry predicts the existence of supersymmetric 
partners for each of the 
particles of the standard model.  The search for these sparticles is a 
principal motivation of the forthcoming Run II of the Fermilab Tevatron 
collider and of the CERN Large Hadron Collider (LHC) program.  A potentially 
important, but heretofore largely overlooked, discovery channel is the 
associated production of a spin-1/2 gaugino
with a spin-1/2 gluino 
or with a spin-0 squark ($\tilde{q}$).  Color-neutral gauginos 
couple with electroweak strength, whereas the colored squarks and gluinos 
couple strongly.  Associated production is therefore a semi-weak process in 
that it involves one somewhat smaller coupling constant than the pair 
production of colored sparticles.  However, in popular models of SUSY 
breaking \cite{gaugemediated,sugrarev}, 
the mass spectrum favors much lighter masses for 
the low-lying neutralinos and charginos than for the squarks and gluinos.  
This mass hierarchy means that the phase space for production of neutralinos 
and charginos, the corresponding partonic luminosities, and the 
production cross sections will be greater than those for gluinos and squarks.  
These advantages are potentially decisive at a collider with limited energy, 
such as the Tevatron. Furthermore, associated production has a clean 
experimental signature.  For example, the lowest lying neutralino is the 
(stable) lightest supersymmetric particle (LSP) in supergravity (SUGRA) 
models \cite{sugrarev}, 
manifest as missing energy in the events, and it is the 
second lightest in gauge-mediated models \cite{gaugemediated}.  
In models with a very 
light gluino \cite{farraraby}, there could be large rates for 
$\gluino \gaugino$ production, with simple signatures, whereas 
$\gluino \gluino$ production suffers from large hadronic jet backgrounds.   

Experimental investigations are facilitated by firm theoretical understanding 
of the expected sizes of the cross sections for production of the
superparticles.  In the case of hadron-hadron colliders,
the large strong coupling strength ($\alphas$) results in potentially
large contributions to cross sections from terms beyond leading order
(LO) in a perturbative quantum chromodynamics (QCD) evaluation of the cross 
section.  For accurate theoretical estimates, it is necessary to extend the 
calculations to next-to-leading order (NLO) or beyond.  NLO contributions 
generally reduce and stabilize dependence on undetermined parameters such 
as the renormalization and factorization scales.  To date, associated 
production has been calculated only in LO \cite{lo}, but NLO results exist for 
hadroproduction of gluinos and 
``light'' squarks\footnote{By light squarks, we refer to the squarks
which are the superpartners of light quarks ($u$, $d$, $s$, $c$, and $b$).
In most models of SUSY breaking these scalars have masses on the order
of a few hundred GeV.} \cite{squarkgluino}, top 
squarks \cite{stop}, sleptons \cite{slepton,gaugino}, and 
gauginos \cite{gaugino}.  Studies have begun to incorporate these NLO results 
into Monte Carlo simulations \cite{scales,newsteve}.

In this Chapter we present the first NLO (in SUSY-QCD) calculation of 
hadroproduction of a $\gluino$ in association with a $\gaugino$, including 
contributions from virtual loops of colored sparticles and particles and 
three-particle final states involving the emission of light real particles.  
We extract the ultraviolet, infrared, and collinear divergences by use of 
dimensional regularization and employ standard \MS renormalization and mass 
factorization procedures.  In the course of computing the virtual 
contributions, we encountered new divergent four-point functions.  The 
contributions from real emission of light particles are treated with a phase 
space slicing method.  We provide predictions for inclusive cross sections at 
Tevatron and LHC energies.
We focus on the $\gluino \gaugino$ 
final state, rather than on the associated production 
of $\tilde{q} \gaugino$, 
because at the energy of the Tevatron the LO cross sections for 
$\gluino \gaugino$ are 3 to 6 times greater than those for 
$\tilde{q} \gaugino$ when  
$m_{\gluino} = m_{\tilde{q}} = 300$ GeV, and 6 to 15 times greater when 
$m_{\gluino} = m_{\tilde{q}} = 600$ GeV.  In obtaining the 
$\tilde{q} \gaugino$ cross 
sections, we sum over five flavors of squarks and anti-squarks.  

\section{Leading Order Cross Sections}

In LO of SUSY-QCD, the associated production of a gluino and a gaugino
proceeds through the subprocess $q\bar{q}\ra\gluino\gaugino$ with a $t$-channel
or a $u$-channel squark exchange. We assume that there is no mixing between
squarks of different generations and that the squark mass eigenstates are
aligned with the squark chirality states, equivalent to the assumption that 
the two squarks of a given flavor are degenerate in mass. We ignore the 
$n_f = 5$ light quark masses in all of the kinematics and couplings.  Under 
these assumptions, the massless incoming quarks and antiquarks have a 
particular helicity, and thus the Feynman diagrams in which a
right-handed squark is exchanged cannot interfere with those mediated
by a left-handed squark. In evaluating the Feynman diagrams involving
Majorana and explicitly charge-conjugated fermions, we follow
the approach of \cite{denner}. In the case of charged gauginos,
only the left-handed squarks participate, whereas neutral gauginos receive
contributions from both left- and right-handed squarks.

The LO matrix element summed (averaged) over the colors and 
helicities of the outgoing (incoming) particles has the analytic form \cite{lo}
\bea
 \overline{|{\cal M}^B|}^{2} = \frac{ 8 \pi\, \alphash}{9} &
 \Biggl[ & \frac{\hat{X}_t\, t_{\gluino}\, t_{\gaugino}}{(t - \msqu{t}^2)^2}
 - \frac{2 \, \hat{X}_{tu}\, s\, m_{\gluino}\, m_{\gaugino}}{(t - \msqu{t}^2)
 (u - \msqu{u}^2)}
 + \frac{\hat{X}_u\, u_{\gluino}\, u_{\gaugino}}{(u - \msqu{u}^2)^2}
 \Biggr] . 
 \label{loeq} 
\eea
Here, $\msqu{t,u}$ is the mass of the squark exchanged in the $t$- and
$u$-channels, and $\alphash =\hat{g}^2_S/4\pi$ is the coupling between
quarks, squarks, and gluinos (at leading order it is equal to the gauge
coupling constant $\alphas$). 
$\hat{X}_{t,tu,u}$ stand 
for the weak couplings of quarks, squarks, and gauginos which will be
explained below, and the quantities
$s$, $t$, and $u$ are the usual Mandelstam invariants at the 
partonic level with 
$t_{\gluino,\gaugino}=t-m_{\gluino,\gaugino}^2$ and 
$u_{\gluino,\gaugino}=u-m_{\gluino,\gaugino}^2$. 

For production of a neutralino of type $\chizero_j$, 
the $\hat{X}$ are given by \cite{chicoupling}
\bea
  \hat{X}_t = \hat{X}_u = \hat{X}_{t u} = 2 \, \left| 
  e \, e_q \, N'_{j \, 1} +
  \frac{e}{\sin \theta_W \, \cos \theta_W}\left( \, T_q -
  e_q \, \sin^2 \theta_W \right) N'_{j \, 2} \right|^2.
\eea
In the expressions above, $e$ is
the electric charge, $\theta_W$ the weak mixing angle, $T_q$ the
third component of the weak isospin for the squark, and $e_{q}$ is the
charge of the quark in units of $e$.  For up-type quarks $e_q = 2 / 3$
and for down-type quarks $e_q = - 1 / 3$.  The matrix $N' \,$ is the
transformation from the interaction to mass eigenbasis defined in
\cite{chicoupling}.
The expressions for production of positive chargino of 
type $\chiplus_j$ are
\bea
  \hat{X}_t &=& \frac{e^2}{\sin^2 \theta_W} |V_{j \, 1}|^2, \\
  \hat{X}_{t u} &=& \frac{e^2}{\sin^2 \theta_W}
  {\rm Re} \, (\, V_{j \, 1} \, U^*_{j \, 1} ) , \nonumber \\
  \hat{X}_u &=& \frac{e^2}{\sin^2 \theta_W} |U_{j \, 1}|^2, \nonumber
\eea
and for the negative chargino $\chiminus_j$ they have the form,
\bea
  \hat{X}_t &=& \frac{e^2}{\sin^2 \theta_W} |U_{j \, 1}|^2 , \\
  \hat{X}_{t u} &=& \frac{e^2}{\sin^2 \theta_W} \,
  {\rm Re} \, (\, V^*_{j \, 1} \, U_{j \, 1}) , \nonumber \\
  \hat{X}_u &=& \frac{e^2}{\sin^2 \theta_W} |V_{j \, 1}|^2, \nonumber
\eea
where $U$ and $V$ are the chargino transformation matrices from
interaction to mass eigenstates defined in \cite{chicoupling}.
As was mentioned above, in the case of chargino production, the
exchanged squark is always left-handed.

\section{Next-to-Leading Order Corrections}

At NLO in SUSY-QCD
the cross section receives contributions from virtual loop diagrams
and from real parton emission diagrams. The virtual contributions arise from
the interference of the Born amplitudes with the related one-loop amplitudes
containing self-energy corrections, vertex corrections, and box diagrams.
We include the full supersymmetric spectrum of strongly interacting particles
in the virtual loops, i.e.\ squarks and gluinos as well as quarks and gluons.

\subsection{Virtual Loop Corrections}

Since the virtual loop contributions are ultraviolet and infrared divergent,
we regularize the cross section by computing the phase space and matrix
elements in $n=4-2\epsilon$ dimensions.
We calculate the traces of Dirac matrices using the ``naive'' $\gamma_5$
scheme in which $\gamma_5$ anticommutes with all other $\gamma_{\mu}$
matrices. This choice is justified for anomaly-free one-loop amplitudes.
The $\gamma_5$ matrix enters the calculation through both the
quark-squark-gluino and quark-squark-gaugino couplings.
We simplify the integration over the internal loop momenta by
reducing all tensorial integration kernels to expressions that are only
scalar in the loop momentum \cite{pasvelt}.
The resulting one-, two-, three-, and some of the four-point functions were 
computed in the context of other physical processes \cite{squarkgluino}.
However, we compute two previously unknown divergent four-point functions; 
these new functions arise because the final state gluino and gaugino 
generally have 
different masses.  We evaluate the scalar four-point functions 
by calculating the absorptive parts with Cutkosky cutting rules and the real 
parts with dispersion techniques.

The ultraviolet (UV) divergences are manifest in the one- and two-point 
functions as poles in $1/\epsilon$.  We remove them by renormalizing the 
coupling constants in the $\overline{\rm MS}$ scheme at the renormalization 
scale $Q$ and the masses of the heavy particles (squarks and gluinos) in the 
on-shell scheme.  The self-energies for external particles are multiplied by 
a factor of $1/2$ for proper wave function renormalization.  A difficulty 
arises from the fact that spin-1 gluons have $n-2$ possible polarizations, 
whereas spin-1/2 gluinos have 2, leading to broken supersymmetry in the 
$\overline{\rm MS}$ scheme.  The simplest procedure to restore supersymmetry 
is with finite shifts in the quark-squark-gluino and quark-squark-gaugino 
couplings~\cite{mismatch}.

In addition to the ultraviolet singularities, the virtual corrections have
collinear and infrared singularities that show up as $1/\epsilon$ or
$1/\epsilon^2$ poles in the derivatives of the two-point function and in the 
three- and four-point functions.
These infrared singularities appear as factors times parts of the Born matrix
elements. They can be separated into $C_F$ and $N_C$ color classes, depending
on the color flow and the Abelian or non-Abelian nature of the correction
vertices. They are cancelled eventually by corresponding soft and collinear
singularities from the real three particle final state corrections.

\subsection{Real Emission Corrections}

The real corrections to the production of gluinos and gauginos arise from 
three particle final-state subprocesses in which additional
gluons and massless quarks and antiquarks are emitted: 
$q\bar{q}\ra\gluino\gaugino g$, $qg\ra\gluino\gaugino q$, and
$\bar{q}g\ra\gluino\gaugino\bar{q}$. 

The $n$-dimensional phase space for $2\ra 3$ scattering may be 
factored into the phase space for $2\ra 2$ scattering and the phase space 
for the subsequent decay of one of the two final state particles with squared 
invariant mass $s_4=(p_1+p_3)^2-m_1^2$ into two particles with momenta
$p_1$ and $p_3$, parametrized in the rest frame of particles
1 and 3 \cite{hquark}.
We follow the procedure of \cite{hquark} and reduce all of the
angular integrals.
to the form
\bea
  I_n^{(k,l)} &=& \int_0^\pi \sin^{1-2\epsilon}(\theta_1)\, d\theta_1\,
  \int_0^\pi \sin^{-2\epsilon}(\theta_2) \,d\theta_2 \\
  & \times & (a + b \cos\theta_1)^{-k}
  ( A + B\cos\theta_1 + C\sin\theta_1\cos\theta_2)^{-l} \nonumber.
\eea
Analytic expressions for the integrals $I_n^{(k,l)}$ for different $k,l$ may
be found in \cite{hquark}. 
The angular integrations involving negative powers
of $t'=(p_b-p_3)^2$ and $u'=(p_a-p_3)^2$, where $p_a$ and $p_b$ are the
four-momenta of the incoming partons, produce poles in $1/\epsilon$ which
correspond to the collinear singularities in which particle 3 is collinear
with particle $a$ or $b$.  
Because these singularities have a universal structure,
they may be removed from the cross section and absorbed into the parton
distribution functions according to the usual mass factorization procedure
\cite{facto}.

In addition to the collinear singularities described above, the corrections
involving real gluon emission also have infrared (IR) singularities arising
when the energy of the emitted gluon approaches zero.  These singularities
appear as poles in $s_4$ in the cross section and must also be extracted so
that they can be combined with corresponding terms in the virtual corrections
and shown to cancel. In order to make this cancellation conveniently, we
slice the gluon emission phase space into hard and soft pieces,
\begin{equation}
  \label{softhard}
  \frac{d^2\hat{\sigma}_{ij}^R}{dt_\gaugino du_\gaugino} =
  \int_0^{\Delta} ds_4 \,\frac{d^3\sigma^S}{dt_\gaugino\, du_\gaugino\, ds_4} +
  \int_\Delta^{{s_4}^{max}} ds_4 \,\frac{d^3\sigma^H}{dt_\gaugino\,
  du_\gaugino \, ds_4},
\end{equation}
where $\Delta$ is an arbitrary cut-off between soft and hard
gluon radiation. When the cut-off is much smaller than the other invariants,
the $s_4$ integration for the soft term becomes simple and can be
evaluated analytically, leading to explicit logarithms of the form
$\log\Delta/m_{\gluino}^2,
\log^2\Delta/m_{\gluino}^2$. The hard term is free 
from infrared and, after mass
factorization, also collinear singularities and can be evaluated numerically
in four dimensions. This procedure leads to an implicit logarithmic dependence 
of the hard term on the cut-off $\Delta$ which cancels the explicit logarithmic
dependence in the soft term.

\section{NLO Inclusive Cross Sections}

To obtain numerical results for the cross sections, we work within a 
particular SUGRA scheme, though the cross sections depend principally on the 
masses of the $\gaugino$ and $\gluino$ and are otherwise fairly independent of 
the details of the SUSY breaking.  
The physical gluino and gaugino masses as well as the gaugino mixing matrices
are calculated from the minimal SUGRA scenario.
We choose
the common scalar and gaugino masses at the GUT scale to be $m_0 = 100$ GeV,
$m_{1/2} = 150$ GeV, trilinear coupling $A_0 = 300$ GeV, and the
ratio of the Higgs vacuum expectation values 
$\tan\beta = 4$. The absolute value
of the Higgs mass parameter $\mu$ is fixed by electroweak symmetry breaking,
and we choose $\mu > 0$. For this set of parameters, 
we find the neutralino masses
$m_{\neutralino_{1-4}}$ to be 55, 104, 283, and 309 GeV with
$m_{\neutralino_3} < 0$ inside a polarization sum. The chargino masses
$m_{\chargino_{1,2}}$ are 102 and 308 GeV and therefore almost degenerate
with the masses of the $m_{\neutralino_2}$ and $m_{\neutralino_4}$, 
respectively.  This is a fairly general feature of the mSUGRA spectrum.

The total hadronic cross section is obtained from the
partonic cross section through the convolution 
\bea
 \sigma^{h_1h_2}(S,Q^2) &=& \sum_{i,j=g,q,\overline{q}}
 \int_{\tau}^1{\rm d}x_1\int_{\tau/x_1}^1{\rm d}x_2 \\
 && f_i^{h_1}(x_1,Q^2) f_j^{h_2}(x_2,Q^2)
 \hat{\sigma}_{ij}(x_1x_2S,Q^2), \nonumber
\label{e4}
\eea
where $\tau = \frac{4m^2}{S}$, with $m = (m_{\gluino} + m_{\gaugino}) / 2$,
and $S$ is the square of the hadronic center-of-mass energy 
For the NLO predictions, we employ the CTEQ4M 
parametrization \cite{cteq4} for the parton densities $f(x,Q^2)$ in
the proton or antiproton and a two-loop approximation for the strong
coupling constant $\alphas$ with $\Lambda^{(5)}=202$ MeV. 
To compute LO quantities we use the CTEQ4L
LO PDF's and the one-loop approximation for $\alphas$
with $\Lambda^{(5)}=181$ MeV.  

In Figures~\ref{xsec1} and \ref{xsec2}
we present predictions for total hadronic cross sections 
at the Tevatron and the LHC, as a function of the physical 
gluino mass.  To obtain these results, we use the average produced mass as the 
hard scale $Q$ in Equation~\ref{e4}, 
$Q = m /2$.  We 
vary the SUGRA parameter $m_{1/2}$ between 100 and 400 GeV 
and keep the other SUGRA parameters fixed at the values listed above.  As 
the gluino mass increases over the 
range shown in the figure, the corresponding 
gaugino mass ranges are
31 to 163 GeV for $\neutralino_1$, 62 to 317 GeV for $\neutralino_2$ and
$\chargino_1$, 211 to 666 GeV for $\neutralino_3$, and 240 to 679 GeV for
$\neutralino_4$ and $\chargino_2$. The chargino cross sections are summed over
production of positive and negative charges. 

\begin{figure}
 \begin{center}
  \epsfysize=4.0in
  \centerline{\epsfbox{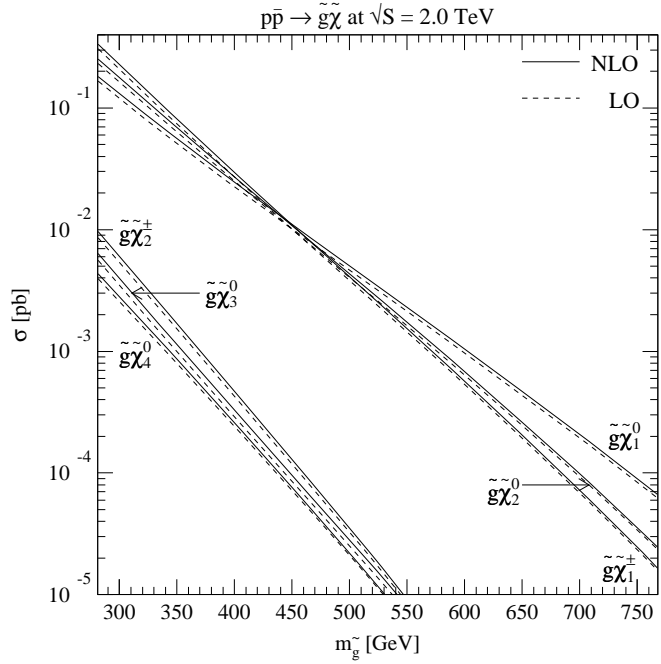}}
 \end{center}
 \caption{Total hadronic cross sections for the associated production
  of gluinos and gauginos at Run II of the Tevatron.
  NLO results are shown as solid curves, and LO results as 
  dashed curves.  We vary the SUGRA scenario as a function of $m_{1/2} \in
  [100;400]$ GeV and provide the cross sections as a function of the
  physical gluino mass $\mg$. The chargino cross sections are summed
  over positive and negative chargino rates.}
\label{xsec1}
\end{figure}

\begin{figure}
 \begin{center}
  \epsfysize=4.0in
  \centerline{\epsfbox{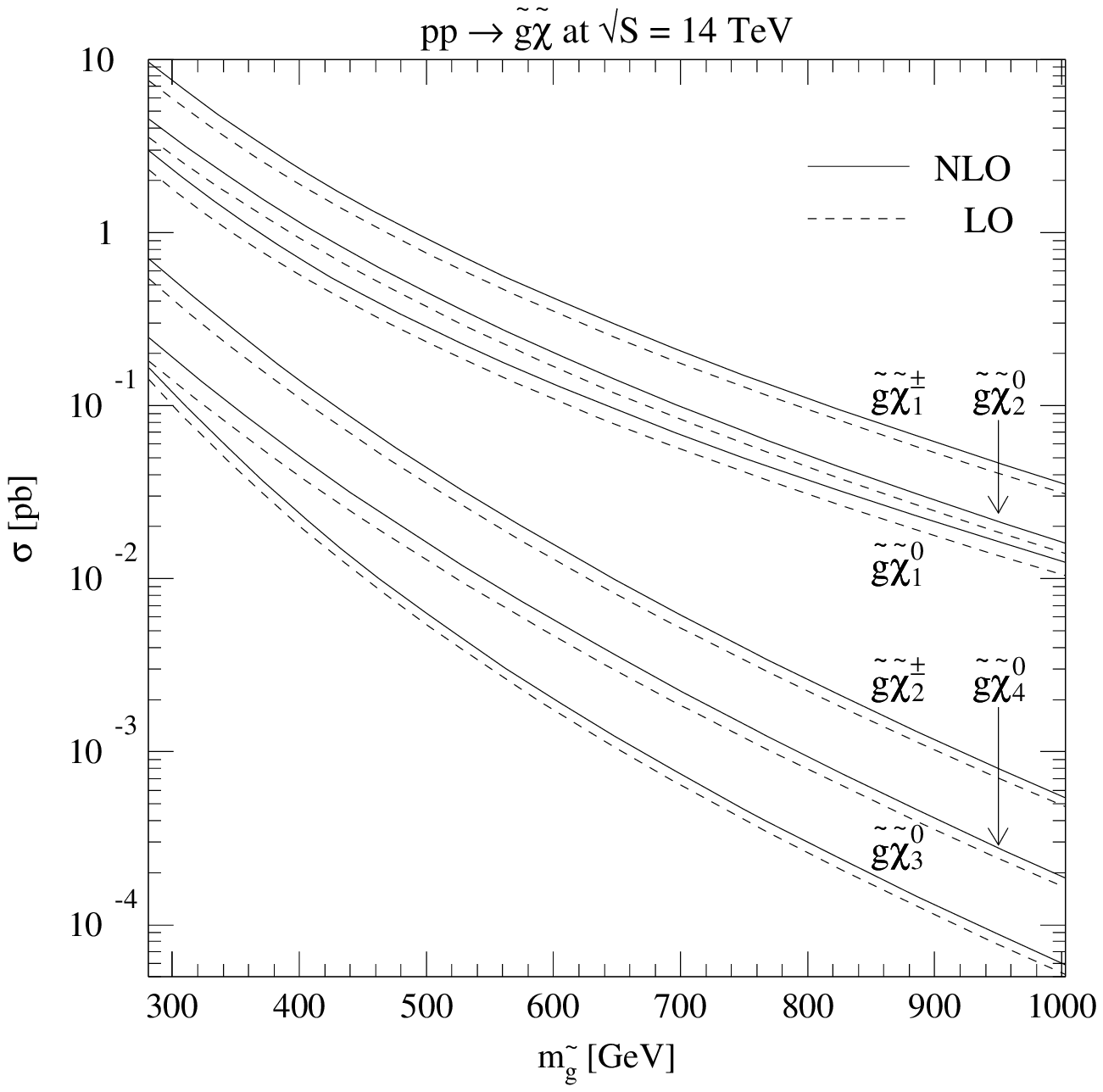}}
 \end{center}
 \caption{Total hadronic cross sections for the associated production
  of gluinos and gauginos at the LHC.
  NLO results are shown as solid curves, and LO results as 
  dashed curves.  We vary the SUGRA scenario as a function of $m_{1/2} \in
  [100;400]$ GeV and provide the cross sections as a function of the
  physical gluino mass $\mg$. The chargino cross sections are summed
  over positive and negative chargino rates.}
\label{xsec2}
\end{figure}

We observe that the cross sections for $\neutralino_2$ and $\chargino_1$ and 
those for $\neutralino_4$ and $\chargino_2$ are very similar in magnitude at 
the Tevatron, as are their respective masses. One might expect the largest 
cross section
for the lightest gaugino $\neutralino_1$. However, its coupling is dominantly
$B$ino-like and smaller than the $W_3$ino-like 
coupling\footnote{The $B$ino and $W_3$ino are the superpartners of the
$B$ and $W_3$ gauge bosons discussed in Chapter~\ref{intro}.}
of $\neutralino_2$ which 
therefore has a larger cross section at small $\mg$ despite its larger
mass. The heavier gauginos $\neutralino_{3,4}$ and $\chargino_2$ are dominantly
higgsino-like and their cross sections are suppressed by more than 
an order of 
magnitude with respect to those of the lighter gauginos.

At the Tevatron, the NLO contributions increase the cross sections by 5 to 
$15\%$ at the hard scattering scale $Q = (m_{\tilde g} + m_{\tilde \chi})/2$, 
depending on the channel considered and the values of the masses.  At 
the LHC, the increases are in the range of 15 to $35\%$.   
The purely NLO $q g$ 
incident channel contributes significantly at the LHC, in addition to 
the $q \bar{q}$ channel, particularly for the lighter gauginos, whereas 
the $q g$ channel plays an insignificant role at the Tevatron.  
In the event sparticles 
are not observed, the predicted increases translate into more restrictive 
experimental mass limits.

The enhancements of the cross sections are modest and, as such, underscore the 
validity of perturbative predictions for the processes considered.  A further 
important benefit of the NLO computation is the considerable reduction in 
theoretical uncertainty associated with variation of the renormalization
and factorization scale $Q$.  For the processes studied here, this dependence 
is typically $\pm 10\%$ at the Tevatron when $Q$ is varied over the interval 
$Q/m$ from 0.5 to 2, compared to $\pm 25\%$ in leading order.  At the LHC, 
the dependences are $\pm 9\%$ at NLO and $\pm 12\%$ at LO.  

We limit ourselves to total cross sections. Differential
distributions in the transverse momentum $p_T$ and the rapidity $\eta$ of
the produced sparticles will be published elsewhere \cite{longpaper}, along 
with figures of scaling functions, renormalization/factorization scale 
dependence, K-factors, and several appendices containing a detailed exposition 
of the calculation. 

\section{Summary}

In summary, we provide NLO predictions of the cross sections for 
the associated production of gauginos and gluinos at hadron colliders.
If supersymmetry exists at the electroweak scale, the cross section for
this process is expected to be large and observable at the Fermilab Tevatron
and/or the CERN LHC. It is enhanced by the large color charge of the gluino
and the (in many SUSY models) small mass of the light gauginos.
The cross sections for $\gluino \neutralino_2$ and 
$\gluino \chargino_1$ production are comparable, and the largest, because
of their
$W$ino-like couplings.  As we have seen,
the NLO predictions are modestly larger than the 
LO values but considerably more stable.

\chapter{Conclusions}
\label{conclusions}

In this work, we have seen that the Standard Model of particle physics,
while fabulously successful at describing high energy physics experiments,
suffers from a number of puzzles that indicate that it is not a fundamental
theory, but should be replaced by something else to describe the 
physics at very short distances.  The primary puzzle confronting particle
physicists today is the understanding of the electroweak symmetry breaking,
responsible for the large masses of the weak bosons and the top quark.
The fact that the top is so much heavier than the other fermions seems to
indicate that it may play some special role.  Its large mass further
indicates that it is
a natural laboratory to test hypotheses concerning the nature of the
symmetry breaking.

If the top does play a special role in nature, one must discover this fact
through careful study of its properties.  In particular, the electroweak
interactions are likely to feel the effect of the true mechanism for the
weak symmetry breaking, and are perhaps the most interesting properties
to examine.  Single top production is a vital means to study these weak
interactions at a hadron collider, and thus we have spent considerable
time describing the physics of single top production, to see how one
can hope to use it as a tool to study the top's electroweak interactions.
We have seen that the three modes of single top production, along with
studies of top decays and top polarization, represent a wealth of
information about the top quark.

We have further studied the bottom quark, whose special partnership with
the top may allow it to inherit some of the top's nonstandard properties.
In particular, we have seen that this can result in an enhancement of the
bottom coupling to scalar particles.  It has been demonstrated that
processes involving associated production of Higgs bosons with bottom
quarks can provide interesting information about a wide class of models
of the weak symmetry breaking, from supersymmetric theories to theories
with dynamical EWSB.

We have also seen that the superpartners of the electroweak gauge bosons,
the charginos and neutralinos can be produced in association with
the gluino, superpartner to the gluon.  This process provides an
interesting means to search for evidence of supersymmetry at hadron
colliders such as the Tevatron and LHC.  In order to obtain a reliable
theoretical prediction for the cross section, one must include higher
orders in SUSY QCD.  We have shown that the NLO corrections to this
process are fairly large, and dramatically increase the stability of
the theoretical prediction, indicating the necessity to include them.

With the advent of the Tevatron Run~II, to be followed in a few years
time by the LHC, we stand on the threshold of a wealth of information
concerning the mechanism of the EWSB.  The processes described above
represent vital means to interpret this information.  Regardless of whether
one favors one particular description or another, it is an exciting time,
as our conceptions of the symmetry breaking are confronted with reality.

\singlespacing

\end{document}